\documentclass[aps, prd, superscriptaddress, preprintnumbers, floatfix, longbibliography,10pt,twocolumn]{revtex4-2}

\usepackage{amsfonts,amsmath,amssymb,amstext}
\usepackage{graphicx,epstopdf}
\usepackage{float,wrapfig}
\usepackage{subfigure, psfrag}
\usepackage{dsfont,bm}
\usepackage{color}
\usepackage[normalem]{ulem}
\usepackage[colorlinks=true,urlcolor=blue,citecolor=blue,linkcolor=blue]{hyperref}
\hypersetup{breaklinks=true}
\usepackage{verbatim}

\newcommand{\df}[1]{\,\delta{\left(#1\right)}}

\usepackage{array}
\newcolumntype{L}[1]{>{\raggedright\let\newline\\\arraybackslash\hspace{0pt}}m{#1}}
\newcolumntype{C}[1]{>{\centering\let\newline\\\arraybackslash\hspace{0pt}}m{#1}}
\newcolumntype{R}[1]{>{\raggedleft\let\newline\\\arraybackslash\hspace{0pt}}m{#1}}

\definecolor{purple}{rgb}{0.8,0,0.6}
\definecolor{darkgreen}{rgb}{0.00,0.6,0.00}

\usepackage[usenames,dvipsnames]{xcolor}

\extrafloats{100}

\begin{document}

\title{Viscoelastic response and anisotropic hydrodynamics in Weyl semimetals}
\date{July 23, 2024}

\author{A.~A.~Herasymchuk}
\affiliation{Faculty of Physics, Taras Shevchenko National University of Kyiv, 64/13 Volodymyrska st., 01601 Kyiv, Ukraine}

\author{E.~V.~Gorbar}
\affiliation{Faculty of Physics, Taras Shevchenko National University of Kyiv, 64/13 Volodymyrska st., 01601 Kyiv, Ukraine}
\affiliation{Bogolyubov Institute for Theoretical Physics, 14-b Metrolohichna st., 03143 Kyiv, Ukraine}

\author{P.~O.~Sukhachov}
\email{pavel.sukhachov@gmail.com}
\affiliation{Center for Quantum Spintronics, Department of Physics, Norwegian University of Science and Technology, NO-7491 Trondheim, Norway}
\affiliation{Department of Physics and Santa Cruz Materials Center, University of California Santa Cruz, Santa Cruz, California 95064, USA}
\affiliation{Department of Physics, Yale University, New Haven, Connecticut 06520, USA}

\begin{abstract}
We study viscoelastic response in Weyl semimetals with broken time-reversal symmetry. The principal finding is that topology and anisotropy of the Fermi surface are manifested in the viscoelasticity tensor of the electron fluid. In the dynamic (interband) part of this tensor, the anisotropy leads to a qualitatively different, compared with isotropic models, scaling with frequency and the Fermi energy. The components of the viscosity tensor determined by the Fermi-surface properties agree in the Kubo and kinetic formalisms; the latter, however, misses the anomalous Hall viscosity originating from filled states below the Fermi surface. The anisotropy of the dispersion relation is also manifested in the acceleration and relaxation terms of the hydrodynamic equations providing means to probe the anisotropy in transport experiments.
\end{abstract}

\maketitle

\section{Introduction}
\label{sec:Introduction}

Experimental discovery of hydrodynamic transport in graphene~\cite{Crossno-Fong:2016,Ghahari-Kim:2016,Krishna-Falkovich:2017,Berdyugin-Bandurin:2018,Bandurin-Falkovich:2018,Ku-Walsworth:2019,Sulpizio-Ilani:2019,Samaddar-Morgenstern:2021,Kumar-Ilani:2021} reignited the interest to electron hydrodynamics first observed in GaAs heterostructures~\cite{Molenkamp-Jong:1994,Jong-Molenkamp:1995}. This is particularly fascinating since the foundation of electron hydrodynamics was laid down in the 1960s in the seminal works of R.~Gurzhi~\cite{Gurzhi:1963,Gurzhi:1968}. Being an interesting regime with strong electron-electron interactions, electron hydrodynamics has several manifestations including the Gurzhi effect~\cite{Gurzhi:1963}, Poiseuille profile of electric current, formation of vortices of electron fluid, etc.; see Refs.~\cite{Lucas-Fong:rev-2017,Narozhny:rev-2019,Narozhny:rev-2022} for a review.

A characteristic and very important property of any fluid is its viscosity. Viscosity is a rank-4 tensor that relates the fluid stress to coordinate derivatives of the fluid velocity. In continuum mechanics, the viscoelastic stress tensor combines the response to strain (elasticity) and the time derivative of strain (viscosity); in what follows, we will use viscoelastic and viscous tensors interchangeably. In systems with rotational invariance, such as most fluids, the structure of the shear viscosity tensor is fixed by symmetry, and viscous response is determined only by two coefficients~\cite{Landau:t6-2013}: bulk and shear viscosities. However, solid-state materials are rarely isotropic where the rotational symmetry is commonly reduced only to certain angles of rotation.

Manifestations of anisotropy of the band dispersion in electron hydrodynamics were considered in Refs.~\cite{Link-Schmalian:2017,Rao-Bradlyn:2019,Offertaler-Bradlyn:2019,Cook-Lucas:2019,Pena-Benitez-Surowka:2019,Varnavides-Narang:2020,Robredo-Bradlyn:2021,Huang-Lucas:2022}. In anisotropic metals, the reduced symmetry can lead to unusual viscosity tensors with additional components.
Using a lattice regularization proposed in Ref.~\cite{Rao-Bradlyn:2019}, the nondissipative Hall viscosity was analyzed in~\cite{Robredo-Bradlyn:2021}, however, other components of the viscosity tensor and the dynamic viscosity received less attention.

In addition to the anisotropy of the band dispersion, certain materials allow for the nontrivial topology of their electron wave functions. Topological aspects of viscoelastic response were addressed in Refs.~\cite{Sun-Wan:2014,Shapourian-Ryu:2015,Cortijo-Vozmediano:2015,Link-Schmalian:2018,Rao-Bradlyn:2019,Offertaler-Bradlyn:2019,Copetti-Landsteiner:2019,Moore-Roy:2020,Varnavides-Narang:2020,Robredo-Bradlyn:2021}. The main attention, however, was paid to nondissipative or Hall responses~\cite{Avron-Zograf:1995,Read:2009,Hughes-Fradkin:2011,Hoyos-Son:2012,Rao-Bradlyn:2019} in mainly two-dimensional (2D) systems. Three-dimensional (3D) topological materials exemplified by Weyl semimetals may also demonstrate nontrivial hydrodynamic response related to their topology~\cite{Gorbar:2017vph,Gorbar:2018vuh,Gorbar:2018sri,Gorbar:2018nmg,Bednik-Syzranov:2019,Sukhachov-Trauzettel:2021,Zhu-Syzranov:2022,Bernabeu-Cortijo:2023a}; experimental signatures of the hydrodynamic behavior in the Weyl semimetal WP$_2$ were reported in Ref.~\cite{Gooth-Felser:2018}. As an example of the topology manifestation in the viscoelasticity tensor in time-reversal symmetry (TRS) broken Weyl materials, we mention the anomalous Hall viscosity determined by the separation of the Weyl nodes in the Brillouin zone~\cite{Cortijo-Vozmediano:2015}. However, the calculation of the complete viscoelastic response that goes beyond the anomalous Hall viscosity, includes both dissipative and nondissipative parts, and accounts for the Weyl node topology in 3D Weyl semimetals as well as their Fermi-surface anisotropy is still lacking.

In this paper, we fill this gap and calculate the viscoelastic response of TRS-broken Weyl semimetals. One of our main findings is that the band structure of Weyl semimetals, exemplified by the separation of the Weyl nodes in the Brillouin zone, and their anisotropy is manifested in both nondissipative (i.e., anomalous Hall) and dissipative components of the viscoelasticity tensor. Using the Kubo framework~\cite{Bradlyn-Read:2012}, we calculate both static and dynamic viscoelastic responses in linearized and two-band models of Weyl semimetals. In our calculations, we use two formulations of the stress tensor:~\footnote{Two formulations of the stress tensor were recently used in Ref.~\cite{Robredo-Bradlyn:2021} for chiral magnetic materials, which realize the so-called spin-1 fermions.} the canonical stress tensor, which follows from the coupling to mechanical strains~\cite{Shapourian-Ryu:2015,Cortijo-Vozmediano:2015}, and the Belinfante-like tensor, which follows from the momentum continuity equation and includes contributions connected with the internal angular momentum~\cite{Link-Schmalian:2018,Rao-Bradlyn:2019}; it also agrees with the result obtained via Noether's second theorem. We found that both definitions of the stress tensor give the same result for the static part of the viscoelastic response but differ, in general, for the dynamic part. Physically, this corresponds to two different types of viscosity and can be probed by different types of observables: thermal transport and acoustic-phonon dispersion for the canonical stress tensor and hydrodynamic transport for the Belinfante-like tensor.

In the kinetic approach, our calculations of the static electron viscosity agree with those for both canonical and Belinfante stress tensors.
However, being sensitive only to the Fermi-surface contributions, the kinetic approach excludes the anomalous Hall viscosity, which has a topological origin; the latter is reproduced in the Kubo approach. The results for the dynamic (interband) viscoelastic response show a more dramatic difference between the two definitions of the stress tensor: internal degrees of freedom are crucial in restoring the standard tensor structure of the viscoelasticity tensor~\cite{Link-Schmalian:2018,Rao-Bradlyn:2019}. In addition to the viscoelasticity tensor, the anisotropy of the dispersion relation of Weyl materials quantified by the separation of the Weyl nodes is also manifested in acceleration and relaxation terms of the hydrodynamic equations. In view of their different dependence on the Fermi energy, these terms can be important in Weyl semimetals with well-separated nodes leading to the suppression of fluid flow along the separation vector.

The viscosity tensor can be mapped out via transport measurements akin to those discussed in Refs.~\cite{Link-Schmalian:2017,Holder-Stern:2019,Varnavides-Narang:2020}. Such transport measurements are readily accessible in graphene~\cite{Krishna-Falkovich:2017,Berdyugin-Bandurin:2018,Ku-Walsworth:2019,Sulpizio-Ilani:2019}; see also Refs.~\cite{Lucas-Fong:rev-2017,Narozhny:rev-2022} and references therein. As an example of such measurements, we mention the study of the current magnitude and its profile in crystals cut along different crystallographic directions. Experimentally accessible signatures of the anisotropy of the viscosity tensor are further discussed in Sec.~\ref{sec:Kinetics}.

The paper is organized as follows. In Sec.~\ref{sec:Kubo}, we summarize the Kubo framework for viscoelastic response and apply it to linearized and two-band models of 3D Dirac and Weyl materials. We rederive the hydrodynamic equations and cross-verify the viscosity tensor in the kinetic framework in Sec.~\ref{sec:Kinetics}. The results are discussed and summarized in Sec.~\ref{sec:Summary}. A few technical details are presented in Appendixes~\ref{sec:App-stress-tensor}, \ref{sec:App-Spectral-function}, \ref{sec:App-Kubo}, \ref{sec:App-zero-kinetic}, and \ref{sec:App-first-kinetic}. Throughout this paper, we use $\hbar=k_{\rm B}=1$.

\section{Kubo approach to viscoelastic response}
\label{sec:Kubo}

\subsection{Viscoelasticity tensor}
\label{sec:Kubo-general}

Besides the linear response to external electromagnetic fields, the Kubo approach can be applied to calculating other responses of the system including the viscoelastic response to static and time-dependent strain. The viscoelasticity tensor is defined via the stress tensor $T_{\mu\nu}$ and the strain transformation generator $J_{\mu\nu}$ (see, Appendix~\ref{sec:App-stress-tensor}) as follows:~\cite{Bradlyn-Read:2012}
\begin{equation}
\begin{aligned}
\eta_{\mu \nu \alpha \beta}(\Omega)&=\frac{1}{\Omega+i0} \left[  \left< [T_{\mu \nu}^{\,\,\,}(0), J_{\alpha \beta}^{\,\,\,}(0)] \right>+i\delta_{\alpha \beta}^{\,\,\,} \left< T_{\mu \nu}^{\,\,\,} \right> \right.\\
&\left.-i  \delta_{\mu \nu} \delta_{\alpha \beta}\kappa^{-1} - i C_{\mu \nu \alpha \beta}(\Omega)\right],
\end{aligned}
\end{equation}
where $C_{\mu \nu \alpha \beta}(\Omega)$ is the Fourier transform of the stress-stress correlation function
\begin{equation}
\begin{aligned}
C_{\mu \nu \alpha \beta}(t-t') &=i \lim_{\varepsilon \to +0} \Theta(t-t') \left< [T_{\mu \nu}^{\,\,\,}(t), T_{\alpha \beta}^{\,\,\,}(t')] \right> \\
&\times e^{-\varepsilon (t-t')}.
\end{aligned}
\end{equation}
In writing the above expressions, we assumed the long-wavelength limit, i.e., the response to uniform strains, which allowed us to set the wave vector of perturbation to zero; this is explicitly manifested in Eq.~(\ref{Kubo_viscoelastic_tensor}) below. Unfortunately, the stress tensor and strain transformation generator are not uniquely defined and different expressions are used in the literature. The key issue is the form and structure of the pseudospin angular momentum, which is connected to the internal degrees of freedom. Furthermore, $\kappa^{-1}=-V\left(\partial P/\partial V \right)_{N}$ is the inverse isentropic compressibility defined as the derivative of pressure with respect to volume of fluid $V$ taken at fixed particle number $N$ and $\Omega$ is frequency.~\footnote{As is discussed in Ref.~\cite{Bradlyn-Read:2012}, it is crucial to subtract the inverse isentropic compressibility in order to obtain finite bulk viscosity.}

By using the spectral function $A(\omega;\mathbf{k})$, see Appendix~\ref{sec:App-Spectral-function} for its definition, in the stress-stress correlator, the viscoelasticity tensor can be rewritten as follows~\cite{Bradlyn-Read:2012}:
\begin{widetext}
\begin{equation}
\label{Kubo_viscoelastic_tensor}
\begin{aligned}
\eta_{\mu \nu \alpha \beta}(\Omega)&=\frac{1}{\Omega+i0} \bigg\{ \left[  \left< [T_{\mu \nu}^{\,\,\,}(0), J_{\alpha \beta}^{\,\,\,}(0)] \right>+i\delta_{\alpha \beta}^{\,\,\,} \left< T_{\mu \nu}^{\,\,\,} \right>  -i  \delta_{\mu \nu} \delta_{\alpha \beta}\kappa^{-1} \right] \\
&- i \int_{-\infty}^{+\infty} d \omega \int_{-\infty}^{+\infty} d \omega' \frac{f(\omega)-f(\omega')}{\omega'-\omega-\Omega -i 0}  \int \frac{d^3 k}{(2\pi)^3}  \text{tr} \left[ T_{\mu \nu}^{\,\,\,} (\mathbf{k}) A(\omega;\mathbf{k}) T_{\alpha \beta}^{\,\,\,} (\mathbf{k}) A(\omega';\mathbf{k}) \right] \bigg\}.
\end{aligned}
\end{equation}
\end{widetext}
Here, $f(\omega)=1/\left[e^{\left(\omega - \mu\right)/T}+1\right]$ is the Fermi-Dirac distribution function, $\mu$ is the chemical potential, $T$ is temperature. Henceforth, to simplify our calculations, we assume the zero-temperature limit $T\to0$ unless otherwise stated.

\begin{figure}
\includegraphics[width=0.75\columnwidth]{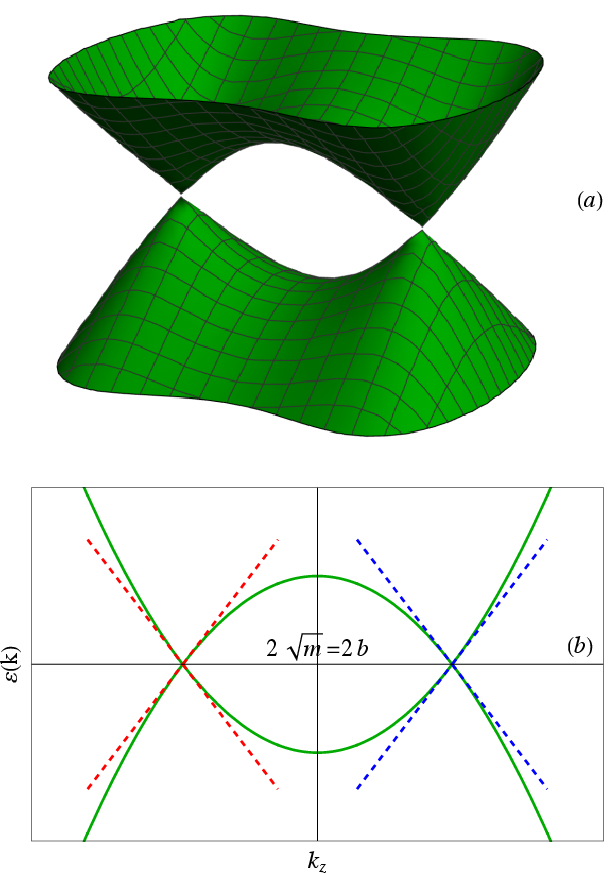}
\caption{Panel (a): The 3D dispersion relation $\varepsilon=\varepsilon\left(k_x,0,k_z\right)$ for the two-band model of Weyl semimetals (\ref{Kubo-Weyl-two-band-H-def}).
Panel (b): The 2D dispersion relation $\varepsilon\left(0,0,k_z\right)$ for the two-band model of Weyl semimetals shown by solid green lines with the dashed lines corresponding to the dispersion relation in the linearized model (\ref{Model-Weyl-lin}). The distance between the Weyl nodes is $2b=2\sqrt{m}$.
}
\label{fig_dispersion}
\end{figure}

In the following few sections, we calculate the viscoelasticity tensor (\ref{Kubo_viscoelastic_tensor}) for linearized models of Dirac and Weyl semimetals as well as more a realistic two-model of Weyl semimetals; the corresponding electron energy dispersions for Weyl semimetals are shown in Fig.~\ref{fig_dispersion}.

\subsection{Dirac semimetals}
\label{sec:Kubo-Dirac}

As a warm-up, let us start with Dirac semimetals whose linearized Hamiltonian for gapless electron quasiparticles reads
\begin{equation}
\label{Kubo-Dirac-H-def}
H=\text{diag}\left(H_{+},H_{-} \right),\,\,\,\,H_{\lambda}=\lambda v_F \bm{\sigma}\cdot\mathbf{k},
\end{equation}
where $\lambda=\pm$ is chirality, $v_F$ is the Fermi velocity, and $\bm{\sigma}$ are the Pauli matrices in the pseudospin space. This Hamiltonian possesses full rotation symmetry, therefore, we use the following strain transformation generator and symmetric \emph{Belinfante stress tensor}~\cite{Belinfante:1940,Rosenfeld:1940}:
\begin{align}
\label{Kubo-Dirac-J}
J_{\alpha \beta }&=-\frac{1}{2}\left\{ x_{\beta },k_{\alpha} \right\}+\frac{i}{8} \left[ \gamma_{\beta} , \gamma_{\alpha}\right],\\
\label{Kubo-Dirac-T}
T_{\mu \nu}&=\frac{ v_F}{2}\left( \gamma_0 \gamma_{\nu} k_{\mu}+\gamma_0 \gamma_{\mu} k_{\nu} \right),
\end{align}
respectively. Here, $\gamma_{\alpha}$ are the standard Dirac-$\gamma$ matrices, $x_{\beta}$ is the coordinate operator, $\left\{,\right\}$ is the anticommutator, and $\left[,\right]$ is the commutator. By calculating the real part of the viscoelasticity tensor in the Kubo approach, Eq.~(\ref{Kubo_viscoelastic_tensor}), we find
\begin{equation}
\label{Kubo-Dirac-eta}
\begin{aligned}
\text{Re}\,\eta_{\mu \nu \alpha \beta}^{\text{(B)}}(\Omega) &=\left[\frac { \Omega^3}{320 \pi  v_F^3  } \Theta\left(\Omega-2\mu\right) +  \frac {\mu^4 }{15\pi v_F^3} \delta (\Omega) \right]\\
&\times \left(\delta_{\mu \alpha} \delta_{\nu \beta}+\delta_{\nu \alpha} \delta_{\mu \beta} -\frac{2}{3} \delta_{\mu \nu} \delta_{\alpha \beta} \right).
\end{aligned}
\end{equation}
For a rotationally-invariant $d$-dimensional system, the viscoelasticity tensor has only two independent components~\cite{Landau:t6-2013}
\begin{equation}
\eta_{\mu \nu \alpha \beta} =\zeta\delta_{\mu \nu}\delta_{\alpha \beta}+\eta^{\text{sh}}\left( \delta_{\mu \alpha}\delta_{\nu \beta}+\delta_{\mu \beta}\delta_{\nu \alpha}-\frac{2}{d}\delta_{\mu \nu}\delta_{\alpha \beta}  \right).
\end{equation}
Comparing with Eq.~(\ref{Kubo-Dirac-eta}), we obtain the trivial bulk viscosity $\zeta=0$ and the following shear viscosity:
\begin{equation}
\eta^{\text{sh}} =\frac { \Omega^3}{320 \pi  v_F^3  } \Theta \left(\Omega-2\mu \right) +  \frac {\mu^4 }{15\pi v_F^3} \delta (\Omega).
\end{equation}
These findings agree with the results obtained in Refs.~\cite{Moore-Roy:2020,Sukhachov-Trauzettel:2021}.

By starting with a lattice model and studying the response to strain, one obtains a different strain transformation generator~\cite{Shapourian-Ryu:2015}
\begin{equation}
J_{\alpha \beta }=-\frac{1}{2}\left\{ x_{\beta },k_{\alpha} \right\},
\end{equation}
which, compared with Eq.~(\ref{Kubo-Dirac-J}), does not contain the pseudospin part. This strain transformation generator leads to a nonsymmetric \emph{canonical stress tensor}
\begin{equation}
\label{Kubo-Dirac-T-canon}
T_{\mu \nu}=v_F \gamma_0 \gamma_{\nu} k_{\mu},
\end{equation}
and, as a result, to a different viscoelasticity tensor
\begin{eqnarray}
\label{Kubo-Dirac-eta-1}
\text{Re}\,\eta_{\mu \nu \alpha \beta}^{\text{(C)}}(\Omega)&=&
\frac{ \Omega^3}{192 \pi v_F^3} \Theta\left(\Omega-2\mu\right) \left(  \delta_{\mu \alpha}  \delta_{\nu \beta} - \delta_{\mu \beta} \delta_{\nu \alpha}   \right)\nonumber\\
&+&\text{Re}\,\eta_{\mu \nu \alpha \beta}^{\text{(B)}}(\Omega),
\end{eqnarray}
cf. Eq.~(\ref{Kubo-Dirac-eta}). Therefore, while the static (intraband) part of the viscoelasticity tensor is the same for symmetric Belinfante and nonsymmetric canonical stress tensors, the dynamic (interband) part is sensitive to the choice of the stress tensor.

Since the Belinfante and canonical stress tensor generically lead to different viscoelastic tensors, cf. Eqs.~(\ref{Kubo-Dirac-eta}) and (\ref{Kubo-Dirac-eta-1}), it is important to discuss the difference between them. Firstly, we note a different physical origin of these stress tensors. As we discuss in detail in Appendix~\ref{sec:App-1-canon}, the canonical stress tensor (\ref{Kubo-Dirac-T-canon}) is defined via mechanical deformations. Therefore, it is often called the phonon stress tensor. On the other hand, the Belinfante stress tensor follows from the momentum continuity equation with additional contributions due to internal angular degrees of freedom, see Appendix~\ref{sec:App-1-continuum}. Both approaches are also discussed and contrasted in Refs.~\cite{Rao-Bradlyn:2019,Robredo-Bradlyn:2021}.

These two different stress tensors correspond to two different forms of viscosity: the phonon viscosity for the canonical stress tensor and ``momentum" viscosity for the Belinfante stress tensor. The former is manifested in thermal transport and acoustic phonon dispersion. The momentum viscosity can be probed via hydrodynamic transport, see the discussion in Sec.~\ref{sec:Kinetics}. Therefore, these two different definitions of the stress tensor are manifested in different types of viscosities and, as a result, different observables.

\subsection{Linearized model of Weyl semimetals}
\label{sec:Kubo-Weyl-lin}

Now we are ready to address the viscoelasticity tensor in TRS-broken Weyl semimetals. We start with a simple linearized model which describes two Weyl nodes separated in momentum space
\begin{equation}
\label{Model-Weyl-lin}
    H=\text{diag}\left(H_{+},H_{-} \right),\,\,\,\,H_{\lambda}=\lambda v_F \bm{\sigma}\cdot \left(\mathbf{k}-\lambda \mathbf{b}\right),
\end{equation}
where, without the loss of generality, we choose the momentum space separation to be along the $z$-direction, $\mathbf{b}=b_z \mathbf{e}_z$.

We first consider the canonical stress tensor. The viscoelasticity tensor can be separated into three parts: the static (intraband) part, the dynamic (interband) part, and the Hall viscosity. The details of calculations and some intermediate results are given in Appendix~\ref{sec:App-Kubo-lin}.

The real part of the static part of the viscoelasticity tensor reads as
\begin{equation}
\label{Kubo-Weyl-eta-stat}
\begin{aligned}
\text{Re}\,\eta_{\mu \nu \alpha \beta}^{\text{stat}}(\Omega)&=  \frac {\mu^2}{3 \pi v_F }  \delta(\Omega) \delta_{\nu \beta}  b_{\mu } b_{\alpha }  +\frac{\mu^4}{15 \pi v_F^3} \delta(\Omega) \\
&\times \left( \delta_{\mu \beta} \delta_{\nu \alpha}+ \delta_{\mu \alpha} \delta_{\nu \beta}- \frac{2}{3} \delta_{\mu \nu} \delta_{\alpha \beta} \right).
\end{aligned}
\end{equation}
The static part in the linearized model gets an additional term proportional to $\propto b_{\mu} b_{\alpha}$, see the first term in Eq.~(\ref{Kubo-Weyl-eta-stat}). Therefore, the separation vector $\mathbf{b}$ modifies the dissipative components of the viscoelasticity tensor. We cross-verify this finding in Sec.~\ref{sec:Kubo-Weyl-two-band} for the two-band model as well as in Sec.~\ref{sec:Kinetics} in the kinetic approach.

The real part of the dynamic viscoelasticity tensor reads as
\begin{equation}
\label{Kubo-Weyl-eta-dyn}
\begin{aligned}
\text{Re}\,\eta_{\mu \nu \alpha \beta}^{\text{(C)},\text{dyn}}(\Omega)&= \Theta\left( \Omega -2\mu \right) \left[ \frac {\Omega }{12 \pi v_F}   \delta_ {\nu \beta}  b_\mu b_\alpha +\frac{\Omega^3}{120 \pi v_F^3} \right. \\
&\times \left. \left(  \delta_{\mu \alpha} \delta_ {\nu \beta}  - \frac{1}{4}    \delta_{\mu \beta} \delta_{\nu \alpha}   - \frac{1}{4} \delta_{\mu \nu} \delta_{\alpha \beta} \right) \right].
\end{aligned}
\end{equation}
Like the static part, it also contains an additional term proportional to $\propto b_{\mu} b_{\alpha}$, see the first term in the square brackets in Eq.~(\ref{Kubo-Weyl-eta-dyn}). As is easy to see, the limit $b \rightarrow 0$ reproduces the results for the Dirac semimetal obtained in the canonical stress tensor approach, cf. Eq.~(\ref{Kubo-Dirac-eta-1}).

The Hall viscosity is determined by the imaginary part of trace $\text{tr} \left[ T_{\mu \nu}^{\,\,\,} (\mathbf{k}) A(\omega;\mathbf{k}) T_{\alpha \beta}^{\,\,\,} (\mathbf{k}) A(\omega';\mathbf{k}) \right]$ in Eq.~(\ref{Kubo_viscoelastic_tensor}); see also Eqs.~(\ref{app-kubo-eta}) and (\ref{app-kubo-tr-Im}). In the limit $\Omega \rightarrow 0$ and at $\mu=0$, we have
\begin{equation}
\label{Kubo-Weyl-eta-Hall-0}
\text{Re}\,\eta_{\mu \nu \alpha \beta}^{\text{(C)},\text{Hall}}(\Omega)= 4\sum_{j=1}^{3} \epsilon_{j \nu \beta} I_{\mu \alpha}^{j},
\end{equation}
where
\begin{equation}
\label{Kubo-Weyl-I12}
I_{\mu \alpha}^{i=1,2}=\left(\delta_{\mu i} \delta_{\alpha 3} +\delta_{\mu 3} \delta_{\alpha i} \right) \frac{b_z}{4 \pi^2 } \left(\Lambda_z^2 -\frac{b_z^2}{3}\right)
\end{equation}
and
\begin{equation}
\label{Kubo-Weyl-I3}
\begin{aligned}
I_{\mu \alpha}^{3}&=\delta_{\mu \alpha} \frac{b_z}{4 \pi^2} \Bigg[ \left(\delta_{\mu 1} +\delta_{\mu 2} \right)\left( -\Lambda_{\perp} \Lambda_z +2\Lambda_{z}^2 +\frac{2b_z^2}{3} \right) \\
&-\delta_{\mu 3} \frac{b_z^2}{6} \Bigg],
\end{aligned}
\end{equation}
where $\Lambda_{\perp}$ and $\Lambda_z$ are momentum cutoffs. Evidently, all components are quadratically divergent except for $\eta_{zxzy}=-b_z^3/(6 \pi^2)$~\footnote{The cutoffs in Eqs.~(\ref{Kubo-Weyl-I12}) and (\ref{Kubo-Weyl-I3}) are connected with the contribution of the states below the Fermi surface. Such terms are absent in lattice models, see, e.g., Ref.~\cite{Robredo-Bradlyn:2021}.}. In the systems with axial symmetry, the other independent component of the Hall viscosity is given by $\eta_{xxxy}$. The obtained results agree with those in Ref.~\cite{Cortijo-Vozmediano:2015}. The expressions for $\Omega \neq 0$ are given in Eqs.~(\ref{viscoelasticity_lin_Hall_Omega_12}) and(\ref{viscoelasticity_lin_Hall_Omega_3}), where the $\eta_{zxzy}$ component of the viscoelasticity tensor is also divergent albeit only logarithmically, $\propto \Omega^2 b_z \ln{(2 v_F \Lambda_z/\Omega)}$.

Now let us take into account the internal degrees of freedom, which lead to the pseudospin angular momentum, and consider the \emph{Belinfante-like stress tensor}.~\footnote{Unlike isotropic Dirac spectrum, see Eq.~(\ref{Kubo-Dirac-T}), rotational invariance is commonly broken in crystals. Therefore, one should consider the modification of the Belinfante symmetrization~\cite{Rao-Bradlyn:2019} or the Belinfante-like stress tensor, which takes into account reduced symmetries of a system.} According to Eq.~(\ref{Model-Weyl-lin}), rotational invariance is preserved only in the plane perpendicular to the Weyl node separation $\mathbf{b}$, hence, the fully symmetric Belinfante stress tensor cannot be realized. Therefore, the strain transformation generator and the Belinfante-like stress tensor read as
\begin{eqnarray}
\label{Kubo-Weyl-J-b}
J_{\alpha \beta}&=&-\frac{1}{2} \left \{ x_{\beta} , k_{\alpha} \right\} + \frac{i}{8} C_{\alpha \beta} \left[ \gamma_{\beta}, \gamma_{\alpha} \right],\\
\label{Kubo-Weyl-T-b}
T_{\mu \nu}&=&v_F \gamma_0 \gamma_{\nu} k_{\mu} +C_{\mu \nu} \frac{v_F}{2}\left( \gamma_0 \gamma_{\mu} k_{\nu}-\gamma_0 \gamma_{\nu} k_{\mu} \right) \nonumber\\
&+&C_{\mu \nu} \frac{v_F}{2}\left( \gamma_0 \gamma_{\mu} \gamma_5 b_{\nu}-\gamma_0 \gamma_{\nu} \gamma_5 b_{\mu} \right),
\end{eqnarray}
where, for $\mathbf{b} =b \mathbf{e}_z$, coefficients $C_{\alpha 3}=C_{3 \alpha}=0$ and $C_{\alpha \beta}=1$ with $\alpha,\beta=1,2$.
We verified that the same result for the stress tensor is obtained directly using Noether's second theorem, see the end of Sec.~\ref{sec:App-1-continuum} and, e.g., Refs.~\cite{Leader-Lorce-AngularMomentumControversy-2014,Freese-Freese-NoetherTheoremsEnergymomentum-2022,Singh-Singh-NonuniquenessEnergymomentumSpin-2024}.

The real part of the static viscoelasticity tensor remains the same as in the case of the canonical stress tensor, see Eq.~(\ref{Kubo-Weyl-eta-stat}). On the other hand, the real part of the dynamic part of the viscoelasticity tensor is modified:
\begin{equation}
\label{Kubo-Weyl-eta-dyn-1}
\begin{aligned}
\text{Re}\,\eta_{\mu \nu \alpha \beta}^{\text{(B)},\text{dyn}}(\Omega)&= \left\{\delta_{\mu \alpha } \delta _{\nu \beta} \left[  1- \frac{5}{8} \left( C_{\alpha \beta }+ C_{\mu \nu } - C_{\alpha \beta }C_{\mu \nu } \right) \right]   \right. \\
&\left.  -\frac{1}{4} \delta_{\mu \beta } \delta _{\nu \alpha} \left[ 1 - \frac{5}{2} \left( C_{\alpha \beta  }+ C_{\mu \nu } - C_{\alpha \beta }C_{\mu \nu } \right) \right]\right. \\
&\left. - \frac{1}{4}  \delta_{\mu \nu} \delta_{\alpha \beta} \right\} \frac{\Omega^3}{120 \pi v_F^3} \Theta \left( \Omega -2 \mu \right) \\
& + \frac{\Omega}{12 \pi v_F} \delta_{\nu \beta}  b_{\mu } b_{\alpha }  \Theta \left( \Omega -2 \mu \right).
\end{aligned}
\end{equation}
As one can see, compared with the viscoelasticity tensor of Dirac semimetals given in Eq.~(\ref{Kubo-Dirac-eta}), there are a few terms related to the additional rotational symmetry of a Weyl semimetal. Among them, we mention the last term in Eq.~(\ref{Kubo-Weyl-eta-dyn-1}) which is linear in $\Omega$ and explicitly depends on the separation vector $\mathbf{b}$.

The Hall viscosity reads
\begin{equation}
\label{Kubo-Weyl-eta-Hall}
\begin{aligned}
\text{Re}\,\eta_{\mu \nu \alpha \beta}^{\text{(B)},\text{Hall}}(\Omega)&=\left( 2- C_{\alpha \beta} \right) \left( 2- C_{\mu \nu} \right) I_{\mu \nu \alpha \beta}\\
&+  C_{\alpha \beta } \left( 2- C_{\mu \nu} \right) I_{\mu \nu \beta \alpha}  \\
&+C_{\mu \nu } \left( 2- C_{\alpha \beta} \right)  I_{\nu \mu \alpha \beta} \\
&+C_{\alpha \beta } C_{\mu \nu }  I_{\nu \mu \beta \alpha},
\end{aligned}
\end{equation}
where $I_{\mu \nu \alpha \beta}= \sum_{j=1}^{3} \epsilon_{j \nu \beta} I_{\mu \alpha}^{j}$ with $I_{\mu \alpha}^{j}$ given in Eqs.~(\ref{Kubo-Weyl-I12}) and (\ref{Kubo-Weyl-I3}). As in the case of the canonical stress tensor, see Eq.~(\ref{Kubo-Weyl-eta-Hall-0}), only $\eta_{zxzy}$ is finite for $\Omega=0$; other nontrivial components are quadratically divergent. According to the structure of coefficients $C_{\mu \nu}$, Hall components $\eta_{zizj}$ and $\eta_{izzj}$ are the same in both approaches, but $\eta_{ijzz}$, $\eta_{ijii}$, and $\eta_{jiii}$ are different. Indeed, for the Belinfante-like stress tensor, we have the following relations for the Hall viscosity tensor components:
\begin{eqnarray}
\text{Re}\,\eta_{i j i i}^{\text{(B),\text{Hall}}}(\Omega)&=& \frac{1}{2}\text{Re}\,\eta_{i j i i}^{\text{(C),\text{Hall}}}(\Omega),\\
\text{Re}\,\eta_{j i i i}^{\text{(B),\text{Hall}}}(\Omega)&=&\text{Re}\,\eta_{i j i i}^{\text{(B),\text{Hall}}}(\Omega)=\frac{1}{2}\text{Re}\,\eta_{i j i i}^{\text{(C),\text{Hall}}}(\Omega),\nonumber\\
\\
\text{Re}\,\eta_{i j z z}^{\text{(B),\text{Hall}}}(\Omega)&=& 0,
\end{eqnarray}
while $\text{Re}\,\eta_{j i i i}^{\text{(C),\text{Hall}}}(\Omega) = 0$ and $\text{Re}\,\eta_{i j z z}^{\text{(C),\text{Hall}}}(\Omega) \neq 0$. For details of derivation, see Appendix~\ref{sec:App-Kubo-lin}.

\subsection{Two-band model of Weyl semimetals}
\label{sec:Kubo-Weyl-two-band}

To cross-verify the obtained anomalous and anisotropic parts of the viscoelasticity tensor, we consider the following two-band model:
\begin{equation}
\label{Kubo-Weyl-two-band-H-def}
H= \bm{\sigma} \cdot \mathbf{d}(\mathbf{k}),
\end{equation}
where $d_x(k_x)=v_F k_x$, $d_y(k_y)=v_F k_y$, and $d_z(k_z)=\gamma \left( k_z^2-m \right)$. This model describes a TRS-breaking Weyl semimetal with two Weyl nodes separated by $2\sqrt{m}$ along the $z$-direction.

\begin{figure}
\includegraphics[width=1\columnwidth]{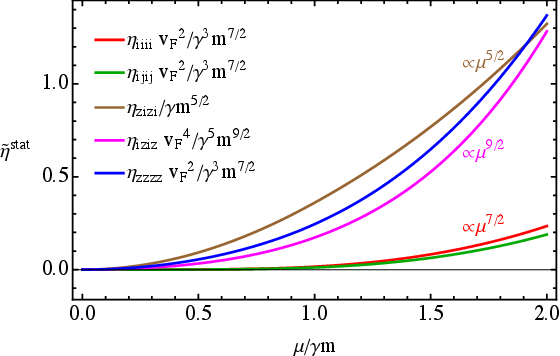}
\caption{The dependence of components of the real part of the static viscoelasticity tensor on $\mu/\gamma m$; see Eq.~(\ref{App-kubo-two-band-stat}) for its explicit form. The notation $\tilde{\eta}^{\text{stat}}_{\mu \nu \alpha \beta}$ stands for dimensionless viscosity, see the legend.
}
\label{fig_viscoelasticity_stat_two_band}
\end{figure}

Let us begin with the case of the canonical stress tensor. The complete result for the real part of the static viscoelasticity tensor is too bulky to be presented in the main text; it is given in Eq.~(\ref{App-kubo-two-band-stat}). We show the dependence of the few components of the viscoelasticity tensor in Fig.~\ref{fig_viscoelasticity_stat_two_band}. As one can see from the figure, the static part of the viscoelasticity tensor $\text{Re }\eta_{\mu \nu \alpha \beta}^{\text{stat}}(\Omega)$ shows no kinklike features at $|\mu| =\gamma m$ where two pockets of the Fermi surface merge. This follows from the fact that parts of the viscoelasticity tensor with $\Theta \left( \gamma m -|\mu| \right)$ are $\propto \left( \gamma m -|\mu| \right)^{5/2}$, hence they have continuous derivatives.

For small Fermi energy, $|\mu| \ll \gamma m$, the viscoelasticity tensor acquires a compact form:
\begin{equation}
\label{Kubo-stat-two-band-mull1}
\begin{aligned}
\text{Re}\,\eta_{\mu \nu \alpha \beta}^{\text{stat}}(\Omega)&\simeq \left[ \frac{\mu^4}{15 \pi v_F^3}  \left(\delta_{\mu \beta} \delta_{\nu \alpha} +\delta_{\mu \alpha} \delta_{\nu \beta} -\frac{2}{3}\delta_{\mu \nu } \delta_{\alpha \beta }  \right) \right. \\
&\left.+\frac{\mu^4}{10 \pi v_F^3} \delta_{\mu 3}  \left( \delta_{\nu 3} -  \delta_{\nu 1} -\delta_{\nu 2} \right)\delta_{\mu \alpha} \delta_{\nu \beta}\right. \\
&\left.+\frac{\mu^2 m}{3 \pi v_F}  \delta_{\mu 3} \delta_{\mu \alpha} \delta_{\nu \beta}\right] \delta(\Omega),
\end{aligned}
\end{equation}
where we also substituted $2\gamma \sqrt{m} \to v_F$. The above viscoelasticity tensor deviates from that in the linearized model, cf. Eq.~(\ref{Kubo-Weyl-eta-stat}), because expansion in $|\mu| \ll \gamma m$ still allows for higher-order anisotropy corrections to contribute to $\text{Re}\,\eta_{\mu \nu \alpha \beta}^{\text{stat}}(\Omega)$.
This result shows that the linearized model of a Weyl semimetal captures only the key qualitative features of the viscoelastic response, such as the scaling with the chemical potential and the separation between the Weyl nodes. Numerical values of the components of the viscoelastic response tensor, however, can be different in different models of semimetals.
The static part of the viscoelasticity tensor is also reproduced in the kinetic approach, see Sec.~\ref{sec:Kinetics}. Note that since $\sqrt{m}$ plays the same role as the momentum space separation $b_z$ in the linearized model, there is a correspondence between the third term in the square brackets in Eq.~(\ref{Kubo-stat-two-band-mull1}) and a term proportional to $\propto b_{\mu} b_{\alpha}$ in the linearized model, cf. Eqs.~(\ref{Kubo-Weyl-eta-stat}) and (\ref{Kubo-stat-two-band-mull1}).

In the opposite limit of the Fermi energy away from the Weyl nodes, $|\mu| \gg \gamma m$, we obtain
\begin{equation}
\label{Kubo-stat-two-band-large-mu}
\begin{aligned}
\text{Re}\,\eta_{\mu \nu \alpha \beta}^{\text{stat}}(\Omega)&\simeq  \df{\Omega} \Bigg\{\frac{2  |\mu|^{7/2}}{45 \pi \gamma^{1/2} v_F^2} \bigg[\delta_{\mu \beta} \delta_{\nu \alpha}  -\frac{4}{5}\delta_{\mu \nu } \delta_{\alpha \beta }\\
&+\left(\delta_{\mu 1}+\delta_{\mu 2} \right) \left(\delta_{\nu 1}+\delta_{\nu 2} \right)  \delta_{\mu \alpha} \delta_{\nu \beta}   \bigg] \\
&+\frac{4 \gamma^{1/2} |\mu|^{9/2} }{77 \pi v_F^4}  \left(\delta_{\mu 1}+\delta_{\mu 2} \right) \delta_{\nu 3}  \delta_{\mu \alpha} \delta_{\nu \beta} \\
&+ \frac{ |\mu|^{5/2}}{21 \pi \gamma^{3/2}}  \delta_{\mu 3} \left(\delta_{\nu 1}+\delta_{\nu 2} \right)  \delta_{\mu \alpha} \delta_{\nu \beta} \\
&+ \frac{2  |\mu|^{7/2}}{15 \pi \gamma^{1/2} v_F^2}  \delta_{\mu 3} \delta_{\nu 3} \delta_{\mu \alpha} \delta_{\nu \beta}
\Bigg\}.
\end{aligned}
\end{equation}
The inherent anisotropy of the electron energy spectrum of the two-band model is clearly imprinted in the viscoelasticity tensor leading to its highly anisotropic form.

\begin{figure}
\includegraphics[width=1\columnwidth]{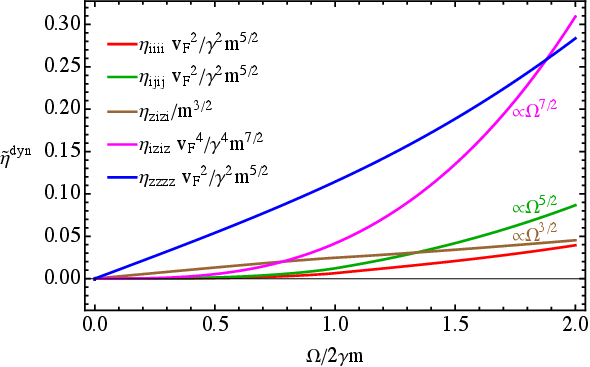}
\caption{The dependence of components of the real part of dynamic viscoelasticity tensor on $\Omega/2\gamma m $; see Eq.~(\ref{App-kubo-two-band-stat}) for its explicit form. The notation $\tilde{\eta}^{\text{dyn}}_{\mu \nu \alpha \beta}$ stands for dimensionless viscosity, see the legend.
}
\label{fig_viscoelasticity_dyn_two_band}
\end{figure}

The full expression for the real part of the dynamic viscoelasticity tensor is given in Eq.~(\ref{App-kubo-two-band-dyn}). We plot a few of its components in Fig.~\ref{fig_viscoelasticity_dyn_two_band}. As one can see, there is no kinklike feature in the dynamical part of the viscoelasticity tensor at $\Omega =2\gamma m$. Indeed, as follows from Eq.~(\ref{App-kubo-two-band-dyn}), the interband parts of the viscoelasticity tensor are $\propto \left( 2 \gamma m - \Omega \right)^{a+1/2}$ with $a\ge 1$. Hence, the derivatives are continuous. The absence of a kinklike feature in the dynamical part of the viscoelasticity tensor differs from that of, e.g., optical conductivity studied in Ref.~\cite{Tabert-Carbotte:2016b}. The latter shows a well-pronounced kink at $\Omega=2\gamma m$ originating from the van Hove singularities of the density of states.

For $\Omega \ll 2 \gamma m$, we have the following expression for the interband part of the viscoelasticity tensor:
\begin{equation}
\label{Kubo-stat-two-band-small-Omega}
\begin{aligned}
\text{Re}\,\eta_{\mu \nu \alpha \beta}^{\text{dyn}}(\Omega)&\simeq  \left[ \frac{\Omega^3 }{120 \pi v_F^3} \left( \delta_{\mu \alpha} \delta_{\nu \beta} - \frac{1}{4} \delta_{\mu \beta} \delta_{\nu \alpha}  - \frac{1 }{4} \delta_{\mu \nu} \delta_{\alpha \beta } \right) \right.\\
&\left.+\delta_{\mu 3} \delta_{\mu \alpha } \delta_{\nu \beta}  \frac{\Omega m}{12 \pi v_F }+\frac{\Omega^3 }{480 \pi v_F^3} \delta_{\mu 3} \delta_{\nu 3}  \delta_{\mu \alpha} \delta_{\nu \beta} \right.\\
&\left. - \frac{\Omega^3 }{80 \pi v_F^3} \delta_{\mu 3}\left(\delta_{\nu1}+ \delta_{\nu2} \right) \delta_{\mu \alpha} \delta_{\nu \beta}  \right] \Theta \left( \Omega - 2 \mu \right),
\end{aligned}
\end{equation}
where we also replaced $2\gamma \sqrt{m} \to v_F$. As with the static part, the viscoelasticity tensor expanded in $\Omega/(2\gamma m)$ in the two-band model is not equivalent to that in the linearized model even in the vicinity of the Weyl nodes.

In the opposite limit of large frequencies $\Omega/(2\gamma m) \gg 1$, we obtain
\begin{equation}
\label{Kubo-stat-two-band-large-Omega}
\begin{aligned}
\text{Re}\,\eta_{\mu \nu \alpha \beta}^{\text{dyn}}(\Omega)&\simeq  \left[  \frac{\gamma^{1/2} (\delta _{\mu 1}+\delta _{\mu 2}) \delta _{\nu 3}  \delta _{\mu \alpha} \delta _{\nu \beta}}{231 \sqrt{2} \pi  v_F^4}\Omega^{7/2}\right. \\
&\left.+\frac{7 (\delta _{\mu 1}+\delta _{\mu 2}) (\delta _{\nu 1}+\delta _{\nu 2}) \delta _{\mu \alpha} \delta _{\nu \beta}}{720 \sqrt{2} \pi  \gamma^{1/2} v_F^2} \Omega^{5/2}\right. \\
&\left.+\frac{5 \delta _{ \mu 3} (\delta _{\nu 1}+\delta _{\nu 2})  \delta _{\mu \alpha} \delta _{\nu \beta} }{336 \sqrt{2} \pi \gamma^{3/2} } \Omega^{3/2} \right.\\
&\left.-\frac{(\delta _{\mu \beta} \delta _{\nu \alpha}+\delta _{\mu \nu} \delta _{\alpha \beta})}{360 \sqrt{2} \pi \gamma^{1/2} v_F^2} \Omega^{5/2}\right. \\
&\left. +\frac{\delta _{\mu 3} \delta _{\nu 3} \delta _{\mu \alpha} \delta _{\nu \beta}}{60 \sqrt{2} \pi \gamma^{1/2} v_F^2} \Omega^{5/2}  \right] \Theta \left( \Omega - 2 \mu \right).
\end{aligned}
\end{equation}
As one can see, the dynamical viscoelasticity tensor acquires dependence on fractional powers of $\Omega$; this is similar to the fractional powers of $\mu$ in Eq.~(\ref{Kubo-stat-two-band-large-mu}).
These results show that both static and dynamic parts of the viscoelasticity tensor can be used to probe the anisotropy of the Fermi surface.

For the Hall viscosity, we find
\begin{equation}
\label{Kubo-stat-two-band-Hall}
\text{Re}\,\eta_{\mu \nu \alpha \beta}^{\text{Hall}}(\Omega)= \sum_{j=1}^{3} \epsilon_{j \nu \beta} I_{\mu \alpha}^{j},
\end{equation}
where
\begin{equation}
\label{Kubo-stat-two-band-Hall-1}
\begin{aligned}
I_{\mu \alpha}^{i=1,2}&=\left(\delta_{\mu 3} \delta_{\alpha i}+\delta_{\mu i} \delta_{\alpha 3} \right) \\
&\times\left( - \frac{\gamma ^2 \Lambda_z^5 }{10 \pi ^2 v_F^2}+\frac{\gamma \Lambda_z^3  \Lambda_{\perp} }{12 \pi ^2 v_F}+ \frac{ \gamma ^2 m \Lambda_z^3}{6 \pi ^2 v_F^2}- \frac{2 \gamma ^2 m^{5/2}}{15 \pi ^2 v_F^2} \right)
\end{aligned}
\end{equation}
and
\begin{equation}
\label{Kubo-stat-two-band-Hall-2}
\begin{aligned}
I_{\mu \alpha}^{3}&=\delta_{\mu \alpha} \bigg[ \left( \delta_{\mu 1} + \delta_{\mu 2}\right) \left( -\frac{\gamma ^2 \Lambda_z^5 }{20 \pi ^2 v_F^2}+\frac{ \gamma ^2 m \Lambda_z^3}{6 \pi ^2 v_F^2}\right.\\
&\left.+\frac{\gamma \Lambda_z^3   \Lambda_\perp }{24 \pi ^2 v_F}-\frac{\gamma ^2 m^2\Lambda_z  }{4 \pi ^2 v_F^2}-\frac{ \gamma m \Lambda_z\Lambda_\perp}{8 \pi ^2 v_F}+\frac{4 \gamma ^2 m^{5/2}}{15 \pi ^2 v_F^2}  \right)\\
&+\delta_{\mu 3}\frac{1}{12\pi^2}\left( \Lambda_z^3 - 2 m^{3/2} \right)\bigg]
\end{aligned}
\end{equation}
with $\Lambda_{\perp}$ and $\Lambda_z$ being momentum cutoffs.

A similar type of divergence is present in the Hall conductivity for the two-band model. Indeed, for $\Omega= \mu=0$, $\sigma_{xy}^{\text{AHE}}$ reads
\begin{equation}
\sigma_{xy}^{\text{AHE}}= \int \frac{d^3 k}{(2\pi)^3} \frac{ v_F^2 d_z }{2\varepsilon_{\mathbf{k}}^3} = \int_{-\Lambda_z}^{\Lambda_z} \frac{dk_z}{8\pi^2} \frac{d_z }{\left| d_z \right|} = \frac{\Lambda_z}{4 \pi^2} - \frac{\sqrt{m}}{2 \pi^2}.\\
\end{equation}
This linear divergence in the above expression originates from the flux of the Berry curvature at $k \rightarrow \infty$. This is a peculiarity of the model at hand, Eq.~(\ref{Kubo-Weyl-two-band-H-def}), and is absent in lattice models, where the flux through the whole Brillouin zone vanishes.
The following subtraction:
\begin{equation}
\sigma_{xy}^{\text{AHE}} \rightarrow \overline{\sigma}_{xy}^{\text{AHE}}=\sigma_{xy}^{\text{AHE}} - \sigma_{xy}^{\text{AHE}} \Big|_{m=0}
\end{equation}
allows us to remove the divergent background term and obtain the anomalous Hall conductivity consistent with that in the linearized model
\begin{equation}
\overline{\sigma}_{xy}^{\text{AHE}}= - \frac{\sqrt{m}}{2 \pi^2}.
\end{equation}
Since the integrands of the Hall conductivity and Hall viscoelasticity have similar structure, see the last term in Eq.~(\ref{Kubo_viscoelastic_tensor}), the background contribution of the same origin should be subtracted from the viscoelastic tensor (\ref{Kubo-stat-two-band-Hall}) too.
The corresponding coefficients $I_{\mu \alpha}^{j}$ in Eqs.~(\ref{Kubo-stat-two-band-Hall-1}) and (\ref{Kubo-stat-two-band-Hall-2}) read
\begin{equation}
\label{Kubo-stat-two-band-I-reg}
\begin{aligned}
\overline{I}_{\mu \alpha}^{i=1,2}&=\left(\delta_{\mu 3} \delta_{\alpha i}+\delta_{\mu i} \delta_{\alpha 3} \right) \left( \frac{ \gamma ^2 m \Lambda_z^3}{6 \pi ^2 v_F^2}- \frac{2 \gamma ^2 m^{5/2}}{15 \pi ^2 v_F^2} \right) \\
&=\left(\delta_{\mu 3} \delta_{\alpha i}+\delta_{\mu i} \delta_{\alpha 3} \right) \left( \frac{ \Lambda_z^3}{24 \pi ^2 }- \frac{ m^{3/2}}{30 \pi ^2} \right),
\end{aligned}
\end{equation}
\begin{equation}
\begin{aligned}
\overline{I}_{\mu \alpha}^{3}&=\delta_{\mu \alpha} \bigg[ \left( \delta_{\mu 1} + \delta_{\mu 2}\right) \left( \frac{ \gamma ^2 m \Lambda_z^3}{6 \pi ^2 v_F^2}-\frac{\gamma ^2 m^2\Lambda_z  }{4 \pi ^2 v_F^2} \right.\\
&\left. -\frac{ \gamma m \Lambda_z\Lambda_\perp}{8 \pi ^2 v_F}+\frac{4 \gamma ^2 m^{5/2}}{15 \pi ^2 v_F^2}  \right)-\delta_{\mu 3}\frac{m^{3/2}}{6\pi^2}\bigg] \\
&=\delta_{\mu \alpha} \bigg[ \left( \delta_{\mu 1} + \delta_{\mu 2}\right) \left( \frac{ \Lambda_z^3}{24 \pi ^2 }-\frac{ m\Lambda_z  }{16 \pi ^2}\right.\\
&\left.-\frac{ m^{1/2} \Lambda_z\Lambda_\perp}{16 \pi ^2 }+\frac{ m^{3/2}}{15 \pi ^2}  \right)-\delta_{\mu 3}\frac{m^{3/2}}{6\pi^2}\bigg],
\end{aligned}
\end{equation}
where we used $2 \gamma \sqrt{m} \rightarrow  v_F$ in the second expressions to compare with linearized model. Like in the linearized model, see Eq.~(\ref{Kubo-Weyl-I3}) and the discussion after it, the only finite component is also $\eta_{zxzy}=-m^{3/2}/(6\pi^2)$. The other components are cubically divergent, unlike the quadratic divergence in the linearized model.

\begin{figure}
\includegraphics[width=1\columnwidth]{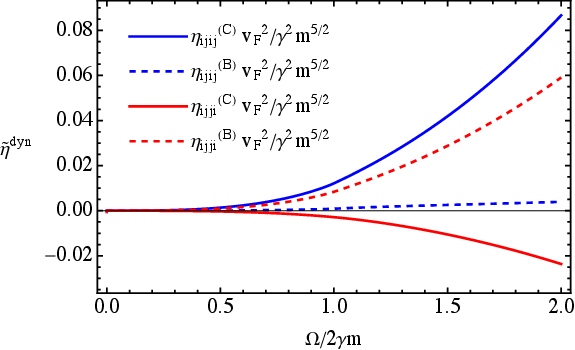}
\caption{The dependence of the components $\eta_{ijij}$ and $\eta_{ijji}$ of the real part of dimensionless dynamic viscoelasticity on $\Omega/(2\gamma m)$ for the canonical (solid lines) and Belinfante-like (dashed lines) stress tensors.}
\label{fig_viscoelasticity_dyn_two_band_C_B}
\end{figure}

For the Belinfante-like stress tensor, the corrections to the dynamic and static parts of the viscoelasticity are given in Eqs.~(\ref{App-kubo-two-band-dyn-correction}) and (\ref{App-kubo-two-band-stat}), respectively. Like in the linearized model, the static part is the same for both definitions of the stress tensor, see Eq.~(\ref{App-kubo-two-band-stat-correction}) for the discussion of the difference. As for the dynamic part, $\eta_{ijij}$ and $\eta_{ijji}$ components differ from the results of the canonical stress tensor case. For $\eta_{ijij}^{\text{(B)}}$, the dependence on $\Omega/(2\gamma m)$ is modified, but the scaling for small and large frequencies $\Omega$ remains the same. In addition, $\eta_{ijji}^{\text{(B)}}$ becomes positive. We compare the components of the dynamic viscoelasticity tensor for Belinfante-like and canonical stress tensors in Fig.~\ref{fig_viscoelasticity_dyn_two_band_C_B}.

The corrections to the Hall part of the viscoelasticity tensor are given in Eq.~(\ref{App-kubo-two-band-Hall-correction}). As in the linearized model, components $\eta_{zizj}$ and $\eta_{izzj}$ are the same for canonical and Belinfante-like stress tensors. On the other hand, components $\eta_{ijzz}$, $\eta_{ijii}$ are different
\begin{eqnarray}
\label{Kubo-two-band-B-vs-C}
\text{Re}\,\eta_{i j i i}^{\text{(B)}}(\Omega)&=& \frac{1}{2}\text{Re}\,\eta_{i j i i}^{\text{(C)}}(\Omega),\\
\text{Re}\,\eta_{j i i i}^{\text{(B)}}(\Omega)&=& \frac{1}{2}\text{Re}\,\eta_{i j i i}^{\text{(C)}}(\Omega),\\
\text{Re}\,\eta_{i j z z}^{\text{(B)}}(\Omega)&=& 0,
\end{eqnarray}
while $\text{Re}\,\eta_{j i i i}^{\text{(C)}}(\Omega) = 0$ and $\text{Re}\,\eta_{i j z z}^{\text{(C)}}(\Omega) \neq 0$. For details of calculations, see Appendix~\ref{sec:App-Kubo-two-band}.

Thus, the nontrivial topology of the Weyl semimetals is manifested in the viscoelastic response leading to anomalous Hall terms. These results agree with the findings in Refs.~\cite{Cortijo-Vozmediano:2016-visco,Cortijo-Vozmediano:2015} extended here to the case of the two-band model.

\section{Kinetic approach to viscosity and hydrodynamic equations}
\label{sec:Kinetics}

In this Section, to gain deeper insight into the role of the band structure in the viscoelastic and transport responses of Weyl semimetals, we derive the hydrodynamic equations of motion, i.e., the continuity and Navier-Stokes equations, for the linearized and two-band models of Weyl semimetals. We focus on the role of the Fermi-surface anisotropy and band structure topology. The latter is quantified by the separation between the Weyl nodes. We show that the anomalous part of the viscoelasticity tensor does not appear in the standard approach based on the Boltzmann equation~\cite{Huang:book,Landau:t10-1995}. Indeed, the stress-stress response used in Sec.~\ref{sec:Kubo} and current-current response are not guaranteed to give the same results for the fluid viscosity;~\footnote{For Galilean-invariant systems, one can derive microscopic expressions for the viscosity tensor by equating the current-current response functions and the coefficients in the current response to a vector potential~\cite{Conti-Vignale:1999}. The connection to a more rigorous approach based on the stress-stress correlation function follows from the fact that the current density is proportional to the momentum density, therefore its time derivative is proportional to the divergence of the stress tensor.} see Ref.~\cite{Principi-Polini:2016} for the corresponding discussion in graphene. As we demonstrate below, while the ``matter" terms, i.e., components of the viscosity tensor determined by the Fermi-surface properties, agree in the Kubo and kinetic formalisms, the topological terms are absent in the latter. We trace this difference to the contribution of the filled states below the Fermi-surface: while these states affect the viscoelastic response, only the Fermi-surface properties determine the viscosity of the electron fluid flow in the kinetic approach. A similar situation is realized for the anomalous Hall effect or Chern-Simons contributions in the electric current; see, e.g., Ref.~\cite{Gorbar:2017wpi} for a detailed discussion.

Since the derivation of the hydrodynamic equation follows the standard scheme and is straightforward, we leave it to Appendixes~\ref{sec:App-zero-kinetic} and \ref{sec:App-first-kinetic}. In what follows, we present the key results and discuss the physics behind them. Among our major findings are the manifestation of the shape (anisotropy) and structure (separation of the Weyl nodes) of the Fermi surface of Weyl semimetals in the hydrodynamic equations.

Using the linearized model of a Weyl semimetal (\ref{Model-Weyl-lin}), we first derive the following set of inviscid hydrodynamic equations:
\begin{equation}
\label{kinetic-zero-hydrodynamic-linear-charge}
\partial_t \rho(t,\mathbf{r}) +  \bm{\nabla}\cdot \left[\mathbf{u} (t,\mathbf{r}) \rho (t,\mathbf{r})+\frac{e^2}{2 \pi^2 } \left[\mathbf{b} \times  \mathbf{E} \right]  \right]=0,
\end{equation}
\begin{equation}
\begin{aligned}
\label{kinetic-zero-hydrodynamic-linear-Euler}
&\partial_t\left[ w (t,\mathbf{r}) \frac{\mathbf{u} (t,\mathbf{r})}{v_F^2} +\nu(t,\mathbf{r}) \left( \mathbf{u} (t,\mathbf{r}) \cdot \mathbf{ b} \right) \mathbf{b}\right]  + \bm{\nabla} P \left(t, \mathbf{r} \right) \\
&+ \rho\left(t,\mathbf{r}\right)  \mathbf{E}= -\frac{1}{\tau}\left[ w(t,\mathbf{r}) \frac{\mathbf{u} (t,\mathbf{r}) }{v_F^2} +\nu(t,\mathbf{r}) \left( \mathbf{u} (t,\mathbf{r}) \cdot \mathbf{ b} \right) \mathbf{b} \right],
\end{aligned}
\end{equation}
\begin{equation}
\label{kinetic-zero-hydrodynamic-linear-energy}
\partial_t \epsilon(t,\mathbf{r}) +\bm{\nabla}\cdot\left[\mathbf{u} (t,\mathbf{r}) w(t,\mathbf{r}) \right]=0,
\end{equation}
where $w(t,\mathbf{r}) = \epsilon(t,\mathbf{r}) +P(t,\mathbf{r}) =4\epsilon(t,\mathbf{r})/3$ is the enthalpy density, $P(t,\mathbf{r}) = \epsilon(t,\mathbf{r})/3$ is the pressure. Assuming $T\to0$, $\epsilon(t,\mathbf{r}) = \mu^4(t,\mathbf{r})/(4\pi^2 v_F^3)$ is the energy density and $\nu(t,\mathbf{r}) = \mu^2(t,\mathbf{r})/(\pi^2 v_F^3)$ is the density of states.

In the case of the two-band model (\ref{Kubo-Weyl-two-band-H-def}), we have
\begin{equation}
\label{kinetic-zero-hydrodynamic-two-band-charge}
\partial_t \rho(t,\mathbf{r}) +  \bm{\nabla}\cdot \left[\mathbf{u} (t,\mathbf{r}) \rho (t,\mathbf{r})+\mathbf{j}^{\text{AHE}} \right]=0,
\end{equation}
\begin{equation}
\label{kinetic-zero-hydrodynamic-two-band-Euler}
\begin{aligned}
&\partial_t \left[  w (t,\mathbf{r}) \frac{\mathbf{u} (t,\mathbf{r})}{v_F^2} + A(t,\mathbf{r})\left( \mathbf{u} (t,\mathbf{r}) \cdot \mathbf{e}_z \right)\mathbf{e}_z \right] + \bm{\nabla} P\left(t,\mathbf{r}\right) \\
&+ \rho\left(t,\mathbf{r}\right)  \mathbf{E}  =- \frac{1}{\tau} \left[ w(t,\mathbf{r}) \frac{\mathbf{u}(t,\mathbf{r})}{v_F^2} +A(t,\mathbf{r}) \left( \mathbf{u} (t,\mathbf{r}) \cdot \mathbf{e}_z \right)\mathbf{e}_z  \right],
 \end{aligned}
\end{equation}
\begin{equation}
\label{kinetic-zero-hydrodynamic-two-band-energy}
\partial_t\epsilon(t,\mathbf{r}) + \bm{\nabla} \cdot \left[  \mathbf{u} (t,\mathbf{r}) w(t,\mathbf{r}) \right]=0,
\end{equation}
where the expressions for $\mathbf{j}^{\text{AHE}}$ and $A(r,\mathbf{r})$ are given in Eqs.~(\ref{App_AHE_term}) and (\ref{App_anisotropic_term}) respectively. Note that the Weyl nodes are separated along the $z$-axis in the two-band model (\ref{Kubo-Weyl-two-band-H-def}); this corresponds to $\mathbf{b}=b \mathbf{e}_z$ in the linearized model.

The general structure of the zero-order hydrodynamic equations is preserved for both linearized and two-band models. However, the anisotropy of the Fermi surface leads to the appearance of terms $\propto \nu(t,\mathbf{r}) \left( \mathbf{u} (t,\mathbf{r}) \cdot \mathbf{ b} \right) \mathbf{b}$ and $\propto A(t,\mathbf{r}) \left( \mathbf{u} (t,\mathbf{r}) \cdot \mathbf{e}_z \right)\mathbf{e}_z$
under in time derivative (acceleration) and dissipative terms of the Euler equation. We note that, in view of weaker compared with $w(t,\mathbf{r})$ dependence of $\nu(t,\mathbf{r})$ and $A(t,\mathbf{r})$ on the chemical potential $\mu(t,\mathbf{r})$, these new terms are expected to play an important role for Fermi energies approaching Weyl nodes. In particular, at $T\to0$, $w(t,\mathbf{r})/(\nu(t,\mathbf{r}) b^2) \sim (p_F/b)^2$; for Weyl materials with well-separated Weyl nodes, $(p_F/b)^2\lesssim1$.

Let us now discuss the electron fluid viscosity, which affects the left-hand side of Eq.(\ref{kinetic-zero-hydrodynamic-two-band-Euler}) via the term $-\eta_{ijkl}\nabla_j\nabla_l u_k(t,\mathbf{r})$. The viscosity tensor $\eta_{ijkl}$ in the linearized model reads
\begin{eqnarray}
\label{kinetic-viscosity-lin-0}
\eta_{ijkl} &=& \frac{4\tau_{ee}}{15} \epsilon(t,\mathbf{r}) \left(\delta_{ik}\delta_{jl} + \delta_{il}\delta_{jk} -\frac{2}{3} \delta_{ij}\delta_{kl}\right) \nonumber\\
&+& \frac{\tau_{ee} v_F^2\nu(t,\mathbf{r})}{3} b_ib_k\delta_{jl},
\end{eqnarray}
see Appendix~\ref{sec:app-kinetic-lin} for the derivation. In the case of the two-band model of Weyl semimetals, the viscosity tensor is given by Eq.~(\ref{kinetic-viscosity-two-band}). By comparing terms at $\delta(\Omega)$ in Eq.~(\ref{Kubo-Weyl-eta-stat}) and at $\tau_{ee}$ in Eq.~(\ref{kinetic-viscosity-lin-0}), we find the exact correspondence between the static (intraband) parts of the dissipative viscoelastic response in the Kubo and kinetic approaches. However, the kinetic approach misses the topological Hall part of the viscoelasticity tensor and is insensitive to the internal degrees of freedom that may affect the Hall viscosity. These results underline important restrictions of the standard kinetic formalism.

Let us now discuss experimentally relevant ways to probe the anisotropic viscosity and dissipation. A direct method would be to study the flow of electron liquid in channels cut along different crystallographic directions. For example, one can consider a $\Gamma$-shaped channel cut from a single crystal. Along one of the arms of the channel, the flow of electrons is directed along the crystallographic direction corresponding to the separation of the Weyl nodes along the flow velocity. The other arm of the channel should be prepared such that the flow velocity is perpendicular to the separation vector. In this case, according to Eqs.~(\ref{kinetic-zero-hydrodynamic-linear-charge}) and (\ref{kinetic-viscosity-lin-0}), the fluid viscosity and dissipation rate will be different in the different arms of the channel. This is manifested in the redistribution of the electric current density in the channel.

To show this, let us use Eq.~(\ref{kinetic-viscosity-lin-0}) and obtain the steady-state Navier-Stokes equation for incompressible fluids at low velocities
\begin{equation}
\begin{aligned}
\label{kinetic-zero-hydrodynamic-NS}
&-\eta_0\Delta \mathbf{u}(\mathbf{r}) -\eta_1\Delta \left(\mathbf{u} (\mathbf{r}) \cdot \mathbf{b}\right) \mathbf{b} \\
&+\frac{1}{\tau} \left[\frac{w}{v_F^2} \mathbf{u}(\mathbf{r}) +\nu \left(\mathbf{u} (\mathbf{r}) \cdot \mathbf{b}\right) \mathbf{b} \right] =\rho \mathbf{E}.
\end{aligned}
\end{equation}
Here, $\eta_0 = \tau_{ee} w/5$, $\eta_1=v_F^2 \tau_{ee}\nu/3$, and, for simplicity, we used the linearized model (\ref{Model-Weyl-lin}). At the surfaces of the channel, we assume the no-slip boundary conditions with vanishing tangential components of velocity $\mathbf{u}_{\parallel}(r_{\perp}=0,L)=\mathbf{0}$. The solution to Eq.~(\ref{kinetic-zero-hydrodynamic-NS}) which captures the cases of flow with $\mathbf{u} \parallel \mathbf{b}$ and $\mathbf{u}\perp \mathbf{b}$ reads
\begin{equation}
\label{kinetic-zero-hydrodynamic-NS-sol}
u_{\parallel}(r_{\perp}) = \frac{v_F^2\tau}{w_{b}} \rho E \left[1 - \frac{\cosh{\left(\frac{L -2r_{\perp}}{2\lambda_{G,b}}\right)}}{\cosh{\left(\frac{L}{2\lambda_{G,b}}\right)}}\right],
\end{equation}
where $w_{b} = w +v_F^2\nu b_{\parallel}^2$, $\lambda_{G,b} = \sqrt{v_F^2\tau (\eta_0 +b_{\parallel}^2\eta_1)/w_b} = \sqrt{15 v_F^2\tau \tau_{ee} [3w +5v_F^2 \nu b^2]/[w +v_F^2 \nu b^2]}$ is the Gurzhi length, and $r_{\perp}$ is the coordinate normal to the channel. As one can see from Eq.~(\ref{kinetic-zero-hydrodynamic-NS-sol}), the chiral shift $b$ affects the profile of the fluid velocity by enhancing the Gurzhi length; the fluid velocity profile becomes more parabolic-like at larger $\lambda_{G,b}$, e.g., when $v_F^2\nu b_{\parallel}^2 \gg w$. Therefore, the separation between the Weyl nodes can be inferred from the curvature of flow profiles.

Another manifestation of the nontrivial Fermi-surface structure of Weyl materials quantified by the distance between the Weyl nodes is suppression of the flow velocity along the separation direction $\mathbf{u}\parallel \mathbf{b}$. Indeed, as one can see from Eq.~(\ref{kinetic-zero-hydrodynamic-NS-sol}), the fluid velocity is suppressed as $\sim b_{\parallel}^2/p_F^2$, which can be a large factor for Weyl semimetals.

The electron fluid viscosity is also manifested in the dynamical response. For example, due to an interplay of viscous and inertial properties of the electron fluid as well as the boundary conditions, a nonuniform, double-peak, profile of the electric current appears in a channel~\cite{Moessner-Witkowski:2018,Sukhachov-Gorbar:2021-time-flow}. The current profile is manifested in a stray magnetic field surrounding the channel. In the case of anisotropic viscosity and $\Gamma$-shaped channel, the double-peak profile may be observed in one arm (smaller viscosity) but not in the other (larger viscosity). Another possible signature would be a different decay rate of hydrodynamic collective modes propagating along different crystallographic directions.

In passing, we note that high-quality channels of Weyl semimetals can be prepared via focused ion beam etching; see Ref.~\cite{Moll-Mackenzie:2016} for the use of this technique for PdCoO$_2$ and Ref.~\cite{Delft-Moll:2020} for the Weyl semimetal WP$_2$.\\

\section{Discussion and Summary}
\label{sec:Summary}

Employing the Kubo formalism and chiral kinetic theory approach, we calculated the viscoelasticity tensor in Weyl semimetals with broken time-reversal symmetry paying special attention to the manifestation of topology of electron states in Weyl semimetals and their Fermi surface anisotropy. In the Kubo formalism, we found that, in addition to the anomalous Hall viscosity discussed in Ref.~\cite{Cortijo-Vozmediano:2015}, the nontrivial band structure of Weyl semimetals affects the dissipative part of the viscoelasticity tensor.

The nondissipative topological part of the viscoelasticity tensor, namely the anomalous Hall viscosity, see Eqs.~(\ref{Kubo-Weyl-eta-Hall-0}) and (\ref{Kubo-stat-two-band-Hall}), shares a similar origin with the Chern-Simons contributions to electric current in the anomalous Hall effect in Weyl semimetals and originates from the filled states below the Fermi surface; see, e.g., Ref.~\cite{Gorbar:2017wpi} for the corresponding discussion. Such contributions are captured in the Kubo approach and do not affect the responses solely determined by the Fermi-surface properties, such as those obtained in the kinetic formalism.

One of the main advances of our work is in clarifying the impact of the nontrivial shape (anisotropy and Weyl node position) of the Fermi surface on viscoelastic properties of linearized and two-band models of Weyl semimetals, see Secs.~\ref{sec:Kubo-Weyl-lin} and \ref{sec:Kubo-Weyl-two-band}, respectively. The Fermi-surface anisotropy is manifested in a rich structure of both static and dynamic parts of viscoelasticity tensors, see Eqs.~(\ref{Kubo-Weyl-eta-stat}),  (\ref{Kubo-Weyl-eta-dyn}), and (\ref{Kubo-Weyl-eta-dyn-1}) for the linearized model and Eqs.~(\ref{Kubo-stat-two-band-mull1}), (\ref{Kubo-stat-two-band-large-mu}), (\ref{Kubo-stat-two-band-small-Omega}), and (\ref{Kubo-stat-two-band-large-Omega}) for the two-band model.

The static (intraband) part of the viscoelasticity tensor contains additional terms determined by the momentum space separation of Weyl nodes, see Eqs.~(\ref{Kubo-Weyl-eta-stat}) and (\ref{Kubo-stat-two-band-mull1}). Despite being related to the distance between the Weyl nodes, these terms originate from the shape of the Fermi surface rather than the nontrivial topology of electron wave functions. In particular, they are captured in both Kubo and kinetic approaches. The anisotropy of the two-band model is manifested in the unusual fractional scaling with the Fermi energy if the latter is large compared with the energy at which the Fermi-surface pockets merge; see Eq.~(\ref{Kubo-stat-two-band-large-mu}).

The dynamic (interband) part of viscosity contains linear and cubic in frequency terms in both linearized and two-band models, see Eqs.~(\ref{Kubo-Weyl-eta-dyn-1}) and (\ref{Kubo-stat-two-band-small-Omega}), respectively. The nonlinearity of the two-band model strongly
modifies dynamic viscosity for larger frequencies exceeding the energy at which the Fermi-surface pockets merge. In this case, the interband viscoelasticity tensor contains fractional powers of frequencies $\Omega^{3/2}$, $\Omega^{5/2}$, and $\Omega^{7/2}$; see Eq.~(\ref{Kubo-stat-two-band-large-Omega}). Such a scaling may be manifested in dynamic responses such as those involving time-dependent flows, see the end of Sec.~\ref{sec:Kinetics} for the corresponding discussion.

In our calculations, we employed two definitions of the stress tensor: the canonical stress tensor, which follows from the coupling to mechanical strains, and the Belinfante-like tensor, which follows from the momentum continuity equation and includes contributions connected with the internal angular momentum. Both definitions of the stress tensor lead to the same result for the static part of the viscoelastic response but differ for the dynamic part. These two stress tensors correspond to different types of viscosity. They are manifested in different types of observables: hydrodynamic transport for the Belinfante-like tensor and thermal transport and acoustic phonon dispersion for the canonical stress tensor.

In addition to viscosity, the anisotropy of the dispersion relation of Weyl materials is also manifested in acceleration and relaxation terms of the hydrodynamic equations, see Eqs.~(\ref{kinetic-zero-hydrodynamic-linear-Euler}) and (\ref{kinetic-zero-hydrodynamic-two-band-Euler}). The new terms are determined by the separation of the Weyl nodes and are parametrically larger compared with other terms in the vicinity of the charge neutrality point. We found no discussion of these terms in the literature. The proposed structure of the viscosity tensor and the new terms in the hydrodynamic equations can be investigated via transport measurements such as those in crystals cut along the different crystallographic directions or in a $\Gamma$-shaped channel discussed at the end of Sec.~\ref{sec:Kinetics}. In particular, we expect a parametrically slower flow along the Weyl node separation direction. Furthermore, the current profile of such flow should have a more pronounced parabolic shape.

Studies of the intra- and interband viscoelastic responses of the Fermi arc surface states or other topological materials such as multifold or spin-1 semimetals~\cite{Bradlyn-Bernevig:2016} are also interesting and will be reported elsewhere. Another possible extension of the present work is to investigate the viscoelastic response of the anisotropic spin-polarized Fermi surfaces of altermagnets~\cite{Smejkal-Jungwirth:2022}.

\begin{acknowledgments}
The work of E.V.G. was supported by the Program ”Dynamics of particles and collective excitations in high-energy physics, astrophysics and quantum macrosystems” of the Department of Physics and Astronomy of the NAS of Ukraine. P.O.S. acknowledges useful communications with A.~Qaiumzadeh and S.~Syzranov. P.O.S. acknowledges support through the Yale Prize Postdoctoral Fellowship in Condensed Matter Theory during his stay at Yale University as well as the Research Council of Norway through Grant No.~323766 and its Centres of Excellence funding scheme Grant No.~262633 “QuSpin” during his stay at NTNU.
\end{acknowledgments}


\appendix

\begin{widetext}

\section{Stress tensor}
\label{sec:App-stress-tensor}

\subsection{Stress tensor and strain transformation generators in continuum models}
\label{sec:App-1-continuum}

In this Section, we derive the stress tensor in the continuum model. Under deformations quantified by the displacement vector $\mathbf{u}(\mathbf{x}, t)$, coordinates transform as $\mathbf{x} \rightarrow \mathbf{x}'= \mathbf{x} + \mathbf{u}(\mathbf{x}, t) \approx \mathbf{x}+\mathbf{e}_{\mu}  \frac{\partial u_{\mu}}{\partial x_{\nu}} x_{\nu}$,
which can be rewritten in terms of homogeneous but time-dependent invertible matrix $\Lambda(t)$ with positive determinant
\begin{equation}
\mathbf{x}'= \Lambda(t) \mathbf{x}.
\end{equation}
The matrix $\Lambda(t)$ is expressed in terms of the strain tensor $\lambda(t)$ as $\Lambda(t)=\exp{(\lambda(t))}=1+ \lambda(t)+o(\lambda)$. This allows one to obtain the following transformed wave function $\psi(\mathbf{x},t) \rightarrow \psi_{\lambda}(\mathbf{x},t)$:
\begin{equation}
\psi_{\lambda}(\mathbf{x},t)=\sqrt{\det \Lambda}\, \psi\left( \Lambda \mathbf{x},t\right)=\psi(\mathbf{x},t) -i \lambda_{\mu \nu} \left( \frac{i}{2} \delta_{\mu \nu} +i x_{\nu}\frac{\partial}{\partial x_{\mu}}\right)\psi(\mathbf{x},t)+o(\lambda) =\psi(\mathbf{x},t)-i \lambda_{\mu \nu} L_{\mu \nu} \psi(\mathbf{x},t)+o(\lambda).
\end{equation}
If a system has additional symmetries, the strain transformation generator $J_{\mu \nu}$ is modified as~\cite{Link-Schmalian:2018}
\begin{equation}
J_{\mu \nu }=L_{\mu \nu }+L_{\mu \nu }^{\text{(int)}}.
\label{strain-transformation-generator}
\end{equation}
In other words, the generators of infinitesimal rotations are the operators of the total angular momentum which includes pseudospin accounting for internal degrees of freedom.

By using the strain transformation
\begin{equation}
S(t)=e^{-i \lambda_{\mu \nu }(t) J_{\mu \nu}}=I-i \lambda_{\mu \nu }(t) J_{\mu \nu}+o\left(\lambda \right),
\end{equation}
the transformed Hamiltonian can be represented as
\begin{equation}
H_{\Lambda}(t) =S(t) H S^{-1}(t)+i \frac{\partial S}{\partial t} S^{-1} =H-i \lambda_{\mu \nu}(t) \left[J_{\mu \nu},H \right]+\frac{\partial \lambda_{\mu \nu}}{\partial t} J_{\mu \nu},
\end{equation}
where $\left[,\right]$ means commutator. The above relation leads to symmetric Belinfante-like stress tensor~\footnote{For isotropic systems, it is equivalent to the Belinfante-Rosenfeld stress-energy tensor~\cite{Belinfante:1940,Rosenfeld:1940}.}
\begin{equation}
T_{\mu \nu} =-\frac{\partial H_{\Lambda}}{\partial \lambda_{\mu \nu}}=i \left[J_{\mu \nu},H \right].
\end{equation}

The same result for the stress tensor can be obtained using Noether's second theorem, see Refs.~\cite{Leader-Lorce-AngularMomentumControversy-2014,Freese-Freese-NoetherTheoremsEnergymomentum-2022,Singh-Singh-NonuniquenessEnergymomentumSpin-2024} for a recent discussion. For the Lagrangian density $\mathcal{L}$, which is a function of the
field $\psi$, invariant with respect to the local coordinate transformation
\begin{equation}
\begin{aligned}
\label{snt_local_transformation}
x_{\mu} \rightarrow x_{\mu}^{\prime}=x_{\mu}+\xi_{\mu}(x),
\end{aligned}
\end{equation}
Noether's second theorem implies
\begin{equation}
\begin{aligned}
\partial_{\nu} \left\{ \mathcal{D}^{\mu\nu} \left[ \psi \right] -g^{\mu\nu} \mathcal{L} \right\}=0,
\end{aligned}
\end{equation}
where
\begin{equation}
\begin{aligned}
 \mathcal{D}^{\mu \nu} \left[ \psi \right]&= -\frac{\partial}{\partial \left( \partial_{\nu} \xi_{\mu} \right) }\left[ \frac{\partial \mathcal{L}}{\partial \psi} \delta \psi+\frac{\partial \mathcal{L}}{\partial \left( \partial_{\lambda} \psi \right)} \delta \left( \partial_{\lambda} \psi \right)  \right].
\end{aligned}
\end{equation}
The variations of the spinor field under the symmetry transformation (\ref{snt_local_transformation}) read
\begin{equation}
\begin{aligned}
\delta \psi&=\frac{i}{4} \left( \partial_{\alpha} \xi_{\beta} \right) \sigma^{\alpha \beta} \psi \\
\delta \bar{\psi} &= -\frac{i}{4} \left( \partial_{\alpha} \xi_{\beta} \right) \bar{\psi} \sigma^{\alpha \beta} \\
\delta \left( \partial_{\mu} \psi \right) &= \frac{i}{4} \left( \partial_{\alpha} \xi_{\beta} \right) \sigma^{\alpha \beta} \left( \partial_{\mu} \psi \right)-\left( \partial_{\mu} \xi^{\lambda} \right) \left( \partial_{\lambda} \psi \right)\\
\delta \left( \partial_{\mu} \bar{\psi} \right) &= - \frac{i}{4} \left( \partial_{\alpha} \xi_{\beta} \right) \left( \partial_{\mu} \bar{\psi} \right) \sigma^{\alpha \beta}  -\left( \partial_{\mu} \xi^{\lambda} \right) \left( \partial_{\lambda} \bar{\psi} \right),\\
\end{aligned}
\end{equation}
where $\sigma^{\mu \nu} = \frac{i}{2} \left[ \gamma^{\mu} , \gamma^{\nu} \right]$.

Let us apply the Noether's second theorem to derive the energy-momentum tensor for the linear model of Weyl semimetal (\ref{Model-Weyl-lin}). The corresponding Lagrangian density is
\begin{equation}
\begin{aligned}
\mathcal{L}\left[ \psi , \bar{\psi} \right] &= \bar{\psi} \left[ \gamma_{\mu} \left( i \partial^{\mu} +\gamma_5 b^{\mu} \right) \right] \psi,\quad \quad b^{\mu}=\left( 0 , \mathbf{b} \right).
\end{aligned}
\end{equation}
Assuming $\mathbf{b} = b \mathbf{e}_{z}$, the Lagrangian $\mathcal{L}$ is invariant under the following local transformation:
\begin{equation}
\begin{aligned}
\xi^{1}(x)&=a^{1}+\omega^1_2 x^{2}\\
\xi^{2}(x)&=a^{2}+\omega^2_1 x^{1}\\
\xi^{3}(x)&=a^{3},\\
\end{aligned}
\end{equation}
where $\omega^2_1=-\omega^1_2$ and $a^{\mu}=\text{const}$. Hence, coefficients $\mathcal{D}^{\mu \nu}$ are given by
\begin{equation}
\begin{aligned}
\mathcal{D}^{\mu \nu}\left[ \psi \right]&=\frac{i}{4}  \bar{\psi} \gamma_{\lambda} \gamma_5 b^{\lambda} \sigma^{\mu \nu} \psi \left( \delta^{\mu}_{1} \delta^{\nu}_{2}+\delta^{\mu}_{2} \delta^{\nu}_{1} \right)\\
\mathcal{D}^{\mu \nu} \left[ \bar{\psi} \right] &= - \frac{i}{4}  \bar{\psi} \sigma^{\mu \nu} \left[ \gamma_{\lambda} \left( i \partial^{\lambda} +\gamma_5 b^{\lambda} \right) \right] \psi\left( \delta^{\mu}_{1} \delta^{\nu}_{2}+\delta^{\mu}_{2} \delta^{\nu}_{1} \right)  \\
\mathcal{D}^{\mu \nu} \left[ \partial \psi \right] &= i \bar{\psi} \gamma^{\nu} \left( \partial^{\mu} \psi \right)  - \frac{1}{4}  \bar{\psi} \gamma^{\lambda}  \sigma^{\mu \nu} \left( \partial_{\lambda} \psi \right)\left( \delta^{\mu}_{1} \delta^{\nu}_{2}+\delta^{\mu}_{2} \delta^{\nu}_{1} \right)  \\
\mathcal{D}^{\mu \nu} \left[ \partial \bar{\psi} \right] &= 0   \\
\end{aligned}
\end{equation}
and the energy-momentum tensor is
\begin{equation}
\begin{aligned}
\hat{T}^{\mu \nu}&=  \mathcal{D}^{\mu \nu} \left[ \psi \right] -g^{\mu \nu} \mathcal{L} = i \bar{\psi} \gamma^{\nu} \left( \partial^{\mu} \psi \right)
-\left\{\frac{1}{4}\bar{\psi} \left[ \gamma^{\lambda} ,  \sigma^{\mu \nu}  \right]\left( \partial_{\lambda} \psi \right)  +\frac{i}{4}  \bar{\psi} \left[ \sigma^{\mu \nu} ,  \gamma_{\lambda} \gamma_5 \right] b^{\lambda}  \psi \right\}
\left( \delta^{\mu}_{1} \delta^{\nu}_{2}+\delta^{\mu}_{2} \delta^{\nu}_{1} \right)  - g^{\mu \nu} \mathcal{L}. \\
\end{aligned}
\end{equation}
By using the field equations for $\psi$ and performing the Fourier transform, we finally obtain the energy-momentum tensor
\begin{equation}
\begin{aligned}
T_{\mu \nu}&= \gamma_{0} \gamma_{\nu} k_{\mu}  + \frac{1}{2} \left[   \gamma_{0} \gamma_{\mu} k_{\nu} - \gamma_{0} \gamma_{\nu}k_{\mu} +  \left( b_{\nu} \gamma_{0} \gamma_{\mu} \gamma_5 -b_{\mu}\gamma_{0}  \gamma_{\nu} \gamma_5 \right)  \right] \left( \delta^{\mu}_{1} \delta^{\nu}_{2}+\delta^{\mu}_{2} \delta^{\nu}_{1} \right).
\end{aligned}
\end{equation}
The result coincides with the stress tensor in Eq.~(\ref{Kubo-Weyl-T-b}).

\subsection{Canonical stress tensor in tight-binding model}
\label{sec:App-1-canon}

Let us provide a different derivation of the stress tensor. We start with the following tight-binding Hamiltonian on cubic lattice:~\cite{Shapourian-Ryu:2015,Cortijo-Vozmediano:2015}
\begin{equation}
\label{App-stress-H}
\hat{H}_{0} =\frac{1}{2} \sum_{i,j} c_{i}^{\dagger} \left( i t \gamma_0 \gamma_{j} -r \gamma_0 \right) c_{i+j}+(m+3 r) \sum_{i} c_{i}^{\dagger}  \gamma_0  c_{i} +\sum_{i,l} b_{l} c_i^{\dagger}\gamma_0 \gamma_l \gamma_5 c_{i}+\text{H.c.},
\end{equation}
where $j$ labels the six next nearest neighbors of the site $i$. Strains lead to the following changes in hopping parameters:
\begin{equation}
\label{App-stress-r}
r \rightarrow r(1-\beta u_{jj}),\\
\end{equation}
\begin{equation}
\label{App-stress-t}
t \gamma_0 \gamma_j \rightarrow t \left(1-\beta \sum_{j'} \frac{\partial u_j}{\partial x_j'} \gamma_0 \gamma_{j'} \right),
\end{equation}
where $u_{\mu \nu}= \partial u_\mu/\partial x_\nu$. By using Eqs.~(\ref{App-stress-r}) and (\ref{App-stress-t}) in Eq.~(\ref{App-stress-H}), we obtain
\begin{equation}
\begin{aligned}
\hat{H}_{u} &= \hat{H}_{0} + \frac{1}{2} \sum_{i,j} c_i^{\dagger} i t \bigg( - \beta u_{jj} \gamma_0 \gamma_j - \beta \sum_{j' \neq j} u_{jj'} \gamma_0 \gamma_{j'} \bigg)  c_{i+j}
+\frac{1}{2} \sum_{i,j} c_i^{\dagger}  r \beta u_{jj} \gamma_0  c_{i+j}\\
&+\bigg(m -r \beta \sum_{j}  u_{jj}\bigg) \sum_{i} c_i^{\dagger}  \gamma_0  c_{i} +\sum_{i,l} b_{l} c_i^{\dagger} \gamma_0 \gamma_{l} \gamma_5 c_{i}+\text{H.c.}
\end{aligned}
\end{equation}
Then, by using the standard definition of the stress tensor
\begin{equation}
\hat{T}_{\mu \nu} = \frac{\delta \hat{H}_{u}}{\delta u_{\mu \nu}} \bigg|_{u_{\mu \nu}=0},
\end{equation}
we find
\begin{equation}
\hat{T}_{\mu \nu} =-\frac{\beta}{2}\sum_{i} c_i^{\dagger} \left( i t \gamma_0 \gamma_{\nu} - r \delta_{\mu \nu} \gamma_0 \right) c_{i+\mu} - r \beta \delta_{\mu \nu}  \sum_{i} c_i^{\dagger}  \gamma_0c_{i}+\text{H.c.}
\end{equation}
By performing the Fourier transform, taking the continuum limit, i.e., setting the lattice constant $a$ to zero, $a\to0$, and replacing $t a \rightarrow v$, we obtain canonical stress tensor for the linearized model
\begin{equation}
\label{App-stress-T-def}
T_{\mu \nu} = \beta v  k_{\mu} \gamma_0 \gamma_{\nu}.
\end{equation}
It corresponds only to the orbital part of the strain transformation generator in (\ref{strain-transformation-generator}), i.e., $L_{\mu\nu}$.

By performing the Fourier transform and projecting out the high-energy states, Hamiltonian (\ref{App-stress-H}) can be reduced to the Hamiltonian of the two-band model: $H=\bm{\sigma}\cdot \mathbf{d}(\mathbf{k})$, see Ref.~\cite{Cortijo-Vozmediano:2015} for details. The corresponding stress tensor is
\begin{equation}
T_{\mu \nu} = \beta k_{\mu} \partial_{\nu} \left( d_{\nu} \right) \sigma_{\nu},
\end{equation}
where $d_x=v k_x$, $d_y=v k_y$, and $d_z=\left(b_z -m\right)-\frac{v^2}{m+b_z}k_z^2$ in the notation of~\cite{Cortijo-Vozmediano:2015}. Thus, we obtained the canonical stress tensor for the two-band model defined in Eq.~(\ref{Kubo-Weyl-two-band-H-def}).~\footnote{Compared with Ref.~\cite{Cortijo-Vozmediano:2015}, we changed sign at $d_z$.}

\section{Spectral function}
\label{sec:App-Spectral-function}

The retarded $(+)$ and advanced $(-)$ Green's functions read\\
\begin{equation}
G\left( \omega \pm i0,\mathbf{k}\right)=\frac{i}{\omega+\mu\pm i0 -H(\mathbf{k})},
\end{equation}
where $\omega$ is frequency (measured in energy units), $\mu$ is the chemical potential, $\mathbf{k}$ is momentum, and $H(\mathbf{k})$ is the Hamiltonian.

The spectral function is defined as the difference between the advanced and retarded Green's functions at vanishing chemical potential
\begin{equation}
A\left( \omega ;\mathbf{k}\right)=\frac{1}{2\pi}\left[G\left( \omega + i0,\mathbf{k}\right)-G\left( \omega - i0,\mathbf{k}\right)\right]_{\mu=0}.
\end{equation}

For the linearized model with broken TRS, see Eq.~(\ref{Kubo-Dirac-H-def}), the spectral function is given by $A(\omega; \mathbf{k})= \sum_{\lambda=\pm} \frac{1+\lambda \tau_z}{2} \otimes A_{\lambda}(\omega;\mathbf{k})$, where $\tau_z$ acts in the nodal space and
\begin{equation}
A_{\lambda}(\omega;\mathbf{k})=\frac{1}{2}\left[\delta(\omega - \varepsilon_{\mathbf{k},\lambda}) +\delta(\omega + \varepsilon_{\mathbf{k},\lambda})\right] +\frac{\lambda  v_{F}}{2\varepsilon_{\mathbf{k},\lambda}} \left[\bm{\sigma} \cdot \left(\mathbf{k}-\lambda \mathbf{b}\right)\right] \left[\delta(\omega - \varepsilon_{\mathbf{k},\lambda}) -\delta(\omega + \varepsilon_{\mathbf{k},\lambda})\right].
\end{equation}
Here, $\varepsilon_{\mathbf{k},\lambda}=v_F |\mathbf{k}- \lambda \mathbf{b}|$ is the dispersion relation of quasiparticles with chirality $\lambda$.

Similarly, for the two-band model with $H(\mathbf{k}) = \left(\bm{\sigma}\cdot \mathbf{d}(\mathbf{k})\right)$, we obtain
\begin{equation}
A(\omega; \mathbf{k}) =\frac{1}{2}\left[\delta(\omega - \varepsilon_{\mathbf{k}}) +\delta(\omega + \varepsilon_{\mathbf{k}})\right] +\frac{1}{2\varepsilon_{\mathbf{k}}}\left[\delta(\omega - \varepsilon_{\mathbf{k}}) -\delta(\omega + \varepsilon_{\mathbf{k}})\right] H(\mathbf{k}),
\end{equation}
where $\varepsilon_{\mathbf{k}}=\left|\mathbf{d} (\mathbf{k}) \right|$.

\section{Viscoelasticity tensor in the Kubo approach}
\label{sec:App-Kubo}

The viscoelasticity tensor is defined in Eq.~(\ref{Kubo_viscoelastic_tensor}). Its real part reads as
\begin{equation}
\label{app-kubo-eta}
\begin{aligned}
\text{Re}\,\eta_{\mu \nu \alpha \beta}(\Omega)
&=\frac{ \text{Im} \,
C_{\mu \nu \alpha \beta}(\Omega)}
{\Omega}+ \pi \delta(\Omega) \left[ \text{Im}
\left< [T_{\mu \nu}^{\,\,\,}(0), J_{\alpha \beta}^{\,\,\,}(0)] \right>
+ \delta_{\alpha \beta}^{\,\,\,} \left< T_{\mu \nu}^{\,\,\,} \right>  - \delta_{\mu \nu} \delta_{\alpha \beta}\kappa^{-1} - \text{Re} \,C_{\mu \nu \alpha \beta}(\Omega) \right] \\
&=\frac{1}{\Omega}  \text{v.p.} \int_{-\infty}^{+\infty} d \omega \int_{-\infty}^{+\infty} d \omega' \frac{f(\omega)-f(\omega')}{\omega'-\omega-\Omega}  \int \frac{d^3k}{(2\pi)^3} \text{Im} \, \text{tr} \left[ T_{\mu \nu}^{\,\,\,} (\mathbf{k}) A(\omega;\mathbf{k}) T_{\alpha \beta}^{\,\,\, } (\mathbf{k}) A(\omega';\mathbf{k}) \right]\\
&+\frac{\pi}{\Omega} \int_{-\infty}^{+\infty} d \omega \left[f(\omega)-f(\omega+\Omega)\right]  \int \frac{d^3 k}{(2\pi)^3} \text{Re}\, \text{tr} \left[ T_{\mu \nu}^{\,\,\, } (\mathbf{k}) A(\omega;\mathbf{k}) T_{\alpha\beta}^{\,\,\, } (\mathbf{k}) A(\omega+\Omega;\mathbf{k}) \right]  \\
&+\pi \delta(\Omega) \, \text{v.p.} \int_{-\infty}^{+\infty} d \omega \int_{-\infty}^{+\infty} d \omega' \frac{f(\omega)-f(\omega')}{\omega-\omega'}  \int \frac{d^3 k}{(2\pi)^3} \text{Re} \, \text{tr} \left[ T_{\mu \nu}^{\,\,\,} (\mathbf{k}) A(\omega;\mathbf{k}) T_{\alpha \beta}^{\,\,\, } (\mathbf{k}) A(\omega';\mathbf{k}) \right]  \\
&+\pi \delta(\Omega) \delta_{\alpha \beta}  \int_{-\infty}^{+\infty} d\omega f(\omega) \int \frac{d^3 k}{(2\pi)^3} \text{tr}\left[  T_{\mu \nu}(\mathbf{k}) A(\omega; \mathbf{k}) \right] -\pi \delta(\Omega )\kappa^{-1} \delta_{\mu \nu} \delta_{\alpha \beta}\\
&+\pi \delta(\Omega) \,
\int_{-\infty}^{+\infty} d\omega f(\omega) \int \frac{d^3 k}{(2\pi)^3} \text{Im}\, \text{tr}\left\{ [T_{\mu \nu}^{\,\,\,}, J_{\alpha \beta}^{\,\,\,} ] A(\omega; \mathbf{k}) \right\}
\\
&=\eta_{\mu \nu \alpha \beta}^{(1)}(\Omega)+\eta_{\mu \nu \alpha \beta}^{(2)}(\Omega)+\eta_{\mu \nu \alpha \beta}^{(3)}(\Omega)+\eta_{\mu \nu \alpha \beta}^{(4)}(\Omega)-\pi \delta(\Omega) \kappa^{-1} \delta_{\mu \nu} \delta_{\alpha \beta} +\eta_{\mu \nu \alpha \beta}^{(5)}(\Omega).
\end{aligned}
\end{equation}
In the subsequent Sections, we calculate $\eta_{\mu \nu \alpha \beta}^{(i)}(\Omega)$ in the linearized and two-band models of Weyl semimetals.

\subsection{Linearized model of Weyl semimetals}
\label{sec:App-Kubo-lin}

In the linearized model, we start with the following strain transformation generator:
\begin{equation}
J_{\alpha \beta}=-\frac{1}{2} \left \{ x_{\beta} , k_{\alpha} \right\} + \frac{i}{8} C_{\alpha \beta} \left[ \gamma_{\beta}, \gamma_{\alpha} \right],
\end{equation}
whose second term includes the generic form of internal angular momentum. The corresponding stress tensor is
\begin{equation}
T_{\mu \nu}=v_F \gamma_0 \gamma_{\nu} k_{\mu} +C_{\mu \nu} \frac{v_F}{2}\left( \gamma_0 \gamma_{\mu} k_{\nu}-\gamma_0 \gamma_{\nu} k_{\mu} \right) +C_{\mu \nu} \frac{v_F}{2}\left( \gamma_0 \gamma_{\mu} \gamma_5 b_{\nu}-\gamma_0 \gamma_{\nu} \gamma_5 b_{\mu} \right).
\end{equation}
The canonical stress tensor (\ref{App-stress-T-def}) is reproduced by taking $C_{\alpha \beta}=0$.

By using the following expression for the imaginary part of the trace:
\begin{equation}
\label{app-kubo-tr-Im}
\begin{aligned}
\text{Im}\, \text{tr} \left[ T_{\mu \nu} (\mathbf{k}) A(\omega;\mathbf{k}) T_{\alpha \beta} (\mathbf{k}) A(\omega';\mathbf{k}) \right] &= \sum_{\lambda} \lambda \frac{v_F^3}{4 \varepsilon_{\mathbf{k},\lambda} }  \left[ \delta \left(\omega +\varepsilon_{\mathbf{k},\lambda}\right) \delta \left(\omega' -\varepsilon_{\mathbf{k},\lambda}\right) - \delta \left(\omega - \varepsilon_{\mathbf{k},\lambda}\right)  \delta \left(\omega' +\varepsilon_{\mathbf{k},\lambda}\right)\right] \\
&\times \bigg[ C_{\alpha \beta } \left( 2- C_{\mu \nu} \right)  \left( k_{j} -\lambda b_{j}\right)  k_{\beta } k_{\mu } \epsilon_{\nu \alpha j}+C_{\mu \nu } \left( 2- C_{\alpha \beta} \right)  \left( k_{j} -\lambda b_{j}\right)  k_{\alpha } k_{\nu } \epsilon_{\mu \beta j}\\
&+C_{\alpha \beta } C_{\mu \nu }  \left( k_{j} -\lambda b_{j}\right) k_{\beta } k_{\nu } \epsilon_{\mu \alpha j}+\left( 2- C_{\alpha \beta} \right) \left( 2- C_{\mu \nu} \right) \left( k_{j} -\lambda b_{j}\right)  k_{\alpha } k_{\mu }  \epsilon_{\nu \beta j} \bigg],
\end{aligned}
\end{equation}
we obtain the first term in Eq.~(\ref{app-kubo-eta}),
\begin{equation}
\label{app-kubo-eta-1}
\begin{aligned}
\eta_{\mu \nu \alpha \beta}^{(1)}(\Omega)&= \frac{1}{\Omega}  \text{v.p.} \int_{-\infty}^{+\infty} d \omega \int_{-\infty}^{+\infty} d \omega' \frac{f(\omega)-f(\omega')}{\omega'-\omega-\Omega}  \int \frac{d^3 k}{(2\pi)^3} \text{Im} \, \text{tr} \left[ T_{\mu \nu}^{\,\,\,} (\mathbf{k}) A(\omega;\mathbf{k}) T_{\alpha \beta}^{\,\,\, } (\mathbf{k}) A(\omega';\mathbf{k}) \right] \\
&=\left( 2- C_{\alpha \beta} \right) \left( 2- C_{\mu \nu} \right) I_{\mu \nu \alpha \beta}+  C_{\alpha \beta } \left( 2- C_{\mu \nu} \right) I_{\mu \nu \beta \alpha}  +C_{\mu \nu } \left( 2- C_{\alpha \beta} \right)  I_{\nu \mu \alpha \beta} +C_{\alpha \beta } C_{\mu \nu }  I_{\nu \mu \beta \alpha}.
\end{aligned}
\end{equation}
Here
\begin{equation}
\label{App_Hall_tensor-0}
I_{\mu \nu \alpha \beta}= \sum_{j=1}^{3} \epsilon_{j \nu \beta} I_{\mu \alpha}^{j}
\end{equation}
and
\begin{equation}
\label{App_Hall_tensor}
I_{\mu \alpha}^{j} =\sum_{\lambda} \lambda \int \frac{d^3 k}{(2\pi)^3} \left[ f \left(-\varepsilon_{\mathbf{k},\lambda}\right) - f \left(\varepsilon_{\mathbf{k},\lambda}\right)\right] \frac{v_F^3\left(k_j - \lambda b_j\right)  k_{\alpha } k_{\mu}}{2 \varepsilon_{\mathbf{k},\lambda} \left(4 \varepsilon_{\mathbf{k},\lambda}^2 -\Omega^2\right)}.
\end{equation}

For $\mu=0$, we obtain
\begin{equation}
\label{viscoelasticity_lin_Hall_Omega_12}
I_{\mu \alpha}^{i=1,2} = \left(\delta_{\mu i} \delta_{\alpha 3} +\delta_{\mu 3} \delta_{\alpha i} \right) \frac{b_z}{12 \pi^2 } \left[ 3\Lambda_z^2 -b_z^2 +\frac{ \Omega^2}{2 v_F^2} \left( \ln \frac{2 v_F \Lambda_z }{\Omega }+\frac{1}{3}\right) \right]
\end{equation}
and
\begin{equation}
\label{viscoelasticity_lin_Hall_Omega_3}
\begin{aligned}
I_{\mu \alpha}^{3}&= \delta_{\mu \alpha} \sum_{\lambda} \frac{v_F^3}{(2\pi)^3}  \int_{0}^{2\pi} d\phi \int_{-\infty}^{+\infty} dk_z\,  \lambda  \left( k_{z} -\lambda b_{z}\right) \left\{ \left(\delta_{\mu 1} +\delta_{\mu 2} \right) \left[ \frac{\Lambda_{\perp}}{4}+\frac{\Omega ^2-4 v_F^2 (k_z-\lambda b_z)^2}{8 \Omega} \right. \right.\\
&\left.\left. \times \left[\frac{i \pi}{2}+ \tanh ^{-1}\left(\frac{2 v_F \left| k_z -\lambda b_z\right|}{\Omega }\right) \right]  -\frac{1}{8} v_F \left| k_z -\lambda b_z\right| \right]  +  \delta_{\mu 3} \frac{v_F^2 k_{z}^2}{4 \Omega } \left[ \frac{i \pi}{2}+  \tanh ^{-1}\left(\frac{2 v_F \left| k_z -\lambda b_z\right|}{\Omega }\right) \right] \right\}\\
&=\delta_{\mu \alpha} \frac{b_z}{4\pi^2} \left\{ \left(\delta_{\mu 1} +\delta_{\mu 2} \right)
\left( -\Lambda_{\perp} \Lambda_z +2 \Lambda_{z}^2 +\frac{2 b_z^2}{3} -\frac{\Omega^2}{6 v_F^2}\right) + \delta_{\mu 3} \left[-\frac{b_z^2}{6} + \frac{\Omega^2}{12 v_F^2}\left( \ln \frac{2 v_F \Lambda_z }{\Omega }-\frac{1}{6} \right) \right] \right\},\\
\end{aligned}
\end{equation}
where $\Lambda_{\perp}$ and $\Lambda_z$ are momentum cutoffs.

Combining the expressions for $\eta_{\mu \nu \alpha \beta}^{(2)}(\Omega)$ and $\eta_{\mu \nu \alpha \beta}^{(3)}(\Omega)$, and using the following expression for the real part of trace:
\begin{equation}
\label{app-kubo-tr-Re}
\begin{aligned}
\text{Re} \, \text{tr} \left[ T_{\mu \nu} (\mathbf{k}) A(\omega;\mathbf{k}) T_{\alpha \beta} (\mathbf{k}) A(\omega';\mathbf{k}) \right]&= \sum_{\lambda} \frac{v_F^2 }{4} \Big\{ \left[ \delta(\omega-\varepsilon_{\mathbf{k},\lambda}) \delta(\omega'+\varepsilon_{\mathbf{k},\lambda})+\delta(\omega+\varepsilon_{\mathbf{k},\lambda}) \delta(\omega'-\varepsilon_{\mathbf{k},\lambda}) \right] \\
&\times \left[ \delta _{\nu \beta} k_{\alpha } k_{\mu } \left(2-C_{\alpha \beta } \right) \left(2-C_{\mu \nu } \right)  +  \delta _{\nu \alpha }  k_{\beta } k_{\mu }  C_{\alpha \beta }\left(2-C_{\mu \nu } \right)  \right. \\
&\left.+
\delta _{\mu \beta } k_{\alpha } k_{\nu } C_{\mu \nu } \left(2-C_{\alpha \beta } \right)  +  \delta _{\mu \alpha } k_{\beta } k_{\nu }  C_{\alpha \beta } C_{\mu \nu } \right] \\
&+ \frac{v_F^2}{\varepsilon_{\mathbf{k},\lambda}^2} \left[\delta(\omega-\varepsilon_{\mathbf{k},\lambda})-\delta(\omega+\varepsilon_{\mathbf{k},\lambda}) \right]\left[\delta(\omega'-\varepsilon_{\mathbf{k},\lambda})-\delta(\omega'+\varepsilon_{\mathbf{k},\lambda}) \right]\\
&\times \big[   \left(2 -C_{\alpha \beta} \right) \left(2 -C_{\mu \nu} \right)  \left(k_{\nu} -\lambda b_\nu \right)  \left(k_{\beta} -\lambda b_\beta \right)  k_{\alpha } k_{\mu }    \\
&+ C_{\alpha \beta } \left(2 -C_{\mu \nu} \right)  \left(k_{\nu} -\lambda b_\nu \right)  \left(k_{\alpha} -\lambda b_\alpha \right)  k_{\beta } k_{\mu } \\
&+ \left(2 -C_{\alpha \beta} \right) C_{\mu \nu }  \left(k_{\mu} -\lambda b_\mu \right)  \left(k_{\beta} -\lambda b_\beta \right)  k_{\alpha } k_{\nu } \\
&+ C_{\alpha \beta } C_{\mu \nu }  \left(k_{\mu} -\lambda b_\mu \right)  \left(k_{\alpha} -\lambda b_\alpha \right)  k_{\beta } k_{\nu } \big] \Big\},
\end{aligned}
\end{equation}
we derive
\begin{equation}
\label{app-kubo-eta-23}
\begin{aligned}
\eta_{\mu \nu \alpha \beta}^{(2)}(\Omega)+\eta_{\mu \nu \alpha \beta}^{(3)}(\Omega)&=\frac{\pi}{\Omega} \int_{-\infty}^{+\infty} d \omega \left[f(\omega)-f(\omega+\Omega)\right] \int \frac{d^3 k}{(2\pi)^3} \text{Re} \, \text{tr} \left[ T_{\mu \nu}^{\,\,\, } (\mathbf{k}) A(\omega;\mathbf{k}) T_{\alpha\beta}^{\,\,\, } (\mathbf{k}) A(\omega+\Omega;\mathbf{k}) \right]  \\
&+\pi \delta(\Omega) \, \text{v.p.} \int_{-\infty}^{+\infty} d \omega \int_{-\infty}^{+\infty} d \omega' \frac{f(\omega)-f(\omega')}{\omega-\omega'}  \int \frac{d^3 k}{(2\pi)^3} \text{Re} \, \text{tr} \left[ T_{\mu \nu}^{\,\,\,} (\mathbf{k}) A(\omega;\mathbf{k}) T_{\alpha \beta}^{\,\,\, } (\mathbf{k}) A(\omega';\mathbf{k}) \right] \\
&=\Theta \left( \Omega -2 \mu \right) \Bigg\{ \frac{\Omega^3}{120 \pi v_F^3} \left[      \delta_{\mu \alpha } \delta _{\nu \beta} \left(  1- \frac{5}{8} \left( C_{\alpha \beta }+ C_{\mu \nu } - C_{\alpha \beta }C_{\mu \nu } \right) \right)  -\frac{1}{4} \delta_{\mu \beta } \delta _{\nu \alpha} \right. \\
&\left. \times \left( 1 - \frac{5}{2} \left( C_{\alpha \beta }+ C_{\mu \nu } - C_{\alpha \beta }C_{\mu \nu } \right) \right)    - \frac{1}{4}  \delta_{\mu \nu} \delta_{\alpha \beta} \right] + \frac{\Omega}{48 \pi v_F} \left[  \left(2 -C_{\alpha \beta } \right) \left(2 -C_{\mu \nu} \right)  \delta_{\nu \beta}  b_{\mu } b_{\alpha }  \right.  \\
& \left.  + C_{\alpha \beta  } \left(2 -C_{\mu \nu} \right) \delta_{\nu \alpha} b_{\mu } b_{\beta }+  \left(2 -C_{\alpha \beta } \right) C_{\mu \nu }  \delta_{\mu \beta} b_{\nu } b_{\alpha }  + C_{\alpha \beta} C_{\mu \nu }  \delta_{\mu \alpha}  b_{\nu } b_{\beta }   \right]  \Bigg\}  \\
&+\frac {\mu^2}{12 \pi v_F }  \delta(\Omega) \left[ \left(2 -C_{\alpha \beta} \right) \left(2 -C_{\mu \nu} \right)  \delta_{\nu \beta}  b_{\mu } b_{\alpha } +C_{\alpha \beta} \left(2 -C_{\mu \nu} \right)  \delta_{\nu \alpha}  b_{\mu } b_{\beta } +\left(2 -C_{\alpha \beta} \right) C_{\mu \nu}  \delta_{\mu \beta}  b_{\nu } b_{\alpha }  \right. \\
&\left.  +C_{\alpha \beta} C_{\mu \nu} \delta_{\mu \alpha}  b_{\nu } b_{\beta }  \right] +\frac{\mu^4}{120 \pi v_F^3} \delta(\Omega) \left( 5 C_{\mu \nu}+5 C_{\alpha \beta }-5 C_{\mu \nu}C_{\alpha \beta }-2\right) \delta_{\mu \beta} \delta_{\nu \alpha}\\
&+\frac{\mu^4}{120 \pi v_F^3} \delta(\Omega) \left( 8-5 C_{\mu \nu}-5 C_{\alpha \beta }+5 C_{\mu \nu}C_{\alpha \beta }\right) \delta_{\mu \alpha} \delta_{\nu \beta}-\frac{\mu^4}{60 \pi v_F^3} \delta(\Omega) \delta_{\mu \nu} \delta_{\alpha \beta}.
\end{aligned}
\end{equation}
Here, we subtracted the contribution of completely filled states in terms with $\df{\Omega}$.

The fourth $\eta_{\mu \nu \alpha \beta}^{(4)}(\Omega)$ and fifth $\kappa^{-1}$ terms in Eq.~(\ref{app-kubo-eta}) are
\begin{equation}
\label{app-kubo-eta-4}
\begin{aligned}
\eta_{\mu \nu \alpha \beta}^{(4)}(\Omega)&= \pi \delta(\Omega) \delta_{\alpha \beta}  \int_{-\infty}^{+\infty} d\omega f(\omega) \int \frac{d^3 k}{(2\pi)^3} \text{tr}\left[  T_{\mu \nu}(\mathbf{k}) A(\omega; \mathbf{k}) \right] =\delta(\Omega) \delta_{\mu \nu}\delta_{\alpha \beta} \frac{\mu^4}{12 \pi v_F^3},\\
\kappa^{-1}&=-V\frac{\partial P}{\partial V}=\left(\epsilon+P\right)\frac{\partial P}{\partial \epsilon }=  \frac{\mu^4}{9 \pi^2 v_F^3},
\end{aligned}
\end{equation}
where we also subtracted the contribution of completely filled states and used
\begin{equation}
\epsilon= \int_{-\infty}^{+\infty} d\omega f(\omega) \int \frac{d^3 k}{(2\pi)^3} \text{tr}\left[  H(\mathbf{k}) A(\omega; \mathbf{k}) \right]= \frac{\mu^4}{4 \pi^2 v_F^3},
\end{equation}
\begin{equation}
\text{tr} \left[ T_{\mu \nu} (\mathbf{k}) A(\omega;\mathbf{k}) \right] = \sum_{\lambda} \frac{v_F^2}{\varepsilon_{\mathbf{k},\lambda}}  \left[\delta(\omega - \varepsilon_{\mathbf{k},\lambda}) -\delta(\omega + \varepsilon_{\mathbf{k},\lambda})\right] \left[k_{\mu} k_{\nu}+C_{\mu \nu} \lambda \left( k_{\mu} b_{\nu}- k_{\nu} b_{\mu} \right) -\lambda k_{\mu} b_{\nu} \right].
\end{equation}
Here, $\kappa$ is the isentropic compressibility at fixed particle number, $V$ and $P$ are volume and pressure, and $\epsilon$ is the energy density.

The last term in the last line in Eq.~(\ref{app-kubo-eta}) reads
\begin{equation}
\label{app-kubo-eta-5}
\begin{aligned}
\eta_{\mu \nu \alpha \beta}^{(5)}(\Omega)&=\pi \delta(\Omega) \, \text{Im} \int_{-\infty}^{+\infty} d\omega f(\omega) \int \frac{d^3 k}{(2\pi)^3} \text{tr}\left\{  [T_{\mu \nu}^{\,\,\,}, J_{\alpha \beta}^{\,\,\,} ] A(\omega; \mathbf{k}) \right\}\\
&= \delta(\Omega) \frac{\mu^4}{12 \pi v_F^3} \left( \frac{C_{\mu \nu}-C_{\alpha \beta}+C_{\mu \nu}C_{\alpha \beta}}{2} \delta_{\mu \alpha} \delta_{\nu \beta}   +  \frac{2-C_{\mu \nu}+C_{\alpha \beta}-C_{\mu \nu}C_{\alpha \beta}}{2} \delta_{\mu \beta} \delta_{\nu \alpha} \right),
\end{aligned}
\end{equation}
where we subtracted the contribution of completely filled states and used
\begin{equation}
\begin{aligned}
\text{tr}\left\{ \left[T_{\mu \nu} (\mathbf{k}),J_{\alpha \beta}\right] A(\omega;\mathbf{k}) \right\} &= \sum_{\lambda} \frac{i v_F^2}{ \varepsilon_{\mathbf{k},\lambda}}  \left[\delta(\omega - \varepsilon_{\mathbf{k},\lambda}) -\delta(\omega + \varepsilon_{\mathbf{k},\lambda})\right] \\
&\times \Big\{ \frac{C_{\mu \nu}}{2}(k_{\mu}-\lambda b_{\mu})  k_{\alpha} \delta_{\nu \beta}  +\frac{C_{\alpha \beta}}{2} \left( 1-\frac{C_{\mu \nu}}{2}\right) \left[ (k_{\beta}-\lambda b_{\beta})  k_{\mu} \delta_{\nu \alpha} -(k_{\alpha}-\lambda b_{\alpha})  k_{\mu} \delta_{\nu \beta} \right] \\
&+\left( 1-\frac{C_{\mu \nu}}{2}\right)(k_{\nu}-\lambda b_{\nu}) k_{\alpha} \delta_{\mu \beta}+\frac{C_{\alpha \beta} C_{\mu \nu}}{4} \left[(k_{\beta}-\lambda b_{\beta})  k_{\nu} \delta_{\mu \alpha} -(k_{\alpha}-\lambda b_{\alpha})  k_{\nu} \delta_{\mu \beta} \right] \\
&+\lambda \frac{C_{\alpha \beta} C_{\mu \nu}}{4} \left[ (k_{\beta}-\lambda b_{\beta})  b_{\nu} \delta_{\mu \alpha} - (k_{\alpha}-\lambda b_{\alpha}) b_{\nu} \delta_{\mu \beta} - (k_{\beta}-\lambda b_{\beta})  b_{\mu} \delta_{\nu \alpha} + (k_{\alpha}-\lambda b_{\alpha})  b_{\mu} \delta_{\nu \beta} \right] \Big\}.
\end{aligned}
\end{equation}

By combining terms in Eqs.~(\ref{app-kubo-eta-1}), (\ref{app-kubo-eta-23}), (\ref{app-kubo-eta-4}), and (\ref{app-kubo-eta-5}), we obtain the viscoelasticity tensor whose different components are given in Eqs.~(\ref{Kubo-Dirac-eta}), (\ref{Kubo-Dirac-eta-1}), (\ref{Kubo-Weyl-eta-stat}), (\ref{Kubo-Weyl-eta-dyn}), (\ref{Kubo-Weyl-eta-dyn-1}), and (\ref{Kubo-Weyl-eta-Hall}).

\subsection{Two-band model of Weyl semimetals}
\label{sec:App-Kubo-two-band}

In this Section, we consider a two-band model of Weyl semimetals with the Hamiltonian $H(\mathbf{k})=\bm{\sigma}\cdot \mathbf{d}(\mathbf{k})$, where $\mathbf{d}=\left\{v_F k_x, v_F k_y, \gamma(k_z^2-m)\right\}$. The strain transformation generator and the stress tensor are
\begin{equation}
J_{\alpha \beta} =-\frac{1}{2} \left \{ x_{\beta} , k_{\alpha} \right\} + \frac{i}{8} C_{\alpha \beta} \left[ \sigma_{\beta}, \sigma_{\alpha} \right]
\end{equation}
and
\begin{equation}
T_{\alpha \beta} =\sigma_i  k_\alpha \partial_{\beta}\left( d_i \right) + \frac{1}{2} C_{\alpha \beta} \left( d_{\beta} \sigma_{\alpha}-d_{\alpha} \sigma_{\beta} \right),
\end{equation}
respectively. For the canonical stress tensor, $C_{\alpha \beta}=0$.

In what follows, we focus on the viscoelasticity tensor for the canonical stress tensor; terms related to the internal degrees of freedom $\propto C_{\alpha \beta}$ will be discussed at the end of this Section.

To calculate the viscoelasticity tensor (\ref{app-kubo-eta}), we use
\begin{equation}
\text{tr} \left[ T_{\mu \nu}^{\,\,\,} (\mathbf{k}) A(\omega;\mathbf{k})  \right] =\frac{1}{\varepsilon_{\mathbf{k}}}  \left[ \delta (\omega-\varepsilon_{\mathbf{k}}) -\delta( \omega+ \varepsilon_{\mathbf{k}})\right] d_i \partial_{\nu}\left( d_i \right) k_{\mu},
\end{equation}
\begin{equation}
\begin{aligned}
\text{Im} \, \text{tr} \left\{ \left[ T_{\mu \nu},J_{\alpha \beta} \right] A(\omega;\mathbf{k})  \right\}  &= \frac{1}{\varepsilon_{\mathbf{k}}}  \left[ \delta (\omega-\varepsilon_{\mathbf{k}}) -\delta( \omega+ \varepsilon_{\mathbf{k}})\right] \left\{ d_i k_{\alpha} \partial_{\beta} \left[ k_\mu \partial_{\nu}\left( d_i \right) \right]+  \frac{C_{\mu \nu}}{2}  \left[ k_\alpha \partial_{\beta}\left( d_\nu \right)d_{\mu}-k_\beta \partial_{\alpha}\left( d_\mu \right)d_{\nu}  \right] \right.\\
&\left. +\frac{C_{\alpha \beta}}{2}  \left[ k_{\mu} \partial_{\nu} \left( d_\beta \right) d_\alpha - k_{\nu} \partial_{\mu} \left( d_\alpha \right) d_\beta \right] +\frac{C_{\mu \nu} C_{\alpha \beta}}{4} \left( \delta_{\nu \beta}d_\alpha d_{\mu} -\delta_{\mu \beta} d_\alpha d_{\nu}+  \delta_{\mu \alpha}d_\beta d_{\nu}-\delta_{\nu \alpha}d_\beta d_{\mu} \right) \right\},
\end{aligned}
\end{equation}
\begin{equation}
\begin{aligned}
\text{Re} \, \text{tr} \left[ T_{\mu \nu} (\mathbf{k}) A(\omega;\mathbf{k}) T_{\alpha \beta} (\mathbf{k}) A(\omega';\mathbf{k}) \right] &= \left[ \delta\left( \omega -\varepsilon_{\mathbf{k}}\right) \delta\left( \omega' +\varepsilon_{\mathbf{k}}\right) + \delta\left( \omega +\varepsilon_{\mathbf{k}}\right) \delta\left( \omega' -\varepsilon_{\mathbf{k}}\right)\right] \bigg\{ k_{\alpha } k_{\mu } \partial_{\beta}\left( d_{i}\right) \partial_{\nu}\left( d_{i}\right)\\
&+ \frac{C_{\alpha \beta}}{2}\left[ k_{\mu } d_{\beta} \partial_{\nu}\left( d_{\alpha}\right)- k_{\mu } d_{\alpha} \partial_{\nu}\left( d_{\beta}\right)  \right] +\frac{C_{\mu \nu}}{2}\left[ k_{\alpha } d_{\nu} \partial_{\beta}\left( d_{\mu}\right)- k_{\alpha } d_{\mu} \partial_{\beta}\left( d_{\nu}\right) \right]\\
&+  \frac{C_{\alpha \beta} C_{\mu \nu}}{4} \left( \delta_{\mu \alpha} d_{\nu} d_{\beta}-\delta_{\nu \alpha} d_{\mu} d_{\beta}-\delta_{\mu \beta} d_{\nu} d_{\alpha}+\delta_{\nu \beta} d_{\mu} d_{\alpha}\right) \bigg\}\\
&+ \frac{1}{\varepsilon_{\mathbf{k}}^2} k_{\alpha } k_{\mu }  d_i  \partial_{\beta}\left( d_{i}\right) d_j \partial_{\nu}\left( d_{j}\right) \left[ \delta\left( \omega -\varepsilon_{\mathbf{k}}\right) - \delta\left( \omega +\varepsilon_{\mathbf{k}}\right) \right]\left[  \delta\left( \omega' -\varepsilon_{\mathbf{k}}\right) - \delta\left( \omega' +\varepsilon_{\mathbf{k}}\right)\right],
\end{aligned}
\end{equation}
and
\begin{equation}
\begin{aligned}
\text{Im} \, \text{tr} \left[ T_{\mu \nu} (\mathbf{k}) A(\omega;\mathbf{k}) T_{\alpha \beta} (\mathbf{k}) A(\omega';\mathbf{k}) \right] &=\frac{ \delta \left( \omega-\varepsilon_{\mathbf{k}} \right) \delta \left( \omega'+\varepsilon_{\mathbf{k}} \right)-\delta \left( \omega+\varepsilon_{\mathbf{k}} \right) \delta \left( \omega'-\varepsilon_{\mathbf{k}} \right)  }{ \varepsilon_{\mathbf{k}} } \bigg[d_i k_{\alpha } k_{\mu } \partial_{\nu}\left( d_{k}\right) \partial_{\beta}\left( d_{l}\right) \epsilon _{k i l} \\
& + \frac{C_{\alpha \beta}}{2} k_{\mu }d_{i} \partial_{\nu}\left( d_{j}\right) \left(  d_{\beta}   \epsilon _{j i \alpha} - d_{\alpha}  \epsilon _{j i \beta}  \right) +\frac{C_{\mu \nu}}{2} k_{\alpha }d_{i} \partial_{\beta}\left( d_{j}\right) \left(  d_{\nu}   \epsilon _{\mu i j} - d_{\mu}  \epsilon _{\nu i j}  \right)   \\
& +  \frac{C_{\alpha \beta} C_{\mu \nu}}{4} d_i \left( \epsilon _{\mu i \alpha} d_{\nu} d_{\beta}-\epsilon _{\mu i \beta} d_{\nu} d_{\alpha}-\epsilon _{\nu i \alpha} d_{\mu} d_{\beta}+\epsilon _{\nu i \beta} d_{\mu} d_{\alpha}\right) \bigg].
\end{aligned}
\end{equation}
For the first term in the last line in Eq.~(\ref{app-kubo-eta}), we obtain
\begin{equation}
\begin{aligned}
\eta_{\mu \nu \alpha \beta}^{(1)}(\Omega)
&= \text{v.p.}  \int \frac{d^3 k}{(2\pi)^3} \frac{2}{\varepsilon_{\mathbf{k}}} \frac{ 1-\Theta\left(|\mu|-\varepsilon_{\mathbf{k}} \right)}{4 \varepsilon_{\mathbf{k}}^2 -\Omega^2} d_i k_{\mu } k_{\alpha} \partial_{\nu}\left( d_{\nu}\right) \partial_{\beta}\left( d_{\beta}\right) \epsilon _{i \nu \beta } =\sum_{j=1}^{3} \epsilon_{j \nu \beta} I^{j}_{\mu \alpha},
\end{aligned}
\end{equation}
where
\begin{equation}
\begin{aligned}
I_{\mu \alpha}^{i=1,2}&= \int \frac{d^3 k}{(2\pi)^3}\frac{2 v_F^2 }{\varepsilon_{\mathbf{k} } }  \frac{\Theta\left(\varepsilon_{\mathbf{k}} -|\mu| \right)}{4 \varepsilon_{\mathbf{k}}^2 -\Omega^2}  \partial_z \left(d_z\right) k_i k_{\alpha } k_{\mu}
\end{aligned}
\end{equation}

\begin{equation}
\begin{aligned}
I_{\mu \alpha}^{3}&= \int \frac{d^3 k}{(2\pi)^3} \frac{2 v_F^2}{\varepsilon_{\mathbf{k}}}   \frac{\Theta\left(\varepsilon_{\mathbf{k}}-|\mu| \right)}{4 \varepsilon_{\mathbf{k}}^2 -\Omega^2}  d_z k_{\alpha } k_{\mu }.
\end{aligned}
\end{equation}
At $\Omega=0$ and $\mu=0$, we find
\begin{equation}
\label{viscoelasticity_two_band_Hall_12}
\begin{aligned}
I_{\mu \alpha}^{i=1,2}&= \left(\delta_{\mu 3} \delta_{\alpha i}+\delta_{\mu i} \delta_{\alpha 3} \right) \int \frac{d^3 k}{(2\pi)^3}\frac{v_F^2  \partial_z \left(d_z\right) }{2 \varepsilon_{\mathbf{k} }^3 } \frac{1}{2} k_\perp^2 k_{z}\\
&=\left(\delta_{\mu 3} \delta_{\alpha i}+\delta_{\mu i} \delta_{\alpha 3} \right)\left(- \frac{\gamma ^2 \Lambda_z^5 }{10 \pi ^2 v_F^2}+\frac{\gamma \Lambda_z^3  \Lambda_{\perp} }{12 \pi ^2 v_F}+ \frac{ \gamma ^2 m \Lambda_z^3}{6 \pi ^2 v_F^2}- \frac{2 \gamma ^2 m^{5/2}}{15 \pi ^2 v_F^2} \right),
\end{aligned}
\end{equation}
\begin{equation}
\label{viscoelasticity_two_band_Hall_3}
\begin{aligned}
I_{\mu \alpha}^{3}&=\delta_{\mu \alpha} \int \frac{d^3 k}{(2\pi)^3} \frac{v_F^2 d_z}{2\varepsilon_{\mathbf{k}}^3} \left[ \left( \delta_{\mu 1} + \delta_{\mu 2}\right) \frac{1}{2} k_{\perp}^2+\delta_{\mu 3}k_z^2\right] =\delta_{\mu \alpha} \bigg[ \left( \delta_{\mu 1} + \delta_{\mu 2}\right) \left(- \frac{\gamma ^2 \Lambda_z^5 }{20 \pi ^2 v_F^2}+\frac{ \gamma ^2 m \Lambda_z^3}{6 \pi ^2 v_F^2}+\frac{\gamma \Lambda_z^3   \Lambda_\perp }{24 \pi ^2 v_F}\right.\\
&\left.-\frac{\gamma ^2 m^2\Lambda_z  }{4 \pi ^2 v_F^2}-\frac{ \gamma m \Lambda_z\Lambda_\perp}{8 \pi ^2 v_F}+\frac{4 \gamma ^2 m^{5/2}}{15 \pi ^2 v_F^2}  \right)+\delta_{\mu 3}\frac{1}{12\pi^2}\left( \Lambda_z^3 -2 m^{3/2} \right)\bigg], \\
\end{aligned}
\end{equation}
where $\Lambda_{\perp}$ and $\Lambda_z$ are momentum cutoffs.

In the interband part of $\eta_{\mu \nu \alpha \beta}^{(2)}(\Omega)+\eta_{\mu \nu \alpha \beta}^{(3)}(\Omega)$, we have
\begin{equation}
\begin{aligned}
\eta_{\mu \nu \alpha \beta}^{\text{(inter)}}(\Omega)
&=\frac{\pi}{\Omega}  \Theta\left(\Omega-2\mu\right)  \int \frac{d^3 k}{(2\pi)^3}  \delta\left( \Omega - 2 \varepsilon_{\mathbf{k}}\right) \bigg\{ k_{\alpha } k_{\mu } \delta_{\nu \beta}\left[\partial_{\nu}\left( d_{\nu}\right)\right]^2 -\frac{1}{\varepsilon_{\mathbf{k}}^2} k_{\alpha } k_{\mu }  d_\beta  \partial_{\beta}\left( d_{\beta}\right) d_\nu \partial_{\nu}\left( d_{\nu}\right) \bigg\}\\
&=\Theta\left(\Omega -2\mu \right)\left(I^{\text{(inter)}}_{1}-I^{\text{(inter)}}_{2} \right).
\end{aligned}
\end{equation}
Here, we used the fact that in the model at hand, $d_i=d_i(k_i)$, and, therefore, $\partial_{\mu} (d_{\nu})=\partial_{\nu} (d_{\nu}) \delta_{\mu \nu}$. The integrals $I^{\text{(inter)}}_{1}$ and $I^{\text{(inter)}}_{2}$ read
\begin{equation}
\begin{aligned}
I^{\text{(inter)}}_{1} &=\frac{1}{32 \pi^2 v_F^2 }  \int_{0}^{2 \pi} d \phi \int_{-\infty}^{+\infty} dk_z  k_{\alpha } k_{\mu } \delta_{\nu \beta}\left[\partial_{\nu}\left( d_{\nu}\right)\right]^2   \Theta \left( \Omega^2 -4 d_z^2 \right) \bigg|_{v_F^2 k_{\perp}^2=\frac{\Omega^2}{4}-d_z^2}=\frac{\delta_{\mu \alpha} \delta_{\nu \beta} }{8 \pi v_F^2 } \Bigg\{ \frac{\left( \delta_{\mu 1}+\delta_{\mu 2} \right) \left( \delta_{\nu 1}+\delta_{\nu 2} \right)}{30} \\
&\times  \left[ \left(3 \Omega ^2+2 \gamma m \Omega -8 \gamma ^2 m^2\right) \sqrt{\frac{2 \gamma m +\Omega}{2 \gamma}}  - \Theta\left(2\gamma m -\Omega \right) \left(3 \Omega^2-2 \gamma m \Omega -8 \gamma ^2 m^2\right)  \sqrt{\frac{2 \gamma m -\Omega}{2 \gamma}}  \right] \\
&+ \frac{\delta_{\mu 3} \delta_{\nu 3}}{5}\left[ \left( 2\gamma m +\Omega \right)^2 \sqrt{\frac{2 \gamma m + \Omega}{2 \gamma}}  - \Theta\left(2\gamma m -\Omega \right) \left( 2\gamma m -\Omega \right)^2\sqrt{\frac{2 \gamma m -\Omega}{2 \gamma}}  \right] \\
& - \frac{\gamma \delta_{\nu 3} \left( \delta_{\mu 1}+\delta_{\mu 2} \right) }{105 v_F^2}\left[ \left( 4\gamma m -5\Omega \right)\left( 2\gamma m +\Omega \right)^2 \sqrt{\frac{2 \gamma m + \Omega}{2 \gamma}}  - \Theta\left(2\gamma m -\Omega \right) \left( 4\gamma m +5\Omega \right)\left( 2\gamma m -\Omega \right)^2  \sqrt{\frac{2 \gamma m -\Omega}{2 \gamma}}  \right] \\
&+ \frac{v_F^2 \delta_{\mu 3} \left( \delta_{\nu 1}+\delta_{\nu 2} \right) }{6 \gamma }\left[ \left( 2\gamma m +\Omega \right) \sqrt{\frac{2 \gamma m +\Omega}{2 \gamma}}  - \Theta\left(2\gamma m -\Omega \right) \left( 2\gamma m -\Omega \right) \sqrt{\frac{2 \gamma m -\Omega}{2 \gamma}} \right]\Bigg\}
\end{aligned}
\end{equation}
and
\begin{equation}
\begin{aligned}
I^{\text{(inter)}}_{2}&=\frac{1}{32 \pi^2 v_F^2 } \int_{0}^{2 \pi} d \phi \int_{-\infty}^{+\infty} dk_z \frac{1}{\varepsilon_{\mathbf{k}}^2} k_{\alpha } k_{\mu }  d_\beta  \partial_{\beta}\left( d_{\beta}\right) d_\nu \partial_{\nu}\left( d_{\nu}\right)   \Theta \left( \Omega^2 -4 d_z^2 \right) \bigg|_{v_F^2 k_{\perp}^2=\frac{\Omega^2}{4}-d_z^2}\\
&=\frac{1}{2 \pi v_F^2 \Omega^2} \Bigg\{\left[\delta_{\mu \beta } \delta_{\nu \alpha} +\delta_{\mu \nu } \delta_{\alpha \beta} +\left( \delta_{\mu 1}+ \delta_{\mu 2} \right) \left( \delta_{\nu 1}+ \delta_{\nu 2} \right) \delta_{\mu \alpha } \delta_{\nu \alpha } \right] \frac{1}{1260} \Bigg[ \left(2 \gamma m +\Omega \right)^2 \sqrt{\frac{2 \gamma m +\Omega}{2 \gamma}} \\
& \times \left( 16 \gamma^2 m^2 -20 \gamma m \Omega +7 \Omega^2 \right)   -\Theta \left( 2 \gamma m -\Omega \right) \left(2 \gamma m -\Omega\right)^2 \sqrt{\frac{2 \gamma m -\Omega }{2 \gamma}} \left( 16 \gamma^2 m^2 +20 \gamma m \Omega +7 \Omega^2 \right)  \Bigg] \\
&+ \delta_{\mu 3} \delta_{\nu 3} \delta_{\mu \alpha} \delta_{\nu \beta }  \frac{\Omega^2}{60} \left[ \left(2 \gamma m +\Omega \right)^2  \sqrt{\frac{2 \gamma m +\Omega}{2 \gamma}}  -\Theta \left( 2 \gamma m -\Omega \right) \left(2 \gamma m -\Omega \right)^2 \sqrt{\frac{2\gamma m -\Omega }{2 \gamma}} \right] \\
& -  \left(\delta_{\mu 1} +\delta_{\mu 2 }\right) \delta_{\nu 3} \delta_{\mu \alpha } \delta_{\nu \beta }    \frac{\gamma}{13860 v_F^2} \Bigg[ \left(  256 \gamma ^3 m^3-320 \gamma ^2 m^2 \Omega +148 \gamma  m \Omega ^2-45 \Omega ^3  \right)\left(2 \gamma m +\Omega \right)^2\\
& \times   \sqrt{\frac{2 \gamma m +\Omega}{2 \gamma}}  - \Theta \left( 2 \gamma m -\Omega \right) \left(  256 \gamma ^3 m^3+ 320 \gamma ^2 m^2 \Omega +148 \gamma  m \Omega ^2+45 \Omega ^3  \right) \left(2 \gamma m -\Omega\right)^2  \sqrt{\frac{2 \gamma m -\Omega }{2 \gamma}} \Bigg] \\
&- \delta_{\mu 3} \left(\delta_{\nu 1} +\delta_{\nu 2 }\right) \delta_{\mu \alpha } \delta_{\nu \beta }   \frac{ v_F^2}{420 \gamma} \left[ \left(2 \gamma m +\Omega\right)^2 \left(4 \gamma m -5\Omega  \right)  \sqrt{\frac{2 \gamma m +\Omega }{2\gamma}} -\Theta \left( 2 \gamma m -\Omega \right) \left(2 \gamma m -\Omega \right)^2 \right.\\
&\left. \times \left(4 \gamma m + 5\Omega  \right)  \sqrt{\frac{2 \gamma m -\Omega}{2 \gamma}} \right] \Bigg\}.
\end{aligned}
\end{equation}
Subtracting the divergent contribution of the filled states, the intraband part of $\eta_{\mu \nu \alpha \beta}^{(2)}(\Omega)+\eta_{\mu \nu \alpha \beta}^{(3)}(\Omega)$ reads
\begin{equation}
\begin{aligned}
\eta_{\mu \nu \alpha \beta}^{\text{(intra)}}(\Omega)&=\pi \delta(\Omega) \, \text{v.p.} \int \frac{d^3 k}{(2\pi)^3} \frac{f(\varepsilon_{\mathbf{k}})-f(-\varepsilon_{\mathbf{k}})}{ \varepsilon_{\mathbf{k}}}  \left\{ k_{\alpha } k_{\mu } \delta_{\nu \beta} \left[ \partial_{\nu}\left( d_{\nu}\right) \right]^2 - \frac{1}{\varepsilon_{\mathbf{k}}^2} k_{\alpha } k_{\mu }  d_\beta  \partial_{\beta}\left( d_{\beta}\right) d_\nu \partial_{\nu}\left( d_{\nu}\right) \right\} \\
&=\delta(\Omega)\left( I^{(\text{intra})}_{1}-I^{(\text{intra})}_{2}\right),
\end{aligned}
\end{equation}
where
\begin{equation}
\begin{aligned}
I^{(\text{intra})}_{1}&= \pi \int \frac{d^3 k}{(2\pi)^3} \frac{1}{\left| \mathbf{d(\mathbf{k}) }\right|} k_{\mu} k_{\alpha} \delta_{\nu \beta} \left( \partial_{\nu} (d_{\nu}) \right)^2 \Theta \left( |\mu| -\left| \mathbf{d} (\mathbf{k}) \right| \right)=\frac{\delta_{\mu \alpha} \delta_{\nu \beta} }{\pi v_F^2 } \Bigg\{ \frac{2\left( \delta_{\mu 1}+\delta_{\mu 2} \right) \left( \delta_{\nu 1}+\delta_{\nu 2} \right)}{105} \Bigg[ 8 \gamma^3 m^{7/2} \\
&+ (|\mu| +\gamma  m)^2 (3 |\mu| -4 \gamma  m) \sqrt{\frac{ \gamma m +|\mu| }{ \gamma}}  - \Theta\left(\gamma m -|\mu|  \right) (|\mu| -\gamma  m)^2 (3 |\mu| +4 \gamma  m) \sqrt{\frac{ \gamma m -|\mu| }{ \gamma}}  \Bigg] \\
&+ \frac{v_F^2 \delta_{\mu 3} \left( \delta_{\nu 1}+\delta_{\nu 2} \right) }{15 \gamma }\left[  -2 \gamma^2 m^{5/2} + (|\mu| +\gamma  m)^2 \sqrt{\frac{ \gamma m +|\mu| }{ \gamma}}  + \Theta\left(\gamma m -|\mu|  \right) (|\mu| -\gamma  m)^2  \sqrt{\frac{ \gamma m -|\mu| }{ \gamma}}   \right] \\
& + \frac{4 \delta_{\mu 3} \delta_{\nu 3}}{35}\left[ -2 \gamma^3 m^{7/2} + (|\mu| +\gamma  m)^3 \sqrt{\frac{ \gamma m +|\mu| }{ \gamma}}  - \Theta\left(\gamma m -|\mu|  \right) (|\mu| -\gamma  m)^3  \sqrt{\frac{ \gamma m -|\mu| }{ \gamma}}    \right] \\
& - \frac{8 \gamma \left( \delta_{\mu 1}+\delta_{\mu 2} \right) \delta_{\nu 3}  }{945 v_F^2}\left[ - 8 \gamma^4 m^{9/2}+ (|\mu| +\gamma  m)^3 (4 \gamma  m -5  |\mu|) \sqrt{\frac{ \gamma m +|\mu| }{ \gamma}}  - \Theta\left(\gamma m -|\mu|  \right) (|\mu| -\gamma  m)^3 \right.\\
&\left. \times (4 \gamma  m +5  |\mu|) \sqrt{\frac{ \gamma m -|\mu| }{ \gamma}}  \right] \Bigg\}
\end{aligned}
\end{equation}
and
\begin{equation}
\begin{aligned}
I^{(\text{intra})}_{2}&= \pi \int \frac{d^3 k}{(2\pi)^3} \frac{1}{\left| \mathbf{d(\mathbf{k}) }\right|} k_{\mu} \frac{d_\nu \partial_\nu (d_\nu)}{\left| \mathbf{d(\mathbf{k}) }\right|}     k_{\alpha}  \frac{d_{\beta} \partial_{\beta} (d_\beta) }{\left| \mathbf{d(\mathbf{k}) }\right|}     \Theta \left( |\mu| -\left| \mathbf{d} (\mathbf{k}) \right| \right)=\\
&=\frac{1}{\pi v_F^2|\mu|} \Bigg\{ \frac{4}{315} \left[\delta_{\mu \beta} \delta_{\nu \alpha} +\delta_{\mu \nu} \delta_{\alpha \beta} +\left( \delta_{\mu1}+ \delta_{\mu2} \right)\left( \delta_{\nu1}+ \delta_{\nu2} \right) \delta_{\mu \alpha} \delta_{\nu \beta} \right]\\
& \times  \left[ 12 |\mu| \gamma^3 m^{7/2} + (|\mu| +\gamma  m )^3 (|\mu| -2 \gamma  m)  \sqrt{\frac{\gamma m + |\mu| }{\gamma }}  -\Theta \left( \gamma m - |\mu| \right) (|\mu| -\gamma  m )^3 (|\mu| +2 \gamma  m)  \sqrt{\frac{\gamma m - |\mu| }{\gamma }} \right]\\
&+\frac{2 v_F^2}{105 \gamma } \delta_{\mu 3} \left(\delta_{\nu 1} +\delta_{\nu 2 }\right) \delta_{\mu \alpha } \delta_{\nu \beta }  \left[ -7 |\mu| \gamma^2 m^{5/2} + (|\mu| +\gamma  m )^3  \sqrt{\frac{\gamma m + |\mu| }{\gamma }}  -\Theta \left( \gamma m - |\mu| \right) ( \gamma  m -|\mu| )^3  \sqrt{\frac{\gamma m - |\mu| }{\gamma }} \right] \\
&+\frac{4}{315 }   \delta_{\mu 3} \delta_{\nu 3} \delta_{\mu \alpha } \delta_{\nu \beta } \left[ -6 |\mu| \gamma^3 m^{7/2} + 3 |\mu| (|\mu| +\gamma  m )^3   \sqrt{\frac{\gamma m + |\mu| }{\gamma }}  -\Theta \left( \gamma m - |\mu| \right) 3 |\mu| (|\mu| -\gamma  m )^3  \sqrt{\frac{\gamma m - |\mu| }{\gamma }} \right]  \\
&+ \frac{8 \gamma}{3465 v_F^2 } \left(\delta_{\mu 1} +\delta_{\mu 2 }\right) \delta_{\nu 3} \delta_{\mu \alpha } \delta_{\nu \beta }  \left[   - 88 \gamma^4  m^{9/2} |\mu| +  \left(5 \mu^2-12 \gamma m |\mu| +16 \gamma ^2 m^2 \right) (|\mu| +\gamma  m )^3 \sqrt{\frac{\gamma m + |\mu| }{\gamma }} \right.\\
&\left.+\Theta\left(\gamma m -|\mu| \right)   \left( 5 \mu^2+ 12 \gamma m |\mu| +16 \gamma ^2 m^2 \right) (|\mu| -\gamma  m )^3 \sqrt{\frac{\gamma m - |\mu| }{\gamma }}  \right] \Bigg\}.
\end{aligned}
\end{equation}

The fourth and fifth terms in the last line in Eq.~(\ref{app-kubo-eta}) are
\begin{equation}
\begin{aligned}
\eta_{\mu \nu \alpha \beta}^{(4)}(\Omega)&
=\pi \delta(\Omega) \delta_{\alpha \beta} \int \frac{d^3 k}{(2\pi)^3}  \frac{1}{\varepsilon_{\mathbf{k}}}  \left[ f \left( \varepsilon_{\mathbf{k}} \right) -f\left( -\varepsilon_{\mathbf{k}} \right) \right] d_i \partial_{\nu}\left( d_i \right) k_{\mu} \\
&=  \frac{2}{105 \pi v_F^2}  \delta(\Omega) \delta_{\mu \nu }\delta_{\alpha \beta} \left[  8 \gamma ^3 m^{7/2}-(\gamma  m + |\mu| )^2  (4 \gamma  m - 3 |\mu|)\sqrt{\frac{\gamma  m+|\mu| }{\gamma }}  \right.\\
&\left. -\Theta(\gamma m -|\mu| )  (\gamma  m-|\mu| )^2 (4 \gamma  m+ 3 |\mu| ) \sqrt{\frac{\gamma  m-|\mu| }{\gamma }} \right]=\pi  \delta(\Omega) \delta_{\mu \nu }\delta_{\alpha \beta}  P,\\
\kappa^{-1}&= P \left( 1+\alpha \right),
\end{aligned}
\end{equation}
where $\alpha= \partial P/\partial \epsilon = \partial P/\partial \mu \left[\partial \epsilon/\partial \mu\right]^{-1}$
and
\begin{equation}
\begin{aligned}
\epsilon&= \int_{-\infty}^{+\infty} d\omega f(\omega) \int \frac{d^3 k}{(2\pi)^3} \text{tr}\left[  H(\mathbf{k}) A(\omega; \mathbf{k}) \right]=\frac{1}{105 \pi^2 v_F^2 }  \Bigg[  -16 \gamma ^3 m^{7/2}+ \left(8 \gamma^3 m^3 - 4 \gamma^2 m^2  |\mu|  +3\gamma m \mu^2 +15  |\mu| ^3 \right) \\
&\times  \sqrt{\frac{\gamma  m+|\mu| }{\gamma }}  + \Theta(\gamma m -|\mu| )   \left(8 \gamma^3 m^3 + 4 \gamma^2 m^2  |\mu|  +3\gamma m \mu^2 -15  |\mu| ^3 \right)  \sqrt{\frac{\gamma  m-|\mu| }{\gamma }}\Bigg].
\end{aligned}
\end{equation}

For the last term in the last line in Eq.~(\ref{app-kubo-eta}), we have
\begin{equation}
\begin{aligned}
\eta_{\mu \nu \alpha \beta}^{(5)}(\Omega)& =\pi \delta(\Omega)  \int \frac{d^3 k}{(2\pi)^3} \frac{f (\varepsilon_{\mathbf{k}}) -f( - \varepsilon_{\mathbf{k}})}{\varepsilon_{\mathbf{k}}} d_i k_{\alpha}  \partial_{\beta} \left[ k_\mu \partial_{\nu}\left( d_i \right) \right]  \\
& = \pi \delta(\Omega)  \int \frac{d^3 k}{(2\pi)^3} \frac{f (\varepsilon) -f( - \varepsilon)}{\varepsilon} \left\{  \left(\delta_{\nu 1}+\delta_{\nu 2} \right) v_F^2 k_\nu k_{\alpha} \delta_{\mu \beta} +\delta_{\nu 3} d_\nu k_{\alpha} \partial_{\beta} \left[ k_\mu \partial_{\nu}\left( d_\nu \right) \right] \right\} \\
&=  \pi  \delta(\Omega) \delta_{\mu \beta }\delta_{\nu \alpha} P +  \delta_{\mu 3 } \delta_{\nu 3} \pi \delta(\Omega) \delta_{\mu \alpha} \delta_{\nu \beta} P-  \frac{ 4 \gamma \left( \delta_{\mu 1 } +\delta_{\mu 2} \right) \delta_{\nu 3}}{945 \pi v_F^4}  \delta(\Omega) \delta_{\mu \alpha} \delta_{\nu \beta} \\
&\times  \left[  64 \gamma ^4 m^{9/2}-\left(32 \gamma^2 m^2- 26\gamma m  |\mu| +5 \mu^2 \right) (\gamma  m+|\mu| )^2 \sqrt{\frac{\gamma  m+|\mu| }{\gamma }}  \right.\\
&\left. -\Theta(\gamma m -|\mu| ) \left(32 \gamma^2 m^2+ 26\gamma m  |\mu| +5 \mu^2 \right) (\gamma  m-|\mu| )^2 \sqrt{\frac{\gamma  m-|\mu| }{\gamma }} \right].
\end{aligned}
\end{equation}

The complete expressions for the static and dynamic parts of the viscoelasticity tensor read
\begin{equation}
\label{App-kubo-two-band-stat}
\begin{aligned}
\text{Re}\,\eta_{\mu \nu \alpha \beta}^{\text{stat}}(\Omega)&=\frac{\delta(\Omega)}{\pi v_F^2 |\mu|}\Bigg\{\left[\delta_{\mu \beta} \delta_{\nu \alpha} +\delta_{\mu \nu} \delta_{\alpha \beta} +\left( \delta_{\mu 1}+ \delta_{\mu 2} \right)\left( \delta_{\nu 1}+ \delta_{\nu 2} \right) \delta_{\mu \alpha} \delta_{\nu \beta} \right] \frac{2}{315} \Bigg[  \left( 4\gamma^2 m^2 -10 \gamma m |\mu| +7 \mu^2 \right)  \\
&\times \left(\gamma m +|\mu|\right)^2 \sqrt{\frac{\gamma m +|\mu|}{\gamma}}  -\Theta \left( \gamma m -|\mu|\right)  \left( 4\gamma^2 m^2 +10 \gamma m |\mu| +7 \mu^2 \right) \left(\gamma m -|\mu|\right)^2 \sqrt{\frac{\gamma m -|\mu|}{\gamma}} \Bigg]\\
&+ \delta_{\mu 3} \delta_{\nu 3} \delta_{\mu \alpha} \delta_{\nu \beta}  \frac{2 \mu^2}{15} \left[ \left(\gamma m +|\mu|\right)^2  \sqrt{\frac{\gamma m +|\mu|}{\gamma}}  -\Theta \left( \gamma m -|\mu|\right) \left(\gamma m -|\mu|\right)^2 \sqrt{\frac{\gamma m -|\mu|}{\gamma}} \right] \\
&- \delta_{\mu 3} \left(\delta_{\nu 1} +\delta_{\nu 2 }\right) \delta_{\mu \alpha} \delta_{\nu \beta}   \frac{ v_F^2}{105 \gamma} \left[ \left(\gamma m +|\mu|\right)^2 \left(2 \gamma m -5|\mu| \right)  \sqrt{\frac{\gamma m +|\mu|}{\gamma}} -\Theta \left( \gamma m -|\mu|\right) \left(\gamma m -|\mu|\right)^2 \right. \\
&\left.\times  \left(2 \gamma m + 5|\mu| \right)  \sqrt{\frac{\gamma m -|\mu|}{\gamma}} \right]- \left( \delta_{\mu 1} +\delta_{\mu 2} \right) \delta_{\nu 3} \delta_{\mu \alpha} \delta_{\nu \beta}   \frac{4 \gamma}{3465 v_F^2} \Bigg[ \left(\gamma m +|\mu|\right)^2  \sqrt{\frac{\gamma m +|\mu|}{\gamma}} \\
& \times \left( 32 \gamma^3 m^3 -80 \gamma^2 m^2 |\mu| +74 \gamma m \mu^2-45 |\mu|^3 \right)  -\Theta \left( \gamma m -|\mu|\right) \left(\gamma m -|\mu|\right)^2  \sqrt{\frac{\gamma m -|\mu|}{\gamma}}  \\
& \times   \left(  32 \gamma^3 m^3 +80 \gamma^2 m^2 |\mu| +74 \gamma m \mu^2+45 |\mu|^3 \right)  \Bigg] \Bigg\}- \pi \delta(\Omega) \delta_{\mu \nu } \delta_{\alpha \beta } \left(1+ \alpha\right) P
\end{aligned}
\end{equation}
and
\begin{equation}
\label{App-kubo-two-band-dyn}
\begin{aligned}
\text{Re}\,\eta_{\mu \nu \alpha \beta}^{\text{dyn}}(\Omega)&=\Bigg\{ \left[ \left(2\gamma m +\Omega \right) \left(49 \Omega ^3 -72 \gamma  m \Omega ^2+48 \gamma ^2 m^2 \Omega-64 \gamma ^3 m^3\right)  \sqrt{\frac{2 \gamma m +\Omega}{2 \gamma}} \right.\\
&\left.+ \Theta\left(2\gamma m -\Omega \right) \left(2\gamma m -\Omega \right) \left(49 \Omega ^3 +72 \gamma  m \Omega ^2+48 \gamma ^2 m^2 \Omega+64 \gamma ^3 m^3  \right)  \sqrt{\frac{2 \gamma m -\Omega}{2 \gamma}}  \right] \\
&\times \frac{\delta_{\mu \alpha} \delta_{\nu \beta}  \left( \delta_{\mu 1}+\delta_{\mu 2} \right) \left( \delta_{\nu 1}+\delta_{\nu 2} \right)}{5040 \pi  \Omega ^2 v_F^2}  - \left[ \left(2 \gamma m +\Omega \right)^2 \left( 16 \gamma^2 m^2 -20 \gamma m \Omega +7 \Omega^2 \right) \sqrt{\frac{2 \gamma m +\Omega}{2 \gamma}} \right. \\
&\left.  -\Theta \left( 2 \gamma m -\Omega \right) \left(2 \gamma m -\Omega\right)^2 \left( 16 \gamma^2 m^2 +20 \gamma m \Omega +7 \Omega^2 \right) \sqrt{\frac{2 \gamma m -\Omega }{2 \gamma}} \right] \frac{\delta_{\mu \beta } \delta_{\nu \alpha} +\delta_{\mu \nu } \delta_{\alpha \beta}}{2520 \pi v_F^2 \Omega^2}  \\
&+ \left[ \left( 2\gamma m +\Omega \right)^2 \sqrt{\frac{2 \gamma m + \Omega}{2 \gamma}}  - \Theta\left(2\gamma m -\Omega \right) \left( 2\gamma m -\Omega \right)^2\sqrt{\frac{2 \gamma m -\Omega}{2 \gamma}}  \right] \frac{\delta_{\mu \alpha} \delta_{\nu \beta} \delta_{\mu 3} \delta_{\nu 3}}{60 \pi v_F^2} \\
&+\left[(2 \gamma  m+\Omega ) \left(16 \gamma ^2 m^2-12 \gamma  m \Omega +25 \Omega ^2\right) \sqrt{\frac{2 \gamma m +\Omega}{2 \gamma}}-\Theta\left( 2\gamma m -\Omega \right) \right.\\
&\left.\times (2 \gamma  m-\Omega )  \left(16 \gamma ^2 m^2+12 \gamma  m \Omega +25 \Omega ^2\right) \sqrt{\frac{2 \gamma m -\Omega}{2 \gamma}}  \right] \frac{\delta_{\mu \alpha} \delta_{\nu \beta} \delta_{\mu 3} \left(\delta_{\nu 1} +\delta_{\nu 2 }\right) }{1680 \pi  \gamma  \Omega ^2} \\
&+\left[\gamma  (2 \gamma  m+\Omega )^3  \left(16 \gamma ^2 m^2-28 \gamma  m \Omega +15 \Omega ^2\right) \sqrt{\frac{2 \gamma m +\Omega}{2 \gamma}} -\Theta\left( 2\gamma m -\Omega \right)  \gamma  (2 \gamma  m-\Omega )^3  \right.\\
&\left.\times \left(16 \gamma ^2 m^2+28 \gamma  m \Omega +15 \Omega ^2\right)  \sqrt{\frac{2 \gamma m -\Omega}{2 \gamma}}  \right] \frac{\delta_{\mu \alpha} \delta_{\nu \beta} \delta_{\nu 3} \left( \delta_{\mu 1}+\delta_{\mu 2} \right)}{3465 \pi  \Omega ^2 v_F^4}  \Bigg\} \Theta \left( \Omega - 2\mu \right),
\end{aligned}
\end{equation}
respectively.

Let us calculate terms determined by the internal degrees of freedom, i.e., terms proportional to $\propto C_{\alpha \beta}$ with $\alpha,\beta=1,2$; we use the in-plane rotational symmetry of the model to fix $C_{\mu 3}=C_{3\mu}=0$. At $\Omega=\mu=0$, the correction for the Hall viscosity reads
\begin{equation}
\label{App-kubo-two-band-Hall-correction}
\text{Re}\,\delta \eta_{\mu \nu \alpha \beta}^{\text{Hall}}(\Omega) =  \left(I_{\mu \nu \alpha \beta}^{(1)}-I_{\mu \nu \beta \alpha}^{(1)} \right)+ \left(I_{\alpha \beta \mu \nu }^{(1)} -I_{\alpha \beta \nu \mu }^{(1)} \right)+\left( I_{\mu \nu \alpha \beta}^{(2)}-I_{\mu \nu \beta \alpha }^{(2)}-I_{\nu \mu \alpha \beta}^{(2)} +I_{\nu \mu \beta \alpha }^{(2)}\right),
\end{equation}
where
\begin{equation}
\begin{aligned}
I_{\mu \nu \alpha \beta}^{(1)}&= C_{\alpha \beta} \int \frac{d^3k}{\left(2\pi\right)^3} \frac{1}{4 \varepsilon_{\mathbf{k}}^3} k_{\mu }d_{i} d_{\beta}\partial_{\nu}\left( d_{\nu}\right)\epsilon_{\nu i \alpha } =C_{\alpha \beta} \Bigg[ \delta_{\mu \beta} \left( \delta_{\beta 1}+\delta_{\beta 2}\right)\left( \delta_{\alpha 1} \delta_{\nu 2} - \delta_{\alpha 2} \delta_{\nu 1}\right) \\
&\times\left(  - \frac{\gamma^2 \Lambda_{z}^5  }{40 \pi ^2 v_F^2}+\frac{\gamma \Lambda_{z}^3   \Lambda_{\perp}  }{48 \pi ^2 v_F}+\frac{\gamma^2 m \Lambda_{z}^3 }{12 \pi ^2 v_F^2}-\frac{ \gamma m \Lambda_{z}   \Lambda_{\perp} }{16 \pi ^2 v_F}-\frac{  \gamma ^2 m^2 \Lambda_{z}}{8 \pi ^2 v_F^2}+\frac{2 \gamma ^2 m^{5/2}}{15 \pi ^2 v_F^2} \right)\\
&+\delta_{\mu \nu} \delta_{\nu 3} \left( \delta_{\alpha 2} \delta_{\beta 1} - \delta_{\alpha 1} \delta_{\beta 2}\right) \left( -\frac{\gamma ^2 \Lambda_{z}^5 }{20 \pi ^2 v_F^2}+\frac{\gamma \Lambda_{z}^3  \Lambda_{\perp} }{24 \pi ^2 v_F}+\frac{\gamma ^2 m \Lambda_{z}^3 }{12\pi ^2 v_F^2}-\frac{ \gamma ^2 m^{5/2}}{15 \pi ^2 v_F^2}\right) \Bigg],
\end{aligned}
\end{equation}
\begin{equation}
\begin{aligned}
I_{\mu \nu \alpha \beta}^{(2)}&=C_{\alpha \beta} C_{\mu \nu} \int \frac{d^3k}{\left(2\pi\right)^3} \frac{1}{8 \varepsilon_{\mathbf{k}}^3} \epsilon_{\mu i \alpha }d_{i} d_{\beta}d_{\nu} =C_{\alpha \beta} C_{\mu \nu}\delta_{\nu \beta} \left( \delta_{\beta 1}+\delta_{\beta 2}\right) \left( \delta_{\alpha 1} \delta_{\mu 2} - \delta_{\alpha 2} \delta_{\mu 1}\right) \\
&\times \left( - \frac{\gamma^2 \Lambda_{z}^5  }{80 \pi ^2 v_F^2}+\frac{\gamma \Lambda_{z}^3   \Lambda_{\perp}  }{96 \pi ^2 v_F}+\frac{\gamma^2 m \Lambda_{z}^3 }{24 \pi ^2 v_F^2}-\frac{ \gamma m \Lambda_{z}   \Lambda_{\perp} }{32 \pi ^2 v_F}-\frac{  \gamma ^2 m^2 \Lambda_{z}}{16 \pi ^2 v_F^2}+\frac{\gamma ^2 m^{5/2}}{15 \pi ^2 v_F^2} \right).
\end{aligned}
\end{equation}
After subtraction discussed in the paragraph before Eq.~(\ref{Kubo-stat-two-band-I-reg}), we have
\begin{equation}
\begin{aligned}
\overline{I}_{\mu \nu \alpha \beta}^{(1)}&=C_{\alpha \beta} \Bigg[ \delta_{\mu \beta} \left( \delta_{\beta 1}+\delta_{\beta 2}\right)\left( \delta_{\alpha 1} \delta_{\nu 2} - \delta_{\alpha 2} \delta_{\nu 1}\right) \left( -\frac{\gamma^2 m \Lambda_{z}^3 }{12 \pi ^2 v_F^2}+\frac{ \gamma m \Lambda_{z}   \Lambda_{\perp} }{16 \pi ^2 v_F}+\frac{  \gamma ^2 m^2 \Lambda_{z}}{8 \pi ^2 v_F^2}-\frac{2 \gamma ^2 m^{5/2}}{15 \pi ^2 v_F^2} \right)\\
&+\delta_{\mu \nu} \delta_{\nu 3} \left( \delta_{\alpha 2} \delta_{\beta 1} - \delta_{\alpha 1} \delta_{\beta 2}\right) \left(-\frac{\gamma ^2 m \Lambda_{z}^3 }{12\pi ^2 v_F^2}+\frac{ \gamma ^2 m^{5/2}}{15 \pi ^2 v_F^2}\right) \Bigg],
\end{aligned}
\end{equation}
\begin{equation}
\begin{aligned}
\overline{I}_{\mu \nu \alpha \beta}^{(2)}&=C_{\alpha \beta} C_{\mu \nu} \delta_{\nu \beta} \left( \delta_{\beta 1}+\delta_{\beta 2}\right) \left( \delta_{\alpha 1} \delta_{\mu 2} - \delta_{\alpha 2} \delta_{\mu 1}\right) \left( -\frac{\gamma^2 m \Lambda_{z}^3 }{24 \pi ^2 v_F^2}+\frac{ \gamma m \Lambda_{z}   \Lambda_{\perp} }{32 \pi ^2 v_F}+\frac{  \gamma ^2 m^2 \Lambda_{z}}{16 \pi ^2 v_F^2}-\frac{\gamma ^2 m^{5/2}}{15 \pi ^2 v_F^2} \right).
\end{aligned}
\end{equation}

The correction in the dynamic part reads
\begin{equation}
\begin{aligned}
\delta \eta_{\mu \nu \alpha \beta}^{(2)}(\Omega)&=\frac{\pi}{\Omega}  \Theta \left( \Omega-2\mu\right)  \int \frac{d^3 k}{(2\pi)^3} \delta\left( \Omega -2 \left| \mathbf{d(\mathbf{k}) }\right| \right)  \left\{ \frac{C_{\alpha \beta}}{2}\left[ k_{\mu } d_{\beta} \partial_{\nu}\left( d_{\alpha}\right)- k_{\mu } d_{\alpha} \partial_{\nu}\left( d_{\beta}\right)  \right] \right.\\
&\left.+\frac{C_{\mu \nu}}{2}\left[ k_{\alpha } d_{\nu} \partial_{\beta}\left( d_{\mu}\right)- k_{\alpha } d_{\mu} \partial_{\beta}\left( d_{\nu}\right) \right] +  \frac{C_{\alpha \beta} C_{\mu \nu}}{4} \left( \delta_{\mu \alpha} d_{\nu} d_{\beta}-\delta_{\nu \alpha} d_{\mu} d_{\beta}-\delta_{\mu \beta} d_{\nu} d_{\alpha}+\delta_{\nu \beta} d_{\mu} d_{\alpha}\right) \right\}.
\end{aligned}
\end{equation}
Here, there are the following two types of integrals:
\begin{equation}
\label{Kubo-two-band-I3}
\begin{aligned}
I^{(\text{inter})}_{3}&=\frac{\pi}{\Omega} \int \frac{d^3 k}{(2\pi)^3} k_{\mu} d_{\alpha} \partial_{\nu} \left( d_{\beta} \right) \delta\left( \Omega -2 \left| \mathbf{d(\mathbf{k}) }\right| \right) =  \frac{\delta_{\mu \alpha} \delta_{\nu \beta} }{240 \pi v_F^2} \left[ \left(\delta_{\mu 1}+\delta_{\mu 2}\right)\left(\delta_{\nu 1}+\delta_{\nu 2}\right)+\delta_{\mu 3}\delta_{\nu 3}\right] \\
&\times\left[ \left(3\Omega^2 +2\gamma m \Omega -8\gamma^2 m^2\right)  \sqrt{\frac{2\gamma m +\Omega }{2\gamma }}-\Theta\left(2 \gamma m -\Omega \right) \left(3\Omega^2 -2\gamma m \Omega -8\gamma^2 m^2\right) \sqrt{\frac{2\gamma m -\Omega }{2\gamma }} \right] \\
&\equiv \delta_{\mu \alpha} \delta_{\nu \beta} \left[ \left(\delta_{\mu 1}+\delta_{\mu 2}\right)\left(\delta_{\nu 1}+\delta_{\nu 2}\right)+\delta_{\mu 3}\delta_{\nu 3}\right] g_{1}(\Omega)
\end{aligned}
\end{equation}
and
\begin{equation}
\begin{aligned}
I^{(\text{inter})}_{4}&=\frac{\pi}{\Omega} \int \frac{d^3 k}{(2\pi)^3}  d_{\mu} d_{\alpha} \delta_{\nu \beta} \delta\left( \Omega -2 \left| \mathbf{d(\mathbf{k}) }\right| \right) =\delta_{\mu \alpha}  \delta_{\nu \beta}  \left\{ \frac{\left(\delta_{\mu 1}+\delta_{\mu 2} \right)}{240 \pi v_F^2} \left[  \left(3\Omega^2 +2\gamma m \Omega -8\gamma^2 m^2\right)  \sqrt{\frac{2\gamma m +\Omega }{2\gamma }}\right.\right.\\
&\left.\left.-\Theta\left(2 \gamma m -\Omega \right) \left(3\Omega^2 -2\gamma m \Omega -8\gamma^2 m^2\right) \sqrt{\frac{2\gamma m -\Omega }{2\gamma }} \right]  + \frac{\delta_{\mu 3}}{480 \pi v_F^2} \left[  \left(3\Omega^2 -8\gamma m \Omega +32\gamma^2 m^2\right)  \sqrt{\frac{2\gamma m +\Omega }{2\gamma }}\right.\right.\\
&\left.\left.-\Theta\left(2 \gamma m -\Omega \right) \left(3\Omega^2 + \gamma m \Omega +32\gamma^2 m^2\right) \sqrt{\frac{2\gamma m -\Omega }{2\gamma }} \right] \right\} \equiv \delta_{\mu \alpha} \delta_{\nu \beta} \left[ \left(\delta_{\mu 1}+\delta_{\mu 2}\right) g_{1}(\Omega)+\delta_{\mu 3} g_{2}(\Omega)\right].
\end{aligned}
\end{equation}
Hence
\begin{equation}
\begin{aligned}
\delta \eta_{\mu \nu \alpha \beta}^{(2)}(\Omega) &=\left( \delta_{\mu \beta} \delta_{\nu \alpha}  - \delta_{\mu \alpha} \delta_{\nu \beta}   \right)  \Theta\left(\Omega-2\mu\right) \Bigg\{ \left(\frac{C_{\mu \nu}}{2}+\frac{C_{\alpha \beta}}{2}\right) \left[ \left(\delta_{\mu 1}+\delta_{\mu 2}\right)\left(\delta_{\nu 1}+\delta_{\nu 2}\right)+\delta_{\mu 3}\delta_{\nu 3}\right] g_{1}(\Omega) \\
&- \frac{C_{\mu \nu} C_{\alpha \beta}}{4} \left[ \left(\delta_{\mu 1}+\delta_{\mu 2}\right) g_{1}(\Omega)+\delta_{\mu 3} g_{2}(\Omega)\right] -  \frac{C_{\mu \nu} C_{\alpha \beta}}{4}  \left[ \left(\delta_{\nu 1}+\delta_{\nu 2}\right) g_{1}(\Omega)+\delta_{\nu 3} g_{2}(\Omega)\right] \Bigg\}.
\end{aligned}
\end{equation}
The rotational symmetry with respect to the $z$-axis allows us to set $C_{3\mu}=C_{\mu3}=0$ and simplify the result,
\begin{equation}
\label{App-kubo-two-band-dyn-correction}
\begin{aligned}
\text{Re}\,\delta \eta_{\mu \nu \alpha \beta}^{\text{dyn}}(\Omega)&= \left( \delta_{\mu \beta} \delta_{\nu \alpha} - \delta_{\mu \alpha} \delta_{\nu \beta}   \right)  \Theta\left(\Omega-2\mu\right) \left(\frac{C_{\mu \nu}}{2}+\frac{C_{\alpha \beta}}{2}- \frac{C_{\mu \nu} C_{\alpha \beta}}{4}\right)  \left(\delta_{\mu 1}+\delta_{\mu 2}\right)\left(\delta_{\nu 1}+\delta_{\nu 2}\right)  g_{1}(\Omega),
\end{aligned}
\end{equation}
where $g_1(\Omega)$ follows from Eq.~(\ref{Kubo-two-band-I3}).

The corrections to the static part read
\begin{equation}
\begin{aligned}
\delta\eta_{\mu \nu \alpha \beta}^{(3)}(\Omega)&= \pi \delta(\Omega) \, \text{v.p.}  \int \frac{d^3 k}{(2\pi)^3}   \frac{\Theta\left( |\mu| -\left| \mathbf{d(\mathbf{k}) }\right| \right)}{\left| \mathbf{d(\mathbf{k}) }\right| } \left[ \frac{C_{\mu \nu}}{2}\left( k_{\alpha } d_{\nu}\delta_{\mu \beta}- k_{\alpha } d_{\mu} \delta_{\nu \beta}\right) \partial_{\beta}\left( d_{\beta}\right) \right.  \\
&\left. +\frac{C_{\alpha \beta}}{2}\left( k_{\mu } d_{\beta} \delta_{\nu \alpha}- k_{\mu } d_{\alpha}\delta_{\nu \beta} \right) \partial_{\nu}\left( d_{\nu}\right)  +  \frac{C_{\mu \nu} C_{\alpha \beta}}{4} \left( \delta_{\mu \alpha} d_{\nu} d_{\beta}-\delta_{\nu \alpha} d_{\mu} d_{\beta}-\delta_{\mu \beta} d_{\nu} d_{\alpha}+\delta_{\nu \beta} d_{\mu} d_{\alpha}\right) \right]
\end{aligned}
\end{equation}
and
\begin{equation}
\begin{aligned}
\delta \eta_{\mu \nu \alpha \beta}^{(5)}(\Omega)&= \pi \delta(\Omega)  \int \frac{d^3 k}{(2\pi)^3} \frac{\Theta\left( |\mu| -\left| \mathbf{d(\mathbf{k}) }\right| \right)}{\left| \mathbf{d(\mathbf{k}) }\right| } \left\{ \frac{C_{\mu \nu}}{2}  \left[ k_\alpha \partial_{\beta}\left( d_\nu \right)d_{\mu}-k_\beta \partial_{\alpha}\left( d_\mu \right)d_{\nu}   \right] \right.\\
&\left.+\frac{C_{\alpha \beta}}{2}  \left[ k_{\mu} \partial_{\nu} \left( d_\beta \right) d_\alpha - k_{\nu} \partial_{\mu} \left( d_\alpha \right) d_\beta \right]  +\frac{C_{\mu \nu} C_{\alpha \beta }}{4}   \left( \delta_{\nu \beta}d_\alpha d_{\mu} -\delta_{\mu \beta} d_\alpha d_{\nu}+  \delta_{\mu \alpha}d_\beta d_{\nu}-\delta_{\nu \alpha}d_\beta d_{\mu} \right) \right\}.
\end{aligned}
\end{equation}
The integrals here are
\begin{equation}
\begin{aligned}
I^{(\text{intra})}_{3}&=\pi \int \frac{d^3 k}{(2\pi)^3} \frac{1}{\left| \mathbf{d(\mathbf{k}) }\right|} k_{\mu} d_{\alpha} \partial_{\nu} \left( d_{\beta} \right)  \Theta \left( |\mu| -\left| \mathbf{d} (\mathbf{k}) \right| \right)= \frac{\delta_{\mu \alpha} \delta_{\nu \beta} }{4 \pi v_F^2} \left[ \left(\delta_{\mu 1}+\delta_{\mu 2}\right)\left(\delta_{\nu 1}+\delta_{\nu 2}\right)+\delta_{\mu 3}\delta_{\nu 3}\right]  \\
&\times \left\{  \frac{8}{105} \left[8\gamma^3 m^{7/2} - (|\mu| + \gamma  m)^2 (4 \gamma  m-3 |\mu| )  \sqrt{\frac{\gamma m +|\mu| }{\gamma }}  -\Theta\left( \gamma m -|\mu| \right) (|\mu| -\gamma  m)^2  \right.\right.\\
&\left.\left. \times (4 \gamma  m+3 |\mu| ) \sqrt{\frac{\gamma m -|\mu| }{\gamma }} \right]  \right\}
\equiv \delta_{\mu \alpha} \delta_{\nu \beta} \left[ \left(\delta_{\mu 1}+\delta_{\mu 2}\right)\left(\delta_{\nu 1}+\delta_{\nu 2}\right)+\delta_{\mu 3}\delta_{\nu 3}\right] f_{1}(\mu)
\end{aligned}
\end{equation}
and
\begin{equation}
\begin{aligned}
I^{(\text{intra})}_{4}&=\pi \int \frac{d^3 k}{(2\pi)^3} \frac{1}{\left| \mathbf{d(\mathbf{k}) }\right|} d_{\mu} d_{\alpha} \delta_{\nu \beta}  \Theta \left( |\mu| -\left| \mathbf{d} (\mathbf{k}) \right| \right)= \frac{\delta_{\mu \alpha}  \delta_{\nu \beta}}{4 \pi v_F^2} \Bigg\{\left(\delta_{\mu 1}+\delta_{\mu 2} \right)  \frac{8}{105} \Bigg[8\gamma^3 m^{7/2} - (|\mu| + \gamma  m)^2 \\
& \times (4 \gamma  m-3 |\mu| ) \sqrt{\frac{\gamma m -|\mu| }{\gamma }} -\Theta\left( \gamma m -|\mu| \right) (|\mu| -\gamma  m)^2 (4 \gamma  m+3 |\mu| ) \sqrt{\frac{\gamma m -|\mu| }{\gamma }} \Bigg]\\
&+ \delta_{\mu 3}\frac{4}{105} \left[-48\gamma^3 m^{7/2} + \left(24 \gamma^2  m^2-8  \gamma  m |\mu| +3\mu^2 \right) (|\mu| + \gamma  m)  \sqrt{\frac{\gamma m + |\mu| }{\gamma }}-\Theta\left( \gamma m -|\mu| \right)\right.\\
&\left. \times  \left(24 \gamma^2  m^2+8  \gamma  m |\mu| +3\mu^2 \right) (|\mu| -\gamma  m) \sqrt{\frac{\gamma m -|\mu| }{\gamma }} \right] \Bigg\}
\equiv \delta_{\mu \alpha} \delta_{\nu \beta} \left[ \left(\delta_{\mu 1}+\delta_{\mu 2}\right) f_{1}(\mu)+\delta_{\mu 3} f_{2}(\mu)\right].
\end{aligned}
\end{equation}
Hence
\begin{equation}
\begin{aligned}
\delta\eta_{\mu \nu \alpha \beta}^{(3)}(\Omega)+\delta\eta_{\mu \nu \alpha \beta}^{(5)}(\Omega)&= \delta(\Omega)\Bigg\{ \frac{C_{\alpha \beta}}{2}\left( \delta_{\mu \beta} \delta_{\nu \alpha}  -\delta_{\nu \beta} \delta_{\mu \alpha} \right) \left[ \left(\delta_{\mu 1}+\delta_{\mu 2}\right)\left(\delta_{\nu 1}+\delta_{\nu 2}\right)+\delta_{\mu 3}\delta_{\nu 3}\right] f_{1}(\mu) \\
& +\frac{C_{\mu \nu}}{2}\left( \delta_{ \mu \beta} \delta_{ \nu \alpha}  -  \delta_{ \nu \beta} \delta_{\mu \alpha }   \right) \left[ \left(\delta_{\mu 1}+\delta_{\mu 2}\right)\left(\delta_{\nu 1}+\delta_{\nu 2}\right)+\delta_{\mu 3}\delta_{\nu 3}\right] f_{1}(\mu)  \\
&+  \frac{C_{\mu \nu} C_{\alpha \beta}}{2}  \left( \delta_{\mu \alpha} \delta_{\nu \beta} -\delta_{\nu \alpha} \delta_{\mu \beta} \right) \left[ \left(\delta_{\nu 1}+\delta_{\nu 2}\right) f_{1}(\mu)+\delta_{\nu 3} f_{2}(\mu) \right]  \\
&  + \frac{C_{\mu \nu} C_{\alpha \beta}}{2}  \left( \delta_{\mu \alpha} \delta_{\nu \beta}-\delta_{\nu \alpha} \delta_{\mu \beta} \right) \left[ \left(\delta_{\mu 1}+\delta_{\mu 2}\right) f_{1}(\mu)+\delta_{\mu 3} f_{2}(\mu) \right]  \Bigg\}.
\end{aligned}
\end{equation}
Employing the rotational symmetry of the system to set $C_{3\mu}=0$, the correction to the static viscoelasticity tensor due to the internal degrees of freedom reads
\begin{equation}
\label{App-kubo-two-band-stat-correction}
\text{Re}\,\delta \eta_{\mu \nu \alpha \beta}^{\text{stat}}(\Omega) = \delta(\Omega) \left(\frac{C_{\mu \nu}}{2} + \frac{C_{\alpha \beta }}{2}-C_{\mu \nu} C_{\alpha \beta}\right)\left( \delta_{\mu \beta} \delta_{\nu \alpha}  -\delta_{\nu \beta} \delta_{\mu \alpha} \right) \left(\delta_{\mu 1}+\delta_{\mu 2}\right)\left(\delta_{\nu 1}+\delta_{\nu 2}\right) f_{1}(\mu).
\end{equation}
It is straightforward to see that for certain strain transformation generators, i.e., $C_{\mu \nu}=1$ or $C_{\mu \nu}=0$ at $\mu,\nu=1,2$, the internal degrees of freedom do not affect the viscoelasticity tensor, $\text{Re}\, \delta \eta_{\mu \nu \alpha \beta}^{\text{stat}}(\Omega)=0$.

\section{Derivation of inviscid hydrodynamic equations}
\label{sec:App-zero-kinetic}

In this Section, we derive the inviscid hydrodynamic equations by using the chiral kinetic theory~\cite{Xiao-Niu:rev-2010,Son:2013,Stephanov:2012,Son-Spivak:2013}. We follow the standard approach discussed in Refs.~\cite{Huang:book,Landau:t10-1995}; see also Refs.~\cite{Gorbar:2017vph,Sukhachov-Trauzettel:2021} for a similar derivation in Weyl semimetals. The viscosity is considered in Sec.~\ref{sec:App-first-kinetic}.

The Boltzmann kinetic equation has the form
\begin{equation}
\frac{\partial f}{\partial t} +\left( \dot{\mathbf{r}} \cdot \frac{\partial f}{\partial \mathbf{r}}\right)+ \left( \dot{\mathbf{k}} \cdot \frac{\partial f}{\partial \mathbf{k}} \right) = I_{\text{col}}\left\{ f \right\},
\end{equation}
where $f=f(t,\mathbf{r},\mathbf{k})$ is the distribution function and $I_{\text{col}}\left\{ f \right\}$ is the collision integral. In the absence of magnetic fields, the semiclassical equations of motion are~\cite{Stephanov:2012,Son:2013}
\begin{equation}
\dot{\mathbf{r}}=\mathbf{v}_{\mathbf{k}}-e\left[ \mathbf{E} \times \mathbf{\Omega}\right] \quad \mbox{and} \quad
\dot{\mathbf{k}}=-e \mathbf{E},
\end{equation}
where $\mathbf{v}_{\mathbf{k}} = \bm{\partial}_{\mathbf{k}} \epsilon_{\mathbf{k}}$ is velocity and $\mathbf{\Omega}$ is the Berry curvature which determines the anomalous velocity.

We assume the hydrodynamic ansatz for the distribution function with the fluid velocity $\mathbf{u}\left( t, \mathbf{r}\right)$~\footnote{The distribution function (\ref{App-hydro-fu}) satisfies the electron-electron collision integral. This form is also known as the Callaway ansatz~\cite{Callaway:1959,Jong-Molenkamp:1995}.}
\begin{equation}
\label{App-hydro-fu}
f^{(\mathbf{u})}_{\eta} (t,\mathbf{r},\mathbf{k})=\frac{1}{e^{\left[  \varepsilon_{\mathbf{k}}-\eta \mu-\left( \mathbf{u} (t,\mathbf{r} )\cdot \mathbf{k}\right) \right]/T}+1},
\end{equation}
where $\eta=\pm$ corresponds to electron ($\eta=+$) and hole ($\eta=-$) states and $T$ is temperature.
In the linear response regime $|\mathbf{u}| \ll v_{F}$, we can expand the distribution function in $|\mathbf{u}|$
\begin{equation}
f^{(\mathbf{u})}_{\eta} (t,\mathbf{r},\mathbf{k})\approx f^{(\mathbf{0})}_{\eta}(t,\mathbf{r},\mathbf{k}) -  \left( \mathbf{u} (t,\mathbf{r}) \cdot \mathbf{k}\right) \partial_{\varepsilon_{\mathbf{k}}} f^{(\mathbf{0})}_{\eta}(t,\mathbf{r},\mathbf{k}).
\end{equation}
For simplicity, we consider the momentum-relaxing scattering in the relaxation-time approximation,
\begin{equation}
I_{\text{col}}\left\{ f_{\eta} \right\}=-\frac{f_{\eta}(t,\mathbf{r},\mathbf{k})-f^{(\mathbf{0})}_{\eta}(t,\mathbf{r},\mathbf{k}) }{\tau}.
\end{equation}

\subsection{Linearized model of Weyl semimetals}

For the linearized model of Weyl semimetals, the kinetic equation for each Weyl node with chirality $\lambda=\pm$ reads
\begin{equation}
\label{kinetic-linear}
\partial_t f_{\lambda,\eta} + \left(\mathbf{v}_{\mathbf{k}}-\eta e\left[  \mathbf{E} \times \mathbf{\Omega}_{\lambda,\eta}\right] \right) \cdot \bm{\nabla} f_{\lambda,\eta} -  \eta e\left( \mathbf{E} \cdot \bm{\partial}_{\mathbf{p}}\right) f_{\lambda,\eta}  =  I_{\text{col}}\left\{ f_{\lambda,\eta} \right\},
\end{equation}
where $\varepsilon_{\mathbf{k},\lambda} =  v_F \left| \mathbf{k} -\lambda \mathbf{b} \right|$, $\mathbf{v}_{\mathbf{k}}=\partial_{\mathbf{k}}\varepsilon_{\mathbf{k},\lambda}=v_F \frac{ \mathbf{k} -\lambda \mathbf{b}}{\left| \mathbf{k} -\lambda \mathbf{b} \right|}$, and $\mathbf{\Omega}_{\lambda,\eta}= \frac{\lambda \eta \left( \mathbf{k}-\lambda  \mathbf{b} \right)}{2 \left| \mathbf{k}-\lambda  \mathbf{b} \right|^3}$.

To obtain hydrodynamic equations, we calculate moments of the Boltzmann equation. For the charge continuity equation, we average Eq.~(\ref{kinetic-linear}) over momentum and sum over Weyl nodes as well as particle and hole states. We obtain
\begin{equation}
\partial_t  \rho(t,\mathbf{r}) +  \bm{\nabla}\cdot \left\{\mathbf{u} (t,\mathbf{r}) \rho (t,\mathbf{r})+\frac{e^2}{2 \pi^2 } \left[\mathbf{b} \times  \mathbf{E} \right] \right\} =0,
\end{equation}
where the charge density is
\begin{equation}
\rho \left(t, \mathbf{r}\right)=\sum_{\lambda=\pm} \sum_{\eta}   \int \frac{d^3 k}{(2\pi)^3} \eta e f^{(\mathbf{0})}_{\lambda,\eta}=-\sum_{\lambda=\pm}\sum_{\eta} \frac{\eta eT^{3}}{\pi^2 v_F^{3}} \text{Li}_{3}\left(-e^{ \eta \mu/T} \right) = \frac{e}{3 \pi^2 v_F^{3}}  \mu \left( \mu^2+\pi^2 T^2 \right),
\end{equation}
where Lin(x) is the polylogarithm function.

The Euler equation is obtained by averaging the Boltzmann equation with the quasiparticle momentum $\mathbf{k}$,
\begin{eqnarray}
\label{App-hydro-lin-Euler}
&&\partial_t  \left[ w(t,\mathbf{r}) \frac{\mathbf{u}(t,\mathbf{r})}{v_F^2} + \nu(t,\mathbf{r}) \left( \mathbf{u} (t,\mathbf{r}) \cdot \mathbf{ b} \right) \mathbf{b} \right]  + \bm{\nabla} P\left(t, \mathbf{r} \right)  + \rho\left(t,\mathbf{r}\right)  \mathbf{E}
= -\frac{1}{\tau}\left[ w(t,\mathbf{r}) \frac{\mathbf{u}(t,\mathbf{r})}{v_F^2} + \nu(t,\mathbf{r}) \left( \mathbf{u} (t,\mathbf{r}) \cdot \mathbf{ b} \right) \mathbf{b} \right],\nonumber\\
\end{eqnarray}
where $w(t,\mathbf{r}) = \epsilon(t,\mathbf{r}) +P(t,\mathbf{r}) = 4\epsilon(t,\mathbf{r})/3$, the energy density is
\begin{equation}
\epsilon = \sum_{\lambda=\pm}\sum_{\eta} \int \frac{d^3 k}{(2\pi)^3}  \varepsilon_{\mathbf{k},\lambda} f_{\lambda,\eta}^{(\mathbf{0})}=-\sum_{\lambda=\pm}\sum_{\eta} \frac{\Gamma(4) T^4}{2\pi^2 v_F^3}\text{Li}_{4}\left( - e^{\eta \mu /T}\right)=\frac{ 1}{4 \pi^2 v_F^3}\left( \mu^4+ 2 \pi^2 \mu^2 T^2+\frac{7}{15}\pi^4 T^4 \right),
\end{equation}
and we introduced the shorthand notation
\begin{equation}
\nu= \frac{1}{ \pi^2 v_F^3}\left(\mu^2+ \frac{1}{3} \pi^2 T^2 \right).
\end{equation}

Finally, for the energy continuity equation, we average the Boltzmann equation with the quasiparticle energy $\varepsilon_{\mathbf{k},\lambda}$,
\begin{equation}
\partial_t \epsilon(t,\mathbf{r}) +\frac{4}{3} \bm{\nabla} \cdot\left[\mathbf{u} (t,\mathbf{r}) \epsilon(t,\mathbf{r}) \right] = 0.
\end{equation}

\subsection{Two-band model of Weyl semimetals}

For the two-band model of Weyl semimetals, see Eq.~(\ref{Kubo-Weyl-two-band-H-def}), the Berry curvature is given by
\begin{equation}
\Omega_{i,\eta}=\eta \epsilon_{imn} \frac{\left( \mathbf{d}\left( \mathbf{k}\right) \cdot \left[ \partial_{k_m} \mathbf{d}\left( \mathbf{k}\right) \times \partial_{k_n} \mathbf{d}\left( \mathbf{k}\right) \right] \right) }{4 \left| \mathbf{d}\left( \mathbf{k}\right) \right|^3} .
\end{equation}

To simplify our results, in what follows, we consider the case of vanishing temperature $T=0$. Then the charge continuity equation acquires the following form:
\begin{equation}
\label{App-hydro-two-band-charge}
\partial_t \rho(t,\mathbf{r}) +  \bm{\nabla} \cdot \left[\mathbf{u} (t,\mathbf{r}) \rho (t,\mathbf{r})+\mathbf{j}^{\text{AHE}} \right]=0,
\end{equation}
where
\begin{equation}
\begin{aligned}
\rho &= \sum_{\eta} \int \frac{d^3 k}{(2\pi)^3} \eta e f^{(\mathbf{0})}_{\eta} =  \frac{e\, \text{sign}(\mu)}{15 \pi^2 v_F^2} \left[ \left( 3 \mu ^2+\gamma m |\mu| -2 \gamma ^2 m^2 \right) \sqrt{\frac{ \gamma  m +|\mu| }{\gamma }} \right.\\
&\left.- \Theta \left( \gamma m -|\mu|\right)  \left( 3 \mu ^2-\gamma m |\mu| -2 \gamma ^2 m^2 \right) \sqrt{\frac{\gamma  m-|\mu| }{\gamma }} \right],
\end{aligned}
\end{equation}
\begin{equation}
\label{App_AHE_term}
\begin{aligned}
\mathbf{j}^{\text{AHE}}&=- e^2 \sum_{\eta} \int \frac{d^3 k}{(2\pi)^3} \left[  \mathbf{E} \times  \mathbf{\Omega}_{\eta}  \right] f^{(\mathbf{0})}_{\eta}  = \frac{e^2}{2 \pi^2 } \left( \mathbf{e}_x E_y -\mathbf{e}_y E_x \right)  \Bigg[ \left( \frac{\Lambda_z}{2}-\sqrt{m} \right) \\
&+  \text{sign}\left(\mu\right) \Bigg(\sqrt{m}- \frac{ \gamma  m +|\mu| }{3 |\mu| } \sqrt{\frac{ \gamma  m +|\mu| }{\gamma }} + \Theta \left( \gamma m -|\mu|\right) \frac{\gamma  m-|\mu|}{3 |\mu| }  \sqrt{\frac{\gamma  m-|\mu| }{\gamma }}  \Bigg) \Bigg],
\end{aligned}
\end{equation}
and we used Faraday's law to construct the full derivative. The Euler equation reads
\begin{equation}
\label{App-hydro-two-band-Euler}
\begin{aligned}
&\partial_t \left[ w (t,\mathbf{r}) \frac{\mathbf{u} (t,\mathbf{r})}{v_F^2} + A(t,\mathbf{r}) \left( \mathbf{u} (t,\mathbf{r}) \cdot \mathbf{e}_z \right) \mathbf{e}_z  \right]  + \bm{\nabla} P\left(t,\mathbf{r} \right)  + \rho\left(t,\mathbf{r}\right)  \mathbf{E}  \\
&= -\frac{1}{\tau} \left[ w (t,\mathbf{r}) \frac{\mathbf{u} (t,\mathbf{r})}{v_F^2} + A(t,\mathbf{r}) \left( \mathbf{u} (t,\mathbf{r}) \cdot \mathbf{e}_z \right) \mathbf{e}_z \right],
\end{aligned}
\end{equation}
where $w =\epsilon+P$ and
\begin{equation}
\begin{aligned}
\epsilon &=\sum_{\eta} \int \frac{d^3 k}{(2\pi)^3} \varepsilon_{\mathbf{k}} f_{\eta}^{(\mathbf{0})}
=\frac{1}{105 \pi^2 v_F^2 }  \left[  -16 \gamma ^3 m^{7/2}+ \left(8 \gamma^3 m^3 - 4 \gamma^2 m^2  |\mu|  +3\gamma m \mu^2 +15  |\mu| ^3 \right) \sqrt{\frac{\gamma  m+|\mu| }{\gamma }}  \right.\\
&\left. + \Theta(\gamma m -|\mu| )   \left(8 \gamma^3 m^3 + 4 \gamma^2 m^2  |\mu|  +3\gamma m \mu^2 -15  |\mu| ^3 \right)  \sqrt{\frac{\gamma  m-|\mu| }{\gamma }}\right],
\end{aligned}
\end{equation}
\begin{equation}
\label{App_anisotropic_term}
\begin{aligned}
P &=\sum_{\eta}\int \frac{d^3 k}{(2\pi)^3} \frac{d_i \partial_i (d_i) k_i}{\left| \mathbf{d(\mathbf{k}) }\right|} f_{\eta}^{(\mathbf{0})} = \frac{2}{105 \pi^2 v_F^2 } \left[ 8 \gamma ^3 m^{7/2} -(\gamma  m + |\mu| )^2  (4 \gamma  m - 3 |\mu|)\sqrt{\frac{\gamma  m+|\mu| }{\gamma }} \right.\\
&\left. -\Theta(\gamma m -|\mu| )  (\gamma  m-|\mu| )^2  (4 \gamma  m+ 3 |\mu|)  \sqrt{\frac{\gamma  m-|\mu| }{\gamma }}\right],
\end{aligned}
\end{equation}
\begin{equation}
A =\frac{\gamma^2 m |\mu|}{30 \pi^2v_F^2 }   \left[ \left(\frac{ \gamma  m +|\mu| }{\gamma } \right)^{3/2}  \left(\frac{5}{\gamma^2 m} +4-6\frac{|\mu|}{\gamma m} \right)- \Theta \left( \gamma m -|\mu|\right)  \left( \frac{\gamma  m-|\mu| }{\gamma }\right)^{3/2}  \left(\frac{5}{\gamma^2 m} +4+6\frac{|\mu|}{\gamma m} \right) \right] .
\end{equation}

Finally,  the energy continuity equation reads
\begin{equation}
\label{App-hydro-two-band-energy}
\partial_t \epsilon(t,\mathbf{r}) + \bm{\nabla}\cdot\left[\mathbf{u}(t,\mathbf{r}) w(t,\mathbf{r}) \right] = 0.
\end{equation}

Equations (\ref{App-hydro-two-band-charge}), (\ref{App-hydro-two-band-Euler}), and (\ref{App-hydro-two-band-energy}) correspond to Eqs.~(\ref{kinetic-zero-hydrodynamic-two-band-charge}), (\ref{kinetic-zero-hydrodynamic-two-band-Euler}) and (\ref{kinetic-zero-hydrodynamic-two-band-energy}) in the main text, respectively.

\section{Derivation of viscosity in the kinetic approach}
\label{sec:App-first-kinetic}

To determine the viscosity of electron fluid, one has to consider the first-order correction to the distribution function $f_{\eta} \left( t, \mathbf{r}, \mathbf{k}\right)$ due to the electron-electron scattering. By following the standard approach~\cite{Huang:book,Landau:t10-1995} and using the relaxation-time approximation, we obtain the following first-order in $\tau_{ee} \ll \tau$ correction to the distribution function:
\begin{equation}
\label{App-first-kinetic-df}
\begin{aligned}
\delta f_{\eta}\left( t,\mathbf{r},\mathbf{k}\right)=-\tau_{ee} &\left\{ \partial_t f^{(\mathbf{u})}_{\eta}+\left(\mathbf{v}_{\mathbf{k}}-\eta e\left[  \mathbf{E} \times \mathbf{\Omega}_{\eta}\right] \cdot \bm{\nabla} f^{(\mathbf{0})}_{\eta}\right)- \eta e\left(\mathbf{E} \cdot \bm{\partial}_{\mathbf{k}} f^{(\mathbf{0})}_{\eta} \right) -
\right.\\
&\left.  - \left(\mathbf{v}_{\mathbf{k}}-\eta e\left[  \mathbf{E} \times \mathbf{\Omega}_{\eta}\right] \cdot \bm{\nabla} \left[ \left( \mathbf{u} (t,\mathbf{r}) \cdot \mathbf{k}\right) \partial_{\varepsilon_{\mathbf{k}}} f^{(\mathbf{0})}_{\eta}
\right]  \right)+  \left( \eta e\mathbf{E} \cdot \bm{\partial}_{\mathbf{k}}
\left[ \left( \mathbf{u} (t,\mathbf{r}) \cdot \mathbf{k}\right) \partial_{\varepsilon_{\mathbf{k}}} f^{(\mathbf{0})}_{\eta}
\right]  \right)  \right\}.
\end{aligned}
\end{equation}
Viscosity enters the Euler equation, see Eq.~(\ref{App-hydro-lin-Euler}) or (\ref{App-hydro-two-band-Euler}), as the following term on its left-hand side:
\begin{equation}
-\eta_{ijkl} \nabla_j \nabla_l u_k(t,\mathbf{r}).
\end{equation}
To obtain such a term, we focus only on terms with two spatial derivatives of $\mathbf{u}(t,\mathbf{r})$. Furthermore, we rewrite the time derivative in Eq.~(\ref{App-first-kinetic-df}) as
\begin{equation}
\partial_t f^{(\mathbf{u})}_{\eta} =-\left(\mathbf{k} \cdot\partial_t \mathbf{u}(t,\mathbf{r}) \right) \partial_{\varepsilon_{\mathbf{k}}} f^{(\mathbf{0})}_{\eta} +\left[\partial_t \epsilon(t,\mathbf{r})\right]  \partial_{ \epsilon } f_{\eta} \sim- \bm{\nabla} \cdot\left[\mathbf{u}(t,\mathbf{r}) w(t,\mathbf{r}) \right]  \partial_{\epsilon}f_{\eta}.
\end{equation}
Then, collecting the relevant terms in Eq.~(\ref{App-first-kinetic-df}), we obtain the following part of the distribution function responsible for the electron viscosity:
\begin{equation}
\label{App-first-kinetic-dtf}
\delta \tilde{f}_{\eta} \left( t,\mathbf{r},\mathbf{k}\right)= \tau_{ee} \left\{ \bm{\nabla}\cdot \left[\mathbf{u}(t,\mathbf{r}) w(t,\mathbf{r}) \right] \partial_{\epsilon} f^{(\mathbf{0})}_{\eta} + \left(\mathbf{v}_{\mathbf{k}}\cdot \bm{\nabla} \left[ \mathbf{u} (t,\mathbf{r}) \cdot \mathbf{k}\right] \right) \partial_{\varepsilon_{\mathbf{k}}} f^{(\mathbf{0})}_{\eta} \right\}.
\end{equation}
The subsequent derivations of the hydrodynamic equations follow the same steps as in Sec.~\ref{sec:App-zero-kinetic}. Viscosity-related terms originate from the second term in the kinetic equation (\ref{kinetic-linear}). After performing the averaging over momenta, we have the following integrals:
\begin{eqnarray}
\label{App-first-kinetic-I1}
I_{(\text{visc})}^{(1)} &=& \tau_{ee} \sum_{\eta} \int \frac{d^3 k}{(2\pi)^3} \mathbf{k}\,\left( \mathbf{v}_{\mathbf{k}} \cdot \bm{\nabla}\right) \left\{ \bm{\nabla} \cdot\left[\bm{u}(t,\mathbf{r}) w(t,\mathbf{r}) \right]  \partial_{\epsilon} f^{(\mathbf{0})}_{\eta} \right\},\\
\label{App-first-kinetic-I2}
I_{(\text{visc})}^{(2)} &=& \tau_{ee} \sum_{\eta} \int \frac{d^3 k}{(2\pi)^3} \mathbf{k}\,\left( \mathbf{v}_{\mathbf{k}} \cdot \bm{\nabla}\right) \left[ \mathbf{v}_{\mathbf{k}}  \cdot \bm{\nabla}\left( \mathbf{u} (t,\mathbf{r}) \cdot \mathbf{k}\right) \partial_{\varepsilon_{\mathbf{k}}}f^{(\mathbf{0})}_{\eta}
\right].
\end{eqnarray}
In the following two Sections, we calculate these integrals in
linearized and two-band models of Weyl semimetals.

\subsection{Linearized model of Weyl semimetals}
\label{sec:app-kinetic-lin}

Let us start with the linearized model (\ref{Model-Weyl-lin}). The integrals in Eqs.~(\ref{App-first-kinetic-I1}) and (\ref{App-first-kinetic-I2}) are
\begin{equation}
\begin{aligned}
I_{(\text{visc})} &=I_{(\text{visc})}^{(1)}+I_{(\text{visc})}^{(2)} =\tau_{ee} \frac{4}{9} \bm{\nabla} \left(\bm{\nabla}\cdot \left[\mathbf{u}(t,\mathbf{r}) \epsilon(t,\mathbf{r}) \right]\right) -\tau_{ee} \frac{4}{15} \Delta \left[\mathbf{u}(t,\mathbf{r}) \epsilon(t,\mathbf{r}) \right] \\
&-\tau_{ee} \frac{8}{15} \bm{\nabla} \left(\bm{\nabla}\cdot \left[\mathbf{u}(t,\mathbf{r}) \epsilon(t,\mathbf{r}) \right]\right) - \tau_{ee} \mathbf{b} \frac{1}{3} \Delta \left[\left(\mathbf{u}(t,\mathbf{r}) \cdot \mathbf{b}\right)  v_F^2 \nu(t,\mathbf{r})\right].
\end{aligned}
\end{equation}
Therefore, linearizing in deviations from local equilibrium $\epsilon(t,\mathbf{r}) =\epsilon$ and $\nu(t,\mathbf{r})=\nu$, we obtain
\begin{equation}
\label{kinetic-viscosity-lin}
\eta_{ijkl}=\frac{4\tau_{ee}}{15} \epsilon \left(\delta_{ik}\delta_{jl} + \delta_{il}\delta_{jk} -\frac{2}{3} \delta_{ij}\delta_{kl}\right)+  \frac{\tau_{ee} v_F^2\nu}{3} b_ib_k\delta_{jl}.
\end{equation}

\subsection{Two-band model of Weyl semimetals}

Let us consider the two-band model of Weyl semimetals defined in Eq.~(\ref{Kubo-Weyl-two-band-H-def}). Leaving only terms with the second-order derivatives of $\mathbf{u}$, integrals (\ref{App-first-kinetic-I1}) and (\ref{App-first-kinetic-I2}) read
\begin{eqnarray}
\label{App-visc-two-band-int-I1}
I_{(\text{visc})}^{(1)}&=& \tau_{ee} w \frac{\partial^2 u_i}{\partial x_i \partial x_j} \sum_{\eta}  \frac{\partial  }{\partial \epsilon} \int \frac{d^3 k}{(2\pi)^3} \mathbf{k} v_{j}  f^{(\mathbf{0})}_{\eta} =\tau_{ee} w \bm{\nabla} \left(\bm{\nabla}\cdot \mathbf{u}\right) \partial_{\epsilon} P,\\
\label{App-visc-two-band-int-I2}
I_{(\text{visc})}^{(2)}&=& - \mathbf{e}_i \frac{\partial^2 u_k}{\partial x_j \partial x_l}  \tau_{ee} \sum_{\eta} \int \frac{d^3 k}{(2\pi)^3}  k_i\, v_{j}   k_k v_{l}  \delta\left( |\mu| -\left| \mathbf{d} (\mathbf{k} \right| \right) \nonumber \\
&=& - \mathbf{e}_i \frac{\partial^2 u_k}{\partial x_j \partial x_l}  \tau_{ee} \sum_{\eta}  \int \frac{d^3 k}{(2\pi)^3}  k_i\,  \frac{d_j \partial_j (d_j)}{\left| \mathbf{d(\mathbf{k}) }\right|}     k_k  \frac{d_l \partial_l (d_l) }{\left| \mathbf{d(\mathbf{k}) }\right|} \delta\left( |\mu| -\left| \mathbf{d} (\mathbf{k}) \right| \right).
\end{eqnarray}
Evaluating the integral in Eq.~(\ref{App-visc-two-band-int-I2}), we obtain
\begin{equation}
\begin{aligned}
I_{(\text{visc})}^{(2)}&=- \mathbf{e}_i \frac{\partial^2 u_k}{\partial x_j \partial x_l}  \frac{\tau_{ee}}{2 \pi^2 v_F^2 |\mu|}\Bigg\{ \left[\delta_{il} \delta_{jk} +\delta_{ij} \delta_{kl} +\left( \delta_{i1}+ \delta_{i2} \right)\left( \delta_{j1}+ \delta_{j2} \right) \delta_{ik} \delta_{jl} \right] \frac{4}{315} \Bigg[  \left( 4\gamma^2 m^2 -10 \gamma m |\mu| +7 \mu^2 \right)  \\
&\times \left(\gamma m +|\mu|\right)^2 \sqrt{\frac{\gamma m +|\mu|}{\gamma}}  -\Theta \left( \gamma m -|\mu|\right)  \left( 4\gamma^2 m^2 +10 \gamma m |\mu| +7 \mu^2 \right) \left(\gamma m -|\mu|\right)^2 \sqrt{\frac{\gamma m -|\mu|}{\gamma}} \Bigg]\\
&+ \delta_{i3} \delta_{j3} \delta_{ik} \delta_{jl}  \frac{4 \mu^2}{15} \left[ \left(\gamma m +|\mu|\right)^2  \sqrt{\frac{\gamma m +|\mu|}{\gamma}}  -\Theta \left( \gamma m -|\mu|\right) \left(\gamma m -|\mu|\right)^2 \sqrt{\frac{\gamma m -|\mu|}{\gamma}} \right]\\
&- \delta_{i3} \left(\delta_{j1} +\delta_{j2 }\right) \delta_{ik} \delta_{jl}  \frac{2 v_F^2}{105 \gamma} \left[ \left(\gamma m +|\mu|\right)^2 \left(2 \gamma m -5|\mu| \right)  \sqrt{\frac{\gamma m +|\mu|}{\gamma}} -\Theta \left( \gamma m -|\mu|\right) \left(\gamma m -|\mu|\right)^2 \right. \\
&\left.\times  \left(2 \gamma m + 5|\mu| \right)  \sqrt{\frac{\gamma m -|\mu|}{\gamma}} \right]- \left( \delta_{i1} +\delta_{i2} \right) \delta_{j3} \delta_{ik} \delta_{jl}  \frac{8 \gamma}{3465 v_F^2} \left[ \left(\gamma m +|\mu|\right)^2  \sqrt{\frac{\gamma m +|\mu|}{\gamma}}  \right. \\
&\left. \times \left( 32 \gamma^3 m^3 -80 \gamma^2 m^2 |\mu| +74 \gamma m \mu^2-45 |\mu|^3 \right)  -\Theta \left( \gamma m -|\mu|\right) \left(\gamma m -|\mu|\right)^2  \right. \\
&\left. \times  \sqrt{\frac{\gamma m -|\mu|}{\gamma}}  \left(  32 \gamma^3 m^3 +80 \gamma^2 m^2 |\mu| +74 \gamma m \mu^2+45 |\mu|^3 \right)  \right] \Bigg\}.
\end{aligned}
\end{equation}

The final result for the viscosity tensor in the two-band model is given by
\begin{equation}
\label{kinetic-viscosity-two-band}
\begin{aligned}
\eta_{ijkl}&= \frac{\tau_{ee}}{\pi^2 v_F^2 |\mu|}\Bigg\{ \left[\delta_{il} \delta_{jk} +\delta_{ij} \delta_{kl} +\left( \delta_{i1}+ \delta_{i2} \right)\left( \delta_{j1}+ \delta_{j2} \right) \delta_{ik} \delta_{jl} \right] \frac{2}{315} \Bigg[  \left( 4\gamma^2 m^2 -10 \gamma m |\mu| +7 \mu^2 \right) \\
&\times \left(\gamma m +|\mu|\right)^2 \sqrt{\frac{\gamma m +|\mu|}{\gamma}}  -\Theta \left( \gamma m -|\mu|\right)  \left( 4\gamma^2 m^2 +10 \gamma m |\mu| +7 \mu^2 \right) \left(\gamma m -|\mu|\right)^2 \sqrt{\frac{\gamma m -|\mu|}{\gamma}} \Bigg]\\
&+ \delta_{i3} \delta_{j3} \delta_{ik} \delta_{jl}  \frac{2 \mu^2}{15} \left[\left(\gamma m +|\mu|\right)^2  \sqrt{\frac{\gamma m +|\mu|}{\gamma}}  -\Theta \left( \gamma m -|\mu|\right) \left(\gamma m -|\mu|\right)^2 \sqrt{\frac{\gamma m -|\mu|}{\gamma}} \right] \\
&- \delta_{i3} \left(\delta_{j1} +\delta_{j2 }\right) \delta_{ik} \delta_{jl}  \frac{ v_F^2}{105 \gamma} \left[ \left(\gamma m +|\mu|\right)^2 \left(2 \gamma m -5|\mu| \right)  \sqrt{\frac{\gamma m +|\mu|}{\gamma}} -\Theta \left( \gamma m -|\mu|\right) \left(\gamma m -|\mu|\right)^2 \right. \\
&\left.\times  \left(2 \gamma m + 5|\mu| \right)  \sqrt{\frac{\gamma m -|\mu|}{\gamma}} \right]- \left( \delta_{i1} +\delta_{i2} \right) \delta_{j3} \delta_{ik} \delta_{jl}  \frac{4 \gamma}{3465 v_F^2} \left[ \left(\gamma m +|\mu|\right)^2  \sqrt{\frac{\gamma m +|\mu|}{\gamma}}  \right. \\
&\left. \times \left( 32 \gamma^3 m^3 -80 \gamma^2 m^2 |\mu| +74 \gamma m \mu^2-45 |\mu|^3 \right)  -\Theta \left( \gamma m -|\mu|\right) \left(\gamma m -|\mu|\right)^2  \right.\\
&\left. \times  \sqrt{\frac{\gamma m -|\mu|}{\gamma}}  \left(  32 \gamma^3 m^3 +80 \gamma^2 m^2 |\mu| +74 \gamma m \mu^2+45 |\mu|^3 \right)    \right] \Bigg\}- \tau_{ee} \delta_{ij} \delta_{kl}  P \left(1+ \partial_{\epsilon}P \right).
\end{aligned}
\end{equation}

\end{widetext}

\bibliography{library-short}

\end{document}